\documentclass{aa}
\usepackage{graphicx}
\usepackage{txfonts}
\usepackage{aalongtable}
\usepackage{natbib}

\def\la{\raise.5ex\hbox{$&      \\lt;$}\kern-.8em\lower 1mm\hbox{$\sim$}}
\def\ma{\raise.5ex\hbox{$&      \\gt;$}\kern-.8em\lower 1mm\hbox{$\sim$}}

\def\kms{$\rm km\, s^{-1}$}
\def\cm3{$\rm cm^{-3}$}

\def\n0{$\rm n_{0}$}	
\def\B0{$\rm B_{0}$}

\def\L12{L$_{12\mu m}$~}
\def\F12{F$_{12\mu m}$~}

\def\fe2{[\ion{Fe}{ii}]}
\def\h2{H$_{2}$}
\def\pp{$\pm$}
\def\ax{$\rm ^a$}
\def\cc{$\rm ^c$}
\def\dd{$\rm ^d$}
\def\ee{$\rm ^e$}
\def\ff{$\rm ^f$}

\def\hh{$\rm ^h$}
\def\kk{$\rm ^k$}
\def\lb{$\lbrack$}
\def\rb{$\rbrack$}
\def\bl{$^{\xi}$}

\begin{document}
   \title{A 0.8-2.4 $\mu$m spectral atlas of active galactic nuclei}

 \titlerunning{NIR Spectral Atlas of AGN}
   \author{R. Riffel \inst{1}, A. Rodr\'{\i}guez-Ardila\inst{2}, 
\thanks{ Visiting Astronomer at the Infrared   Telescope Facility, which is operated by the University of Hawaii under Cooperative Agreement no. NCC 5-538 with the National Aeronautics and Space Administration, Science Mission Directorate, Planetary Astronomy Program.}
 \and M. G. Pastoriza\inst{1}  }

   \offprints{riffel@ufrgs.br}

   \institute{Departamento de Astronomia, Universidade Federal do Rio Grande do Sul. 
              Av. Bento Gon\c calves 9500, Porto Alegre, RS, Brazil.\\
              \email{riffel@ufrgs.br}\\
	      \email{miriani.pastoriza@ufrgs.br}
         \and
              Laborat\'{o}rio Nacional de Astrof\'{i}sica - Rua dos Estados Unidos 154,
Bairro das Na\c{c}\~{o}es .
CEP 37504-364, Itajub\'{a}, MG, Brazil\\
             \email{aardila@lna.br}
             }


   \abstract
   {}
   {We present a near-infrared spectral atlas of 47 active galactic nuclei (AGN) of all degrees of activity in the 
   wavelength interval of 0.8-2.4 $\mu$m, including the fluxes of the observed emission lines. We analyze the 
   spectroscopic properties of the continuum and emission line spectra of the sources.}
   { In order to exclude aperture and seeing effects we used near-infrared spectroscopy 
    in the short cross-dispersed mode (SXD, 0.8$-$2.4~$\mu$m), taking the {\it JHK}-bands spectra simultaneously.}
   {We present the most extensive NIR spectral atlas of AGN to date. This atlas offers a suitable database for 
   studying the continuum and line emission properties of these objects in a region full of interesting features.
   The shape of the continuum of QSOs and 
    Sy~1's are similar, being essentially flat in the H and K bands, while a strong variation is found in the J band. 
    In Seyfert 2 galaxies,  the continuum in the F$\lambda\, \times\, \lambda$ 
    space smoothly decreases in flux from 1.2$\mu$m redwards in almost all sources. In J, it smoothly  rises 
    bluewards in some sources, while in others a small decrease in flux is observed. 
   The spectra are dominated by strong emission features of \ion{H}{i}, 
   \ion{He}{i}, \ion{He}{ii}, [\ion{S}{iii}] and by conspicuous forbidden 
   lines of low and high ionization species. Molecular lines of \h2\
   are common features of most objects. The absence of \ion{O}{i} and \ion{Fe}{ii}
   lines in Seyfert~2 galaxies and the smaller FWHM of these lines relative to 
   that of \ion{H}{i} in the Seyfert~1 give observational support to the fact that they 
   are formed in the outermost portion of the broad-line region. The[\ion{P}{ii}] and coronal 
   lines are detected for all degrees of activity. The \fe2 12570\AA /16436\AA\ line ratio becomes 
   a reliable reddening indicator for the narrow-line region of Seyfert galaxies.}
   {}
   \keywords{near-infrared atlas -- AGN -- emission lines }

   \maketitle
%

\section{Introduction}

From the spectroscopic point of view, active galactic nuclei (AGNs)
have been poorly studied in the near-infrared (NIR) spectral region, particularly
in the interval between 1~$\mu$m and 2.4~$\mu$m. This region has
been systematically absent in most surveys mainly because it does not 
fall within the spectral coverage of optical CCD detectors or 
infrared satellites (i.e., ISO, Spitzer). As a result, very little is 
known about the spectroscopic properties of AGNs in a transition zone that 
contains interesting features, in both the continuum and emission
lines that can help to put firm constraints on the physical 
properties of the nuclear emitting gas and its environment.
    
With the new generation of IR arrays and their improved sensitivity, 
it is now possible to carry out spectroscopy at moderate resolution
on faint and extended targets, such as galaxies and quasars.
In addition, with the availability of 
cross-dispersed spectrographs offering simultaneous wavelength coverage
in the interval 0.8$-$2.4~$\mu$m, 
it is now possible to study the NIR region avoiding the aperture and seeing effects 
that usually affect {\it JHK} spectroscopy done in long-slit mode and
single-band observations.
  
There is manifold interest in the NIR range. At $\sim$1.1~$\mu$m
({\it J}-band), the nuclear continuum emission that dominates the UV 
and optical spectral energy distribution of quasars and Seyfert~1 galaxies 
no longer dominates \citep{bar87,kab05}. At the same time, 
reprocessed nuclear emission by hot dust starts becoming an important 
source of continuum emission, mainly from the {\it K}-band and longer
wavelengths \citep{bar87,rom06,gli06}. Moreover, because the NIR is less 
affected by extinction than the optical, the detection of highly reddened 
objects with buried AGN activity, usually associated to starburst and 
ultra-luminous infrared galaxies, increases. A better understanding of 
the AGN-starburst connection can then be made. Last but not least,
NIR spectroscopy on AGNs of the local universe allows the construction
of spectral templates to study the commonest features and the
physical processes that originate them. These templates, in turn, 
are essential for understands the true nature of high-redshift objects 
discovered using Spitzer, for instance. In this sense, \citet{gli06}
recently published an NIR template for AGNs, made
from observations of 27 quasars in the redshift range 
0.118 $<$ {\it z} $<$ 0.418. They studied the emission lines in 
that region, revealing the Pashen series lines, as well as oxygen
helium and forbidden sulfur emission. 
 
With the above in mind, here we present the most extensive
spectroscopic atlas in the 8000~$\AA -$24000~\AA\ region to date 
for a sample composed of 47 AGNs in the redshift range 
0.0038 $<$ {\it z} $<$ 0.549. It is aimed at constructing a 
homogeneous database for these objects at good S/N and spectral
resolution, allowing the study of 
the continuum and line emission properties of the individual
sources and the comparison of these properties among the different types of AGN. 
Moreover, most of the sources 
have no previous spectroscopic information in the literature
covering the whole NIR interval. Therefore, this atlas is also intended 
to fill the existing gap in the SED observations of known sources and at the
same time to increase the number of spectral features common to AGN that can be 
used to put additional constraints on the modelling of the physical 
properties of the nuclear gas emission. 
  
This paper is structured as follows. In Sect.~\ref{s1} we describe the
sample selection, observations and data reduction process. In Sect.~\ref{s2}
we present the results.  Comments about the main features found in the
spectra are in Sect.~\ref{indiv}. The final remarks are presented in 
Sect.~\ref{s4}. Throughout the text, a Hubble constant of 75~\kms ~Mpc$^-1$
will be employed. 


\section{Observations}\label{s1}
\subsection{Sample selection}

The 47 AGNs that compose our sample are divided into 7 quasars,
13 narrow-line Seyfert~1 galaxies, 12 classical Seyfert 1s, and 15 
Seyfert~2s. Note that the above classification was based on 
published optical spectroscopy of these sources made by different 
authors. In addition, 4 starburst galaxies were included for comparison
purposes, giving a total of 51 spectra available. The dominance
of Type~1 objects is not by chance. Originally, 
we were aimed to select Type~1 objects because
most NIR spectroscopy published previously was done on samples
dominated by Seyfert~2 galaxies/LINERS \citep{gvh94,vgh97,sosa01} and very little was
known about the NIR spectra of Type~1 sources, except probably for those
works on some individual sources and for the recent NIR
spectroscopy on quasars \citep{gli06}. Moreover,
to avoid the effects of strong blending produced by the broad
components of the permitted lines that could mask or dilute weak
emission lines, emphasis was 
given to some narrow-line Seyfert 1 galaxies (NLS1). This sub-sample
was selected on the basis of their singular
behavior in the ultraviolet and/or soft X-ray energy
bands. The list of \citet{bbf96} was used to this purpose.  
We then increased our sample with classical 
Seyfert~1 and~2 galaxies. The selection of these
objects was based on the CfA sample \citep{hb92}. 
Finally, our list of objects was complemented with 
quasars selected from the Palomar Bright quasar survey
(PG) of \citet{sg83}.  

The main criterion in the selection of the final sample was to 
include, as much as possible, well-known studied
sources in the optical/UV and X-ray regions that would allow us to establish
correlations between the NIR emission and that in other wavelength
intervals. Other criteria, such as the $K$-band magnitude, 
limited to $K<$12, was also applied in order to keep the
exposure time under reasonable values to reach S/N$>$50 in the continuum
emission in that band. After compiling a list of 102 AGN that 
matched the above conditions, objects that have a declination $<$-35$^{\rm o}$
or were already extensively studied in the NIR region, were cut out from
the list. The final output was a list of 48 AGNs, plus the additional
three starburst galaxies, included for comparison purposes.

Based on the above, although our sample of AGNs is not
complete in any sense, we consider it as representative of the 
class of AGNs in the local universe (most sources have {\it z}$<$0.1),
because it is composed of well-known and studied objects in other 
wavelength bands. Note that most of the targets 
have already been studied in the NIR by imaging techniques.

Columns~2 and~4 of Table~\ref{obslogs} list the final sample of objects and
the corresponding redshift, respectively. The latter value was taken 
from the NED database and confirmed by the position of the most intense
lines in the individual spectra. Errors of less than 1\% were found
between our redshift determination and that published in the NED.

\subsection{Observations and data reduction}

The NIR spectra were obtained at the
NASA 3\,m Infrared Telescope Facility (IRTF) from April/2002 to June/2004.
The SpeX spectrograph \citep{ray03}, was used in the short cross-dispersed mode 
(SXD, 0.8$-$2.4~$\mu$m). A complete journal of observations is in 
Table~\ref{obslogs}. The galaxies are listed in order of right ascension.
In all cases, the detector employed consisted of a 1024$\times$1024
ALADDIN 3 InSb array with a spatial scale of 0.15''/pixel. A 0.8''$\times$15'' slit was employed 
giving a spectral resolution of 360 \kms. This value
was determined both from the arc lamp spectra and the sky line spectra and was found 
to be constant with wavelength within 3\%. During the different
nights, the seeing varied between 0.7''$-$1''. Observations were done nodding in an ABBA 
source pattern with typical integration times from 120\,s to 180\,s per frame and total 
on-source integration times between 35 and 50 minutes.  Some sources
were observed on multiple nights. In these cases, these data
were combined, after reduction, to form a single
spectrum. During the observations, an A0\,V star was observed near each target to
provide a telluric standard at similar airmass. It was also used to flux calibrate
the corresponding object.

The spectral reduction, extraction and wavelength  calibration procedures were
performed using SPEXTOOL, the in-house software developed and provided
by the SpeX team for the IRTF community \citep{cvr04}\footnote{SPEXTOOL is available from the
IRTF web site at http://irtf.ifa.hawaii.edu/Facility/spex/spex.html}.
No effort was made to extract spectra at positions
different from the nuclear region even though some objects show
evidence of extended emission. 

The 1-D spectra were then corrected for telluric absorption and flux calibrated
using Xtellcor \citep{vcr03}, another in-house software developed by the IRTF team. 
Finally, the different orders of each galaxy spectrum were merged to 
form a single 1-D frame. It was later corrected for redshift, 
determined from the
average $z$ measured from the positions of [S\,{\sc iii}]
0.953$\mu$m, Pa$\delta$, He\,{\sc i}~1.083$\mu$m, Pa$\beta$, and 
Br$\gamma$.  A Galactic extinction
correction as determined from the {\it COBE/IRAS} infrared maps of
\citet{sch98} was applied. The value of the
Galactic {\it E(B-V)} used for each galaxy is listed in Col.~5
of Table~\ref{obslogs}. Final reduced spectra in laboratory wavelengths, in the intervals
0.8$-$1.35~$\mu$m (left panels), 1.35$-$1.8~$\mu$m (middle panels),
and 1.8$-$2.4~$\mu$m (right panels), are plotted in Figs.~\ref{indiv1}
to~\ref{indiv8}. Because the blue region of SpeX includes the wavelength interval
0.8$\mu$m--1.03$\mu$m, which does not belong to the standard {\it J}-band,
in the rest of this text we will refer to that region as the {\it z}-band,
following the SpeX naming convention of the different orders.  

\section{Results}\label{s2}

\subsection{Final Reduced Spectra}
Final reduced spectra are presented from Figs.~\ref{indiv1} to \ref{indiv8}, 
sorted in  the order of increasing right ascension. For each galaxy, the left panel 
displays the $z$+$J$ bands, the middle panel the $H$ band, and the right panel 
the $K$ band. 
The abscissa represents the monochromatic flux in units of 
$\rm 10^{-15}\, erg \, cm^{-2} \, s^{-1}$~\AA$^{-1}$. For reference, we marked 
(dotted lines) the brightest emission lines, usually \lb\ion{S}{iii}] 9531\,$\AA$, Pa$\delta$, 
\ion{He}{i} 10830\,$\AA$, [\ion{P}{ii}] 11886\,$\AA$, \fe2\ 12570\,$\AA$, Pa$\beta$ 
(left panel), [\ion{Si}{x} 14300\,$\AA$, \fe2\ 16436\,$\AA$ (middle panel), Pa $\alpha$ 
\h2\ 19570\,$\AA$, \h2\ 21213\,$\AA$, and Br$\gamma$ (right panel).
The two high redshift sources Ton\,0156 and 3C\,351 were drawn in a separate 
panel (Fig.~\ref{indiv8}) because the blue edge of their spectra starts at $\sim$5800~\AA, 
in laboratory wavelengths.

Emission line fluxes for each object of the sample were measured by fitting a Gaussian 
function to the observed profile and then 
integrating the flux under the curve. The LINER software \citep{pow93} was used
to this purpose. The results are listed in Tables~\ref{flsy1_1} to \ref{flsy2}. 
We consider 3$\sigma$ level errors. For the large 
majority of our targets, these measurements represent the most complete lists of NIR 
fluxes made up to date in AGNs. The line fluxes of Mrk\,1210 are already reported by
\citet{mza05} and that of Mrk\,766 in \citet{rcv05}. 

\subsection{The continuum spectra}\label{s3}

The NIR spectra of AGN have been studied mostly via broad-band photometry. 
One of the most important results reported is that the continuum shape is correlated 
with the Seyfert type, in the sense that flatter spectral energy distributions (SEDs) 
tend to be found in Sy~1's and steeper ones in Sy~2's, in accordance with the unified
model \citep[e.g][and references therein]{alh03,alh01}. 
However, no systematic study of the continuum characteristics in a representative
sample of AGN have been made yet by means of spectroscopy. Previous works, 
concentrated on individual or on a small sample of objects, report a  continuum described well by a 
broken power-law, with a flattening of the continuum slope at $\sim$1.1$\mu$m 
\citep{th095,tho96,rudy00,rudy01,ara02}. 
It means that there would be a minimum in the continuum emission
around 1.1$\mu$m, probably associated to the red end of the optical power-law
distribution associated to the central engine and the onset of the emission due to
reprocessed nuclear radiation by dust \citep{bar87,rudy00}.  \citet{boisson02}, in a {\it H}-band 
spectroscopic study of 5 AGN, report that the Sy~2 nuclear spectra are dominated by stars, while evidence 
for dilution of the nuclear stellar components by hot dust and/or power-law AGN are found in Sy~1. 

From what is said above, the main goal of this section is to characterize  
the NIR continuum observed in our sample and compare it to the different 
types of AGNs and to other data in the literature. To this 
purpose we normalized to unity the continuum emission of all spectra at
$\lambda$\,12230\,$\AA$, except for quasars, where the normalization was done at 
$\lambda$\,11800\,$\AA$.  The region around these two positions are free
of emission lines. The normalization point for quasars is different than for
the other objects because in the former, after the spectra are converted to
rest frame, the first position falls in a region of bad atmospheric transmission. 
In order to help in the comparison, we grouped the
spectra according to the type of nuclear activity. For each type
of AGN, the data were sorted according to the spectral shape, 
from the bluest ones (top) to the reddest ones (bottom). These plots are shown 
from Figs.~\ref{plsy1} to \ref{plsb}. 

Overall, it is easy to see from the normalized spectra (Figs.~\ref{plnls} to \ref{plsb}) 
that the continuum shape of quasars, NLS1s and Sy~1's are rather 
similar in the {\it H} and {\it K}-bands, where it is essentially flat or
decreases smoothly in flux with wavelength. In contrast, in the $z$+$J$ bands, 
the continuum shape varies from that which remains nearly flat, as in Mrk~334 
and Mrk~124, to that displaying a strong blue continuum from 1.2$\mu$m bluewards, 
as is the case for most quasars, such as Mrk~509 and NGC~4151. 
In most cases, it seems to be a break in the continuum
form at $\sim$1.1$\mu$m.  At first sight, when looking at the nearby sources, 
one is tempted to state that the blue NIR excess is very similar in 
form and strength to the so called small blue bump (SBB) that is usually 
observed from 4000~\AA\ bluewards in the optical spectra of Seyfert~1 galaxies 
and quasars. The SBB, modelled in detailed by \citet{wnw85}, was described in 
terms of \ion{Fe}{ii}, \ion{Mg}{ii}, and high-order Balmer lines and the 
Balmer continuum. However, the blue end of the NIR region does not contain that large number of \ion{Fe}{ii} 
emission features, as in the optical, able to create an excess of emission 
over the underlying continuum. Likely, the Pashen continuum and high-order 
Pashen lines can contribute to the NIR bump.  

The quasars Ton~156 and 3C~351, which at rest wavelengths include a large
portion of the optical region in the {\it z}-band, provide us with important 
clues for studying the actual shape and extension of the blue NIR excess. Clearly, the 
continuum emission in these two high-redshift sources decreases steadily
in flux with wavelength with no appreciable change in the steepness from 
the optical up to $\sim$1.2$\mu$m, where a clear break in the continuum
shape is observed, then becoming flat redwards. This situation can be easily
extended to the nearby sources, as the location of the turnover point is 
rather similar in most objects, being the steepness in the blue 
continuum the only difference among the different objects. This confirms that the break
at $\sim$1.1$\mu$m is a common characteristic of Type~1 sources. What
we have called the ``blue NIR excess'' could simply be the red end of the
continuation of the power-law optical continuum typical of Type~1
sources for the NIR. Our results agrees with the findings of \citet{gli06},
who report a broken power-law function in the interval 0.57$\mu$m--2.23$\mu$m
with the breaking point at 1.085$\mu$m, to describe the continuum of
a NIR composite quasar spectrum. We should add here that Type~1 sources
display a nearly featureless continuum in the NIR, with only a few sources
showing absorption features. In only a few sources, the 2.3$\mu$m CO bandheads
are relatively prominent. Arp~102B and NGC~1097, which were classified as
Seyfert~1s because of their broad double-peaked Balmer lines, are among the
Type~1 sources with conspicuous stellar features. CO absorption lines are
also seen in the {\it H}-band, with equivalent widths of just
a few Angstroms. 

The exception to all the trends for Seyfert~1 galaxies mentioned above is 
Mrk\,1239, whose continuum emission is outstanding because it is dominated 
by a strong bump of emission peaking at 2.2~$\mu$m, with a strength not 
reported before in an AGN. In this object, 
the continuum does not becomes flatter at 1.1$\mu$m, as in most Seyfert~1s but rather steep,
reaching a maximum of emission at 2.2~$\mu$m and then declining again in flux
with wavelength. This extreme case was the subject of a separate publication 
by \citet{rom06}. They found that a blackbody of T$\sim$1200~K was needed to
account for the strong excess of emission over a featureless continuum
wich a power-law form. The blackbody component was interpreted in terms of very
hot dust ($T_{\rm d}$=1200~K) near its sublimation temperature, very likely 
located both in the upper layers of torus and close to the apex 
of a hypothetical polar scattering region in this object. It is worth mentioning
that Mrk~766 and Mrk~478 display an emission bump similar in form to
that of Mrk~1239, although wich much lower intensity.

In contrast to Seyfert~1 galaxies, none of our Seyfert 2s display the blue
rise of the continuum shortward of 1.1~$\mu$m. Moreover, all objects show
prominent absorption lines and bands in {\it H} and {\it K}. Indeed,
the 2.3~$\mu$m CO bandheads are present in all sources but NGC~1275 and NGC~262. 
In {\it J}, most Seyfert 2s display  an absorption band at 1.1~$\mu$m, not
reported before in AGNs, and these we tentatively associated with CN \citep{mar05}.
According to \citet{mar05}, that band, prominent in the NIR region,
is indicative of thermal-pulsing AGB stars with ages $\sim$1~Gyr.
The association of young stellar population and the CN feature can be
strengthened if we consider that the three starburst galaxies of the
sample display this absorption (see below). The contribution of stellar
population to the observed continuum is further supported by the detection of
CaII triplet absorption features in the large majority of these objects.
  
Overall, the continuum emission of Type~2 objects can be divided into two groups 
based on its shape: one that decreases in flux with wavelength across the NIR
and that can in a first approach be approximated by a power-law function.
Twelve out of 16 Seyfert~2s belong to this category. Another is dominated by a red
continuum, with the flux increasing with wavelength up to 1.2~$\mu$m.
From that point redwards, the flux decreases with wavelength. Two
objects, namely Mrk~1066 and NGC~2110, share these characteristics.
At this point we should comment on the continuum in NGC~7674, which does not
fit in any of the above two categories. From 0.8~$\mu$m up to $\sim$1.4~$\mu$m,
the continuum decreases in flux with wavelength as in most Seyferts~2s. 
In {\it H} and {\it K}, however, it displays a clear excess of emission, similar 
to that reported for Mrk~1239. It should also be noted that, although NGC~7674
is classified as a Seyfert 2 from its optical spectrum, in the NIR
region it displays broad emission components in the permitted lines, similar to
what is observed in classical Seyfert 1s.    

Finally, the continuum emission of the Starburst galaxies, from 1.3~$\mu$m
redwards, is rather similar for the three objects analyzed, decreasing
smoothly in flux with wavelength. For NGC~3310 and NGC~7714, this
same behavior is found in the blue portion of the spectrum. No upturns
or breaks are found in the NIR. In contrast, the continuum in NGC~1614,  is 
strongly reddened in the interval 0.8~$\mu$m--1.2~$\mu$m,
becoming flat in the region between 1.2~$\mu$m--1.3~$\mu$m. Also,
this source displays the most prominent absorption lines of the three
galaxies. The CN absorption feature at 1.1~$\mu$m is also conspicuous in
the three objects. The detection  of this feature in the spectra of 
Seyfert~2 galaxies that display prominent circumnuclear starburst 
activity, such as  Mrk~1066, suggest that it can be a useful
tracer of young stellar populations.

We conclude this section by noting that the continuum in the NIR displays
significant differences between Type~1 and Type~2 sources. In the former,
the continuum can be characterized by a broken power-law, with the break
located almost invariably at $\sim$1.1~$\mu$m. Shortwards to the break,
the continuum is blue, and its steepness can be associated to the
spectral index of the power-law that dominates the optical continuum 
emission. Redwards, the continuum is rather flat or else displays a smooth
decrease in flux with wavelength. Overall, the composite power-law
continuum is featureless, although absorption lines can be identified 
in some sources. The continuum emission of Type~2 sources, on the other 
hand, can be grouped into two classes: one that follows a single
power-law function across the NIR and another displaying
a red spectrum bluewards of 1.2~$\mu$m and then decreasing
steeply in flux with wavelength. The objects in the latter category
display prominent absorption bands of CO and CN. They likely
are dominated by circumnuclear starburst activity as told from the
similarity with the spectra of genuine starburst galaxies. A quantitative 
approach of the analysis of the continuum emission is beyond the scope of 
this paper, but is left for a future publication.

\subsection{The NIR emission line spectrum}

The 51 NIR spectra presented in this work offer a prime opportunity for identify
the most common emission features found in AGNs in a region 
not yet observed in such details. For completeness, the emission
line fluxes of these lines, listed in Tables~\ref{flsy1_1} to~\ref{flsy2}, 
form the largest and most complete database in the
interval 0.8$\mu$m--2.4$\mu$m published so far for these objects.
 
From our data, it is easy to see that, independent of the Seyfert class,
NIR AGN spectra are dominated by strong emission features 
of \ion{H}{i}, \ion{He}{i}, \ion{He}{ii}, and [\ion{S}{iii}]. Moreover,
conspicuous forbidden low-ionization lines of ions such as [\ion{Fe}{ii}], [\ion{S}{ii}],
and [\ion{C}{i}], as well as molecular H$_{2}$ lines are detected in the large 
majority of objects. Also detected in an important fraction of
the targets are coronal lines of [\ion{S}{viii}], [\ion{S}{ix}], 
[\ion{Si}{vi}], [\ion{Si}{x}], and [\ion{Ca}{viii}]. This set of lines need ionization energies of up 
to 360~eV for the production of the parent ion. Their detection
is considered an unambiguous signature of nuclear activity, and it increases
the number of coronal line species available to study the origin, location,
and physical conditions of the gas that emits them.
Overall, the fluxes listed in Tables~\ref{flsy1_1} to~\ref{flsy2} can be
use to add firm constraints to model the physical state of the emission gas, 
both from the broad line and narrow line regions.

In this section we will describe the commonest NIR emission lines detected 
in the galaxy sample according to the Seyfert type. To start with, and
summarizing what is said in the paragraph above,
Figure~\ref{hist_blg} shows the frequency with which the most important NIR 
emission lines appear in the different spectra. A detailed discussion of 
the main spectral characteristics observed in each 
source can be found in Sec.~\ref{indiv}.
\setcounter{figure}{13}
\begin{figure}
\centering
\includegraphics[width=9cm]{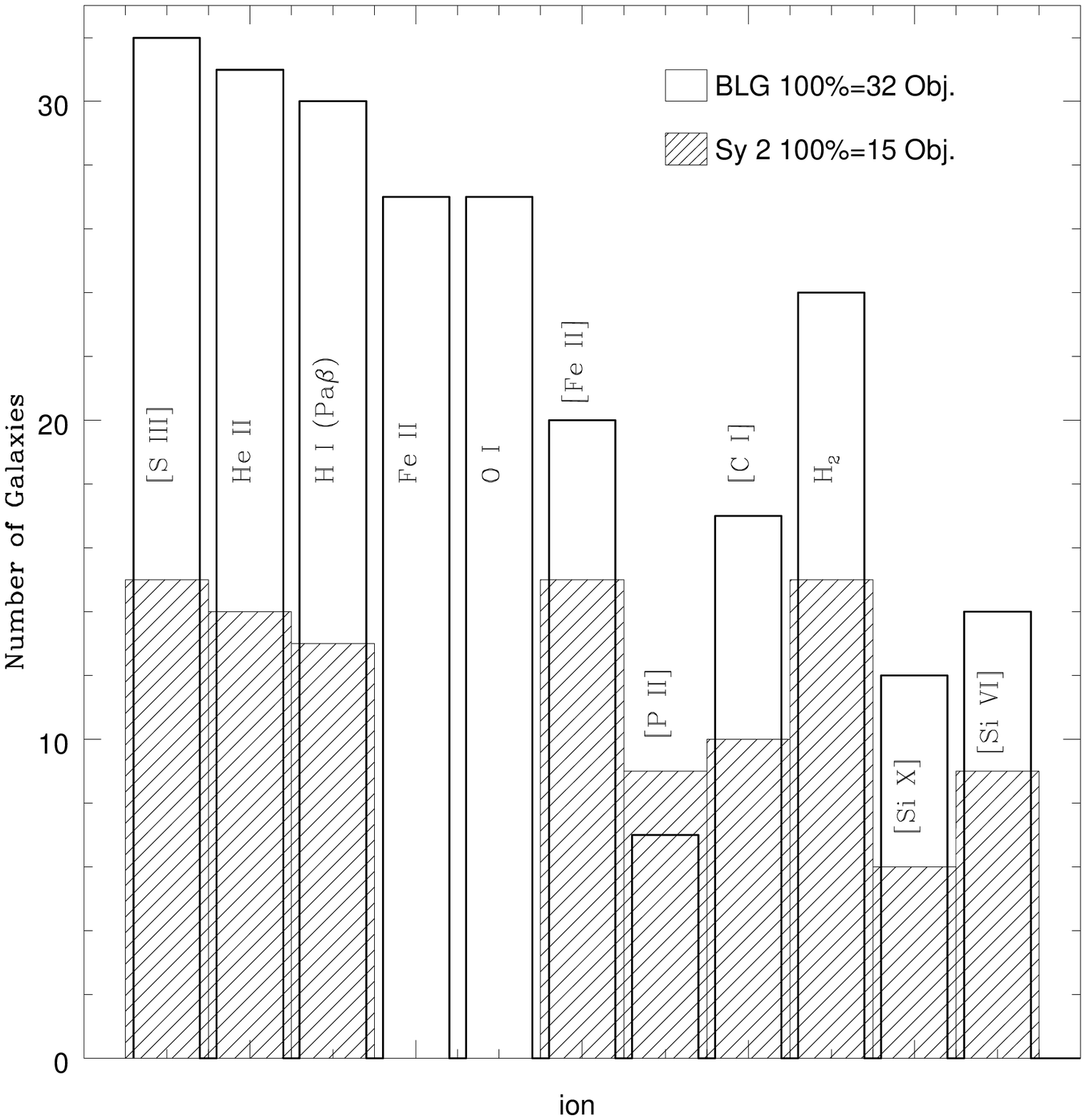}
\caption{Histogram showing statistics of the commonest NIR emission lines.}
\label{hist_blg}
\end{figure}

\subsubsection{Seyfert~1 galaxies}\label{ssc1}

\begin{figure}
\centering
\includegraphics[width=8cm]{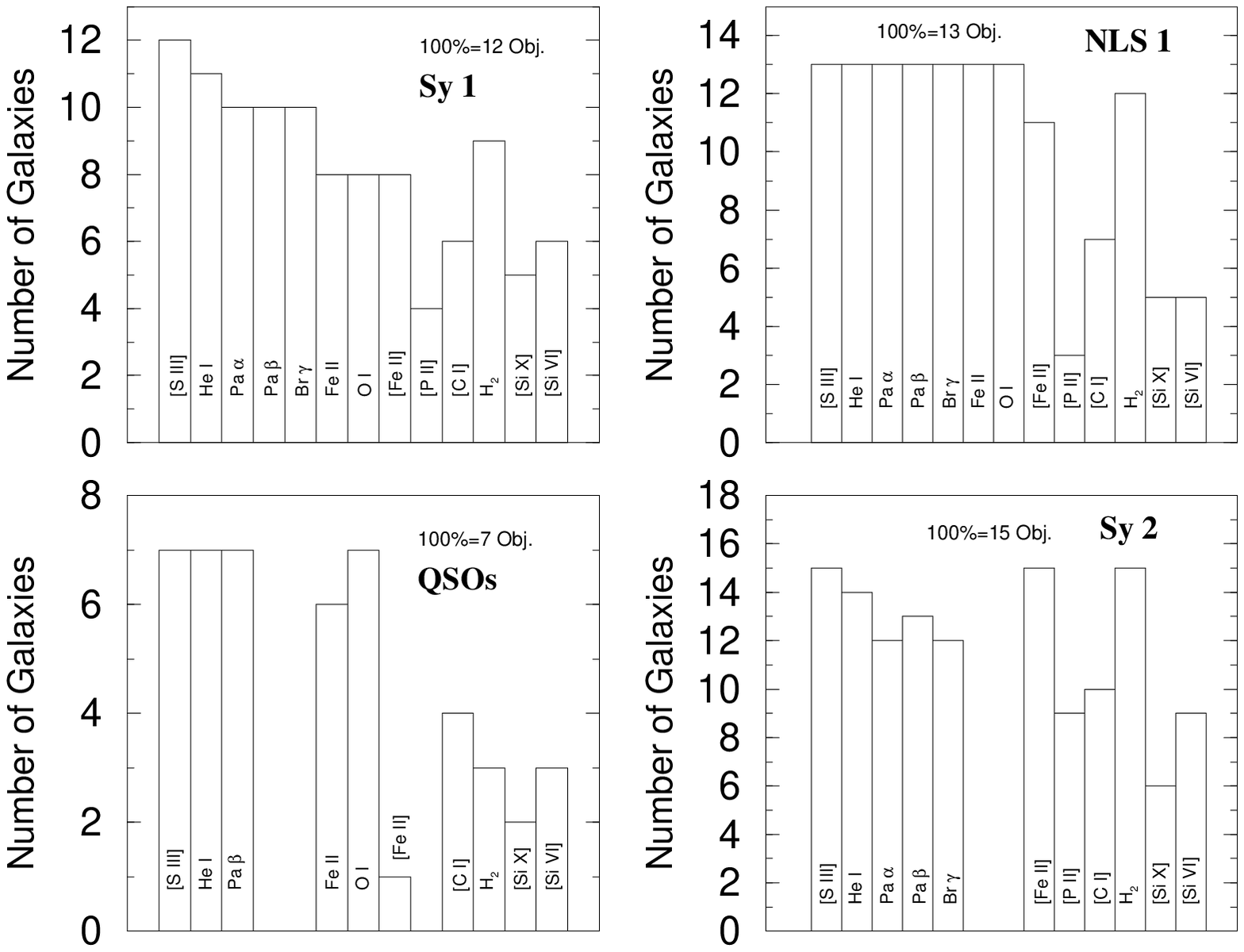}
\caption{Histogram, showing statistics of the commonest NIR emission lines, according 
to each group of nuclear activity.}
\label{hist_ind}
\end{figure}

According to \citet{Osterbrock89}, the emission-line spectrum of 
Seyfert~1 galaxies is characterized by permitted 
broad \ion{H}{i}, \ion{He}{i} and \ion{He}{ii} lines, with FWHM of order 
of 5000~km\,s$^{-1}$, and narrow permitted and forbidden emission lines 
with FWHMs of $\sim$500~km\,s$^{-1}$. Lines with similar characteristics 
are also observed in the NIR, as can be seen in Fig.~\ref{hist_ind}  
and Fig.~\ref{plsy1}.

We found that the forbidden [\ion{S}{iii}] $\lambda\lambda$\,9069, 9531\,$\AA$ 
lines are present in all the Sy~1 galaxies of our sample. The permitted 
\ion{He}{i} $\lambda$\,10830~\AA\ line is detected in 91\% of the sources. 
\ion{H}{i} emission lines such as Pa$\alpha$, Pa$\beta$, Pa$\gamma$ and 
Br$\gamma$ are common to 83\% of the spectra. Moreover, exclusive BLR 
signatures like the \ion{Fe}{ii} and \ion{O}{i} lines were detected in 
in 67\% of the Sy~1 galaxies. Forbidden low ionization species were 
also detected in the Sy~1 spectra. 
The commonest are \fe2\ $\lambda\lambda$\,12570, 16436\,$\AA$, which 
are present in 67\% of the galaxies. Mrk\,334, NGC\,7469, NGC\,3227 and 
NGC\,4151 display [\ion{P}{ii}] $\lambda$\,11886\,$\AA$ line, 
corresponding to 33\% of the Sy~1 sample. The carbon emisson line 
[\ion{C}{i}] $\lambda$\,9850~\AA\ is identified 
in 50\% of the sources. The molecular H$_2$~2.121~$\mu$m line 
is observed in 75\% of the objects. Finally, the coronal line 
[\ion{Si}{vi}] $\lambda$\,19641\,$\AA$ is present in 50\% of the galaxies, while
[\ion{Si}{x}] $\lambda$\,14300~\AA\ is common in 42\% of the objects. 

\subsubsection{Narrow-line Seyfert~1 galaxies}

The Narrow-line Seyfert~1 galaxies are a peculiar group of Sy~1 sources 
first identified by \citet{oter85}. Among other properties, they are
characterized by optical spectra displaying broad permitted lines with 
FWHM $<$ 2000~\kms\ and strong \ion{Fe}{ii} emission. Our NLS1 subsample
of objects, composed of 13 galaxies, is the largest set of AGN belonging
to this category already observed in the NIR region and published in the 
literature, allowing the study of the most important emission features 
detected in their spectrum. Moreover, our NLS1 list is composed of 
well-studied objects in other spectral regions.  

As can be observed in Figs.~\ref{hist_ind} and \ref{plnls}, the most 
conspicuous emission lines identified in the spectra are the first
three lines of the Pashen series (Pa$\alpha$, Pa$\beta$, and Pa$\gamma$) and 
the \ion{He}{i} $\lambda$\,10830~\AA\ line, all of which are observed in all objects. 
In addition, exclusive BLR features of \ion{O}{i}, \ion{Fe}{ii}, and \ion{Ca}{ii}, 
free from contamination of the NLR, are common to all the NLS1 galaxies. 
The presence of these three features, in particular \ion{Fe}{ii}, 
represents a firm advantage of the NIR region compared to the optical in the
study of that emission. The large number of \ion{Fe}{ii} multiplets and its
proximity in wavelength in the optical leads to the formation of a 
pseudo-continuum that usually hampers the detection of individual
\ion{Fe}{ii} lines, even in NLS1. In the NIR, the larger separation
in wavelength among the different \ion{Fe}{ii} multiples in combination
with the small FWHM of broad features displayed by NLS1 allows the
identification of iron lines that put firm constraints on the 
mechanisms that creates them. This is the case, for example, for 
the \ion{Fe}{ii} lines located in the 9200~\AA\ region, detected in
the majority of the NLS1 sources (see Tables~\ref{flsy1_1} to~\ref{flsy1_2}), 
and these are considered as primary cascading lines following Ly$\alpha$ 
fluorescence \citep{sp98,sp03}. 

Aside from the lines mentioned above, the forbidden [\ion{S}{iii}] 
$\lambda$\,9531~\AA\ line is also detected in all the NLS1 galaxies.
Other conspicuous features, such as \fe2\ and molecular hydrogen, are  
found in 85\% and 92\% of the galaxies, respectively. Three of the NLS1s, 
(Ark\,564, 1H1934-063, and Mrk\,766), display [\ion{P}{ii}] 
$\lambda\lambda$\,11460,11886\,$\AA$ lines representing 23\% of the sample. 
The forbidden [\ion{C}{i}] $\lambda$\,9850\,$\AA$ line is clearly identified in 
54\% of the objects. As in the Sy~1 galaxies, the coronal lines 
[\ion{Si}{x}] $\lambda$\,14300\,$\AA$ and [\ion{Si}{vi}] $\lambda$\,19641\,$\AA$
are observed in 38\% of the  sample.

\subsubsection{Quasi stellar objects}

Overall, the emission line spectrum of quasars are similar
to that of Seyfert~1s and NLS1s (see Fig.~\ref{plqso}). The only
appreciable difference is in the intensity of the forbidden lines, which are
weak or absent in a large fraction of the objects studied. It must be
recalled, however, that the small number of targets (7) 
only allow us to establish trends about the frequency of the
most important emission features. The advantage here is that our 
statistics can be compared with the results found by \citet{gli06},
who studied a larger sample of quasars in the NIR. We recall
that the Glikman et al. sample is composed of more distant
quasars than ours. 

As expected, the NIR spectrum of quasars is dominated by broad permitted 
lines of \ion{H}{i}, \ion{He}{i}~1.083~$\mu$m, \ion{O}{i}, and \ion{Fe}{ii}. 
These features are identified in all objects except in 3C\,351, which
lacks \ion{Fe}{ii}. This can be a dilution effect if we consider that
3C\,351 displays extremely broad permitted lines, with FHWM reaching
$\sim$12000~\kms. Any weak-to-moderate \ion{Fe}{ii} emission that broad 
would either be diluted in the continuum or heavily blended with nearby features turning them 
very difficult to isolate and identify. The lack of \ion{Fe}{ii} can also be explained on physical 
grounds. It is well known from the work of \citet{bg92} that steep
radio sources display weak or no \ion{Fe}{ii} emission and that would be
the case of 3C\,351.  

Regarding the detection of signatures revealing the presence of a NLR, 
it is interesting to note that the forbidden [\ion{S}{iii}] 
$\lambda\lambda$9069, 9531\,$\AA$ is found in all
quasars. In addition, [\ion{C}{i}] is clearly identified in 57\% of the objects.
The high ionization line [\ion{Si}{x}]~$\lambda$\,14300\,$\AA$ 
is detected in two sources (PG\,1612 and PG\,1126), representing a frequency
of 28\%. [\ion{Si}{vi}]~$\lambda$\,19641\,$\AA$ is the commonest coronal line. It was 
detected in 43\% of the objects. Similarly, molecular hydrogen is 
clearly present in PG\,1448, PG\,1612, and PG\,1126, corresponding to 43\% of the galaxies. We should note
that the spectra of the high redshift QSOs, Ton\,0156 and 3C\,351, display 
the presence of H$\alpha$. The measured line fluxes of these two objects are
presented in Table~\ref{flux_qso}. 

Our results agree very closely with those reported by
\citet{gli06}. The only lines that appear in our data, but seems to
be missed in their composite quasar spectrum correspond to the
[\ion{C}{i}] and the coronal lines. This, however, needs to be looked at with
caution because the spectrum that they present corresponds to a composite one instead of individual sources. 
\setcounter{table}{5}
\begin{table}[h]
\begin{scriptsize}
\renewcommand{\tabcolsep}{0.70mm}
\caption{\label{flux_qso} Observed fluxes for the two high redshift QSOs in units of $\rm 10^{-15}\, erg \, cm^{-2} \, s^{-1}$.
The fluxes of the permitted lines are the total flux of the line.}    
\centering
\begin{tabular}{lccc|ccccc}
\hline\hline
\noalign{\smallskip}
Ion  & $\rm \lambda_{lab}\,(\AA)$& Ton\,0156 &  3C\,351 & Ion  & $\rm \lambda_{lab}\,(\AA)$& Ton\,0156 &  3C\,351\\
\noalign{\smallskip}
\hline \noalign{\smallskip}
\ion{H}{i}	     & 6563	& 200.12\pp2.11 &935.04\pp24.70$^{\delta}$&\ion{H}{i}	     & 10049	& 11.04\pp1.13  &     -        \\
\ion{O}{i}+\ion{Ca}{ii}&8486    &  8.57\pp2.14  & 72.11\pp4.16            & \ion{Fe}{ii}      & 10500	 &  1.87\pp0.45  &   -         \\
\lb\ion{S}{iii}\rb   & 9069	&     -         & 4.22\pp0.16             & \ion{He}{ii}      & 10830	 & 16.42\pp1.03  & 52.01\pp3.40 \\
\lb\ion{S}{iii}\rb   & 9531	&     -         & 11.33\pp0.43            & \ion{H}{i}        & 10938	 & 9.59\pp0.70   & 18.74\pp3.37 \\     
\ion{Fe}{ii}         & 9127     &  0.57\pp0.14  &      -                  & \ion{O}{i}        & 11287	 &  2.53\pp0.63  &		\\ 
\ion{Fe}{ii}         & 9177     &  1.05\pp0.26  &      -                  & \ion{H}{i}        & 12820	 & 14.16\pp0.79  &  22.09\pp4.85 \\
\noalign{\smallskip}
\hline
\multicolumn{6}{l}{$\delta$ Blend with [\ion{N}{ii}] 6548$\AA$ and [\ion{N}{ii}] 6583$\AA$.} \\
\end{tabular}
\end{scriptsize}
\end{table}

\subsubsection{Seyfert~2 galaxies}

The spectrum of Sy~2  galaxies is dominated by strong emission features of permitted
and forbidden lines, with FWHM rarely exceeding $\sim$600~\kms. By far, the strongest emission 
lines observed are [\ion{S}{iii}]~9531\,$\AA$ and \ion{He}{i}~1.083~$\mu$m,
detected in allmost all the sources  (see Figs.~\ref{hist_ind} and \ref{plsy2}).
Permitted \ion{H}{i} is clearly identified in 87\% of the objects. Low 
ionization lines of \fe2\ and molecular \h2\ are found in all spectra. Phosphorus and carbon are
also identified. At least one of the phosphorus forbidden 
transitions either [\ion{P}{ii}]$\lambda$\,11460\,$\AA$ or [\ion{P}{ii}] $\lambda$11886\,$\AA$ is detected 
in  60\% of the Sy~2 sample. [\ion{C}{i}] is detected in 67\% of the Sy~2s. 
Forbidden high ionization lines are also detected. The [\ion{Si}{x}] $\lambda$\,14300\,$\AA$ line 
is common to 40\% of the objects, and the [\ion{Si}{vi}] $\lambda$\,19641\,$\AA$ line is found in 60\% of 
the spectra. Broad permitted lines of \ion{H}{i} were found in NGC~7674 and Mrk~993, leading
us to consider that they are obscured Seyfert 1 objects. Both sources display 
broad emission lines in polarized lines \citep{mg90}.

\subsubsection{Starburst galaxies}\label{ssc2}

For comparison purposes, four starburst galaxies, namely NGC\,34, NGC\,1614, NGC\,3310, and 
NGC\,7714 were include in our survey, been the last three genuine SB, while NGC\,34 has an 
ambiguous classification (see Sect.~\ref{indiv}).
Their spectra are dominated by unresolved permitted lines of Pa$\alpha$, Pa$\beta$, Pa$\gamma$, 
Br$\gamma$ and \ion{He}{i} $\lambda$10830 $\AA$ (see Fig.~\ref{plsb}). The forbidden emission 
lines of [\ion{S}{iii}] $\lambda\lambda$9069, 9531\,$\AA$, and [\ion{Fe}{ii}] 
$\lambda\lambda$12570, 16436\,$\AA$ are also conspicuous in the three objects.
molecular hydrogen lines are clearly visible in the data. No high ionization
lines were found. All three objects display a remarkably similar emission
line spectrum, nearly indistinguishable from each other. The continuum
emission is different in NGC\,1614 and NGC\,34.

\subsection{Reddening in Seyfert galaxies by means of NIR line ratios}
The flux ratios between hydrogen emission lines are very often used as 
diagnostics of the reddening affecting the emitting gas of an AGN.  
This approach, however, is subject to large uncertainties
when trying to determine the extinction in Type~1 sources, either because the 
\ion{H}{i} lines are strongly blended with nearby features, as 
in classical Seyfert~1, or because of the intrinsic difficulties in deblending
the contribution from the NLR and the BLR in NLS1s. Another major problem
is the fact that the intrinsic line ratios may depart significantly
from Case B because of the high-density environment of the BLR and
radiation transfer effects \citep{collin82,Osterbrock89}. The alternative is to use
forbidden line ratios, but this method is very limited because
the lines involved need to be from the same ion, have a large separation 
in wavelength, and must share the same upper limit, so that the line 
ratio is insensitive to the temperature over a wide range of densities
been only function of the transition probabilities. 

For all the above, the NIR region opens a new window to explore
this issue. First,  observational  studies based on small samples 
of objects indicate that the line ratios of   Pa$\beta$/Br$\gamma$ are
 not only comparable in both BLR and NLR  but also consistent with Case 
 B recombination,  confirming that this ratio  is less affected by collisional 
 effects than by optical lines \citep{rhee00}, which is expected because the NIR 
 lines have smaller optical depths. Second, suitable pairs of forbidden lines 
can be found, allowing an alternative route for determining the 
extinction, as is the case of \fe2\ lines at 1.257~$\mu$m and 1.643~$\mu$m
\citep[see, for example,][]{ara04}.
 
\begin{figure*}
   \centering
  \includegraphics[width=12cm]{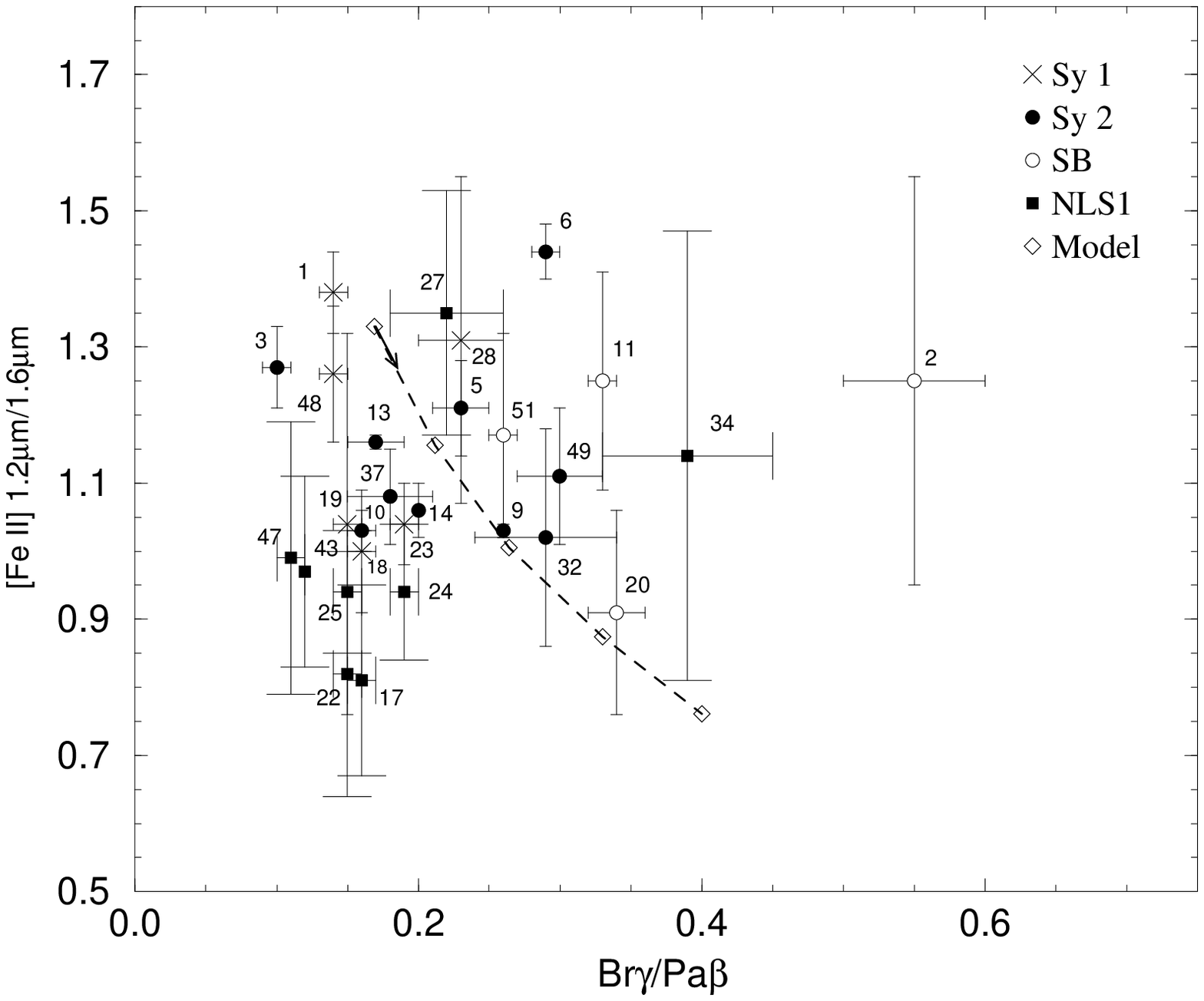}
\caption{Reddening diagram involving the \fe2 12570$\AA$/16436$\AA$ line ratio and Br$\gamma$/Pa$\beta$. 
Stars represent the Seyfert~1 galaxies, filled circles the Seyfert~2 galaxies, 
open circles represent the starburst galaxies, and filled boxes are the NLS1 galaxies of our sample. The dashed
line corresponds to a reddening sequence, from E(B-V)=0 (diamond with arrow) up to E(B-V)=2. The diamonds 
are the theoretical values reddened in steps of E(B-V)=0.5 mag, assuming the \citet{ccm89} law.}
 \label{feh}
   \end{figure*}

In order to see if the reddening determined by means of
the \ion{H}{i} line ratios and that found from the
[\ion{Fe}{ii}] are similar, we have plotted 
the ratios Br$\gamma$/Pa$\beta$ vs [\ion{Fe}{ii}] 1.2$\mu$m/1.6$\mu$m in Fig.~\ref{feh}.
In the calculation, we assumed that the intrinsic ratios are
0.17 and 1.33, respectively \citep{hs87,bpr98}. Note that
for Seyfert~1 galaxies, we used the total flux of the lines
to avoid uncertainties introduced in the deblending of the
broad and narrow components, mainly in the NLS1. The dashed
line corresponds to a reddening sequence, from E(B-V)=0
(diamond with arrow) up to E(B-V)=2, in steps of
E(B-V)=0.5 mag. The \citet{ccm89} [CCM] reddening law was
employed for this purpose. 

A first inspection on Fig.~\ref{feh} shows
that Seyfert~2s tend to have a much narrower distribution
in the [\ion{Fe}{ii}] flux ratio than do Seyfert~1s. Also,
Seyfert 2s tend to lie close to the locus of points of
the reddening curve, with E(B-V) in the interval 0.25-1 mag,
implying that the regions emitting the \ion{H}{i} and
[\ion{Fe}{ii}] lines are affected by similar amounts
of extinction. Seyfert~1s, in contrast, appear to be
divided into two groups. One is populated predominantly by
broad-line Seyfert~1s, which display extinction values near
to zero for both ratios, and the second group, composed
mostly of NLS1s, displays high values of extinction for the
[\ion{Fe}{ii}] gas but close to zero for the \ion{H}{i} region.
Moreover, a few Seyfert~1s have lower Br$\gamma$/Pa$\beta$ ratios
than the intrinsic Case B.

Keeping in mind that the total flux of the Br$\gamma$ and Pa$\beta$ 
lines plotted in Fig.~\ref{feh} for the Seyfert~1s is likely to be dominated by the
one emitted by the  BLR component, we propose that the lack of 
significant reddening for the \ion{H}{i} gas in these objects can be 
explained if Case B intrinsic values are ruled out for the NIR lines, 
as happens in the optical region. Density and radiation transport effects 
modify them so that they are not a reliable source
of information for the reddening. The alternative is that the 
region emitting the \ion{H}{i} lines, particularly the BLR,
is little or not affected by dust. This hypothesis is highly
plausible, as the environment of the BLR is rather turbulent
and very close to the central source making the environment unfavorable for dust grain survival. 

\begin{figure*}
\centering
\includegraphics[width=17cm]{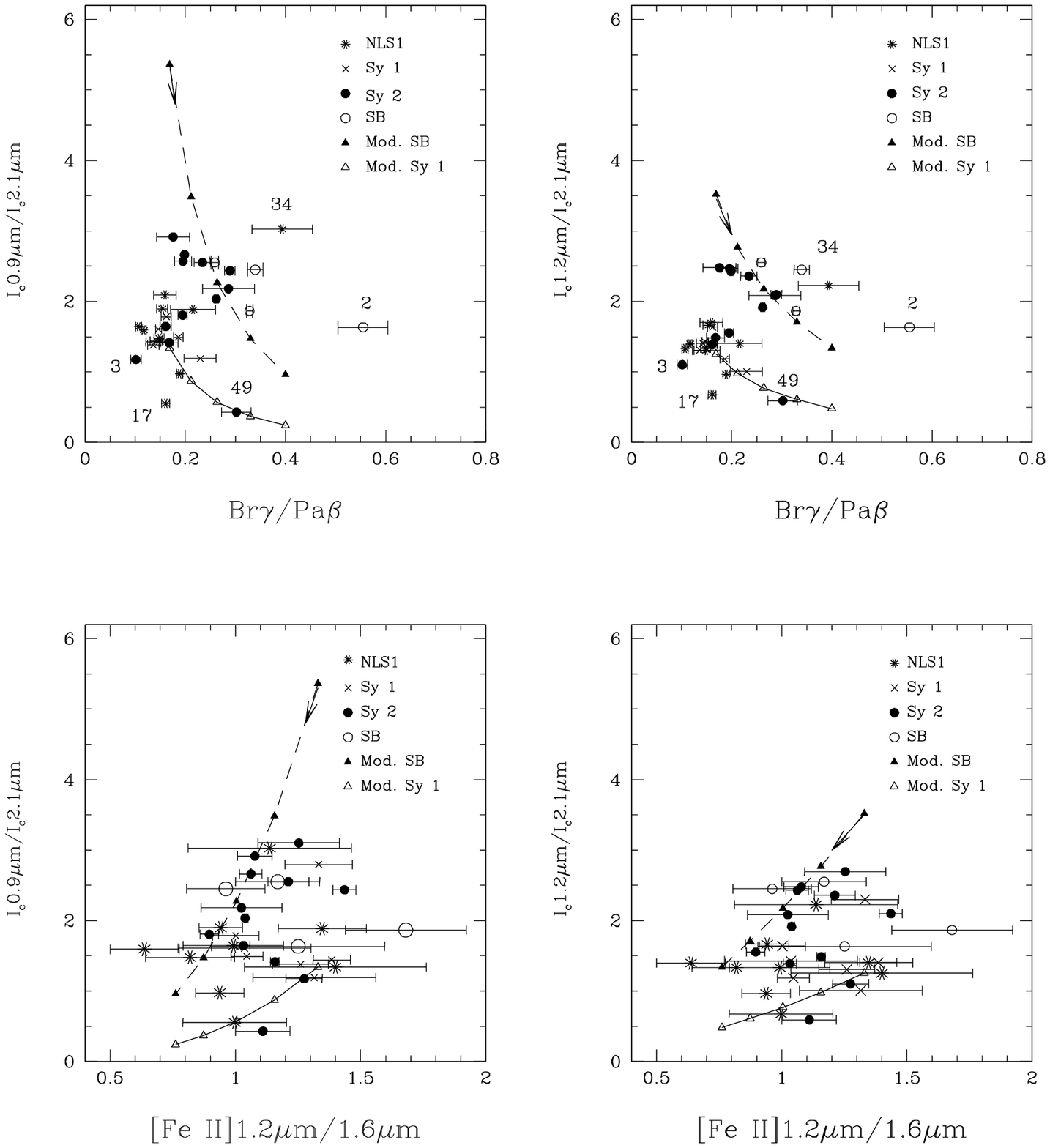}
\caption{Plot of the reddening indicators $\rm Pa\beta/Br\gamma$ (top) 
and \fe2\ 12570$\AA$/16436$\AA$ (bottom) vs the flux ratio of continuum 
emission integrated in windows of $\sim$100~\AA, free from line emission 
contributions. Stars are NLS1 galaxies, crosses are the Sy~1, 
filled circles represent Sy~2s, open circles are the SB of our sample. 
Filled triangles are the intrinsic values of the line ratios 
($\rm Pa\beta/Br\gamma$ and \fe2 12570$\AA$/16436$\AA$) 
and the dereddened  continuum ratios of the SB galaxy NGC\,3310. 
These triangles joined by a  dashed line represent 
the reddening curve, in steps of E(B-V)= 0.5 mag (Mod. SB). 
The open triangles represents a reddening sequence starting from the 
continuum ratios measured in Mrk\,493 (Mod. Sy~1), $\rm I_c 0.9\mu m$ 
represents the mean continuum in the range 9700-9800$\rm \AA$, 
$\rm I_c 1.2\mu m$ represents the mean continuum in the range 12230-12330 
$\rm \AA$, and $\rm I_c 2.1\mu m$ 
for the range 20900-21000 $\rm \AA$. The measured continuum fluxes are presented 
in Table~\ref{cont}. For more details see text.}
\label{red_cont}
\end{figure*}

In Sect.~\ref{s3} we already noted that the NIR continuum
within the same type of AGN was rather homogeneous, particularly
in the {\it H} and {\it K}-bands, being
the major appreciable difference the steepness of the 
continuum in the $z$+$J$ band. Is that steepness related 
to a measurable parameter such as extinction?
In order to investigate if such a relationship can be
established, we plotted in Fig.~\ref{red_cont} the reddening
indicators $\rm Pa\beta/Br\gamma$ (top) and \fe2\ 12570$\AA$/16436$\AA$
(bottom) vs NIR color indices derived from the flux ratio of continuum 
emission integrated in windows of $\sim$100~\AA. The regions
chosen for integration are free of line emission contribution and are 
meant to be representative of the form of the continuum
across the NIR region. The measured continuum fluxes are presented in
Table~\ref{cont}. 

As reference, we compared the observed ratios in Fig.~\ref{red_cont}
with two reddening sequences: one that starts from points representing
the intrinsic values of the line ratios ($\rm Pa\beta/Br\gamma$ and
\fe2 12570$\AA$/16436$\AA$) and the dereddened  continuum
ratios taken from the SB galaxy NGC\,3310, assumed to be
representative of a continuum typical of a young stellar population. 
The value of E(B-V) for the deredening was taken from \citet{rrp05}.
For this object, the dereddened continuum was then reddened
in steps of E(B-V)=0.5 mag (Mod. SB, filled triangles joined
by a dashed line) up to a E(B-V)=2 mag. The CCM reddening law was employed.
We also plot a reddening sequence for the Type~1 galaxies in Fig.~\ref{red_cont} (open triangles 
connected by a solid line, Mod. Sy 1), using as zero points in the abscissa axis the continuum ratios
measured in Mrk\,493, an NLS1 galaxy whose continuum is considered to be affected by extinction 
and stellar population very little or not at all  \citep[see for example][]{cre02}.

A first inspection to the two upper panels of Fig.~\ref{red_cont}
allows us to state that \ion{H}{i} ratios constrain the
reddening in Seyfert 1 objects poorly, while they are useful
diagnostics for Seyfert 2 galaxies. This conclusion is based on
the fact that the former type of objects are concentrated in the region close to the
point corresponding to E(B-V)=0. The two upper plots also confirm that
the continuum emission of Seyfert~1s is rather homogeneous
from object to object as the ratio between the continuum and
line emission display little scatter. In contrast, Seyfert~2 seems to
be divided into two groups. One follows the theoretical
reddening curve, suggesting that their continuum emission can be
reproduced by means of a reddened starburst component, and another whose \ion{H}{i}
ratios and continuum emission seem to be dominated by emission
from the central engine, as these objects share similar continuum
and emission line ratios of Seyfert 1s. The two outliers, identified
with the numbers 2 (NGC~34) and 34 (Mrk~291), may represent
extreme cases of highly reddened sources. For instance, NGC~34
is a luminous infrared galaxy with strong water megamaser emission,
suggesting both strong thermal emission by dust and starburst 
activity. Mrk~291 is a NLS1 galaxy whose emission line spectrum 
is closer to that of a Seyfert~2. 

The two lower panels of Fig.~\ref{red_cont}, which involves
the reddening sensitive line ratio \fe2\ 12570$\AA$/16436$\AA$,
confirm  that an important fraction of the Seyfert~2s of our
sample display a continuum emission that is dominated by
reddened stellar emission, most likely emitted by circumnuclear
starburst activity. The \fe2\ provides a reliable measurement
of the NLR extinction, as most points are close to the
reddening sequence. Seyfert~1 galaxies display a large scatter
in reddening for the NLR (measured through the \fe2\ ratio),  
although the slope of the continuum
emission varies little. It means that the continuum emission
it little or not affected by dust. The fact that some Seyfert~1s
display continuum flux ratios compatible with a highly reddened
starburst component may be artificial. Rather, these objects
can have an important stellar contribution to the observed
continuum emission, where part of the \fe2\ may be emitted.

Overall, the panels shown in Fig.~\ref{red_cont} reveal the
complex nature of the NIR continuum in AGNs, but proves to be
useful in detecting objects with important starburst activity.
Moreover, they show that there are strong differences in the
form of the NIR continuum emission between Seyfert~1s and 2s
objects.

\begin{table}
\begin{scriptsize}
\renewcommand{\tabcolsep}{0.70mm}
\caption{\label{cont} Mean continuum fluxes on selected ranges, 
in units of $\rm 10^{-15}\, erg \, cm^{-2} \, s^{-1}$.}    
\centering
\begin{tabular}{lcccc}
\hline\hline
\noalign{\smallskip}
\noalign{\smallskip}
               &  Cont. range   & Cont. range       & Cont. range       & Cont. range    \\ 
Source         & 9700-9800$\rm \AA$ & 12230-12330$\rm \AA$  & 16600-16700$\rm \AA$  & 20900-21000$\rm \AA$ \\
\hline \noalign{\smallskip}
  Mrk\,334     &   1.56  & 	  1.52  & 	 1.35	& 	1.09  \\ 
  NGC\,34      &   2.82  & 	  2.91  & 	 2.60	& 	1.78  \\ 
  NGC\,262     &   0.86  & 	  0.81  & 	 0.76	& 	0.73  \\ 
  Mrk\,993     &   1.46  & 	  1.26  & 	 0.91	& 	0.51  \\ 
  NGC\,591     &   1.21  & 	  1.12  & 	 0.83	& 	0.47  \\ 
  Mrk\,573     &   1.61  & 	  1.39  & 	 1.00	& 	0.66  \\ 
  NGC\,1097    &   2.73  & 	  2.38  & 	 1.76	& 	1.05  \\ 
  NGC\,1144    &   0.74  & 	  0.63  & 	 0.43	& 	0.24  \\ 
  Mrk\,1066    &   2.14  & 	  2.01  & 	 1.58	& 	1.05  \\ 
  NGC\,1275    &   1.88  & 	  1.58  & 	 1.32	& 	1.14  \\ 
  NGC\,1614    &   6.41  & 	  6.84  & 	 5.57	& 	3.67  \\ 
  MCG-5-13-17  &   2.75  & 	  2.29  & 	 1.69	& 	1.08  \\ 
  NGC\,2110    &   2.69  & 	  2.83  & 	 2.54	& 	1.91  \\ 
 ESO\,428-G014 &   3.72  & 	  3.39  & 	 2.48	& 	1.40  \\ 
  Mrk\,1210    &   0.76  & 	  0.65  & 	 0.54	& 	0.42  \\ 
  Mrk\,124     &   0.68  & 	  0.70  & 	 0.76	& 	0.71  \\ 
  Mrk\,1239    &   3.32  & 	  4.05  & 	 5.11	& 	5.99  \\ 
  NGC\,3227    &   5.39  & 	  4.95  & 	 4.16	& 	3.02  \\ 
  H1143-192    &   1.68  & 	  1.49  & 	 1.25	& 	1.05  \\ 
  NGC\,3310    &   2.98  & 	  2.60  & 	 1.85	& 	1.06  \\ 
  PG1126-041   &   1.94  & 	  1.89  & 	 1.80	& 	1.72  \\ 
  NGC\,4051    &   3.89  & 	  3.51  & 	 2.98	& 	2.63  \\ 
  NGC\,4151    &   8.69  & 	  6.90  & 	 6.22	& 	5.83  \\ 
  Mrk\,766     &   2.35  & 	  2.32  & 	 2.42	& 	2.41  \\ 
  NGC\,4748    &   1.76  & 	  1.54  & 	 1.23	& 	0.93  \\ 
  Ton\,0156    &   0.23  & 	  0.18  & 	  -	& 	-     \\ 
  Mrk\,279     &   1.32  & 	  0.98  & 	 0.82	& 	0.70  \\ 
  NGC\,5548    &   0.78  & 	  0.66  & 	 0.65	& 	0.65  \\ 
  PG1415+451   &   0.59  & 	  0.49  & 	 0.46	& 	0.40  \\ 
  Mrk\,0684    &   1.76  & 	  1.28  & 	 0.93	& 	0.69  \\ 
  Mrk\,478     &   1.92  & 	  1.82  & 	 1.85	& 	1.69  \\ 
  NGC\,5728    &   0.95  & 	  0.91  & 	 0.72	& 	0.44  \\ 
  PG\,1448+273 &   1.05  & 	  0.78  & 	 0.61	& 	0.52  \\ 
  Mrk\,291     &   0.40  & 	  0.30  & 	 0.21	& 	0.13  \\ 
  Mrk\,493     &   1.46  & 	  1.37  & 	 1.31	& 	1.09  \\ 
  PG\,1519+226 &   0.46  & 	  0.43  & 	 0.44	& 	0.40  \\ 
  NGC\,5929    &   1.50  & 	  1.27  & 	 0.94	& 	0.51  \\ 
  NGC\,5953    &   3.60  & 	  3.13  & 	 2.17	& 	1.16  \\ 
  PG\,1612+261 &   0.48  & 	  0.45  & 	 0.46	& 	0.45  \\ 
  Mrk\,504     &   0.50  & 	  0.41  & 	 0.32	& 	0.24  \\ 
  3C\,351      &   1.36  & 	  1.16  & 	 1.15	& 	 -    \\ 
  Arp\,102B    &   1.10  & 	  0.90  & 	 0.64	& 	0.39  \\ 
  1H\,1934-063 &   1.71  & 	  1.50  & 	 1.27	& 	1.07  \\ 
  Mrk\,509     &  35.80  & 	 27.31  & 	22.58	&      20.24  \\ 
  Mrk\,896     &   1.20  & 	  1.10  & 	 1.01	& 	0.84  \\ 
   1H2107-097  &   1.81  & 	  1.60  & 	 1.37	& 	1.25  \\ 
   Ark\,564    &   1.90  & 	  1.54  & 	 1.30	& 	1.16  \\ 
  NGC\,7469    &   3.88  & 	  3.67  & 	 3.29	& 	2.81  \\ 
  NGC\,7674    &  10.52  & 	 14.51  & 	19.83	&      24.52  \\ 
  NGC\,7682    &   1.13  & 	  1.08  & 	 0.82	& 	0.44  \\ 
  NGC\,7714    &   4.27  & 	  3.40  & 	 2.38	& 	1.33  \\ 

\noalign{\smallskip}
\hline
\end{tabular}
\end{scriptsize}
\end{table}


\section{Notes on individual objects} \label{indiv}

In this section we describe the most important spectral features found in
the AGN sample. It is motivated by the fact that for a large fraction of
objects  (44/51), no previous NIR spectra
covering the JHK bands simultaneously are available in the literature. In
fact, only NGC\,4151 \citep{th095}, Mrk\,478 \citep{rudy01}, Ark564 \citep{ara02, ara02c},
1H1934-063 \citep{ara00, ara02} , Mrk766 \citep{rcv05}, Mrk\,1210 \citep{mza05}, and Mrk1239
\citep{rom06} have been observed before in this interval.

\begin{itemize}

\item {\bf Mrk\,334}. This Seyfert galaxy from the CfA catalog, classified as 1.8, is located in an 
interacting system in an advanced stage of merger. It has a tidal arm visible in both the
$J$ and $H$ images \citep{mar01}. The {\it HST} images in the F606W filter reported
by \citet{pm02} reveal knots of emission quite near the nucleus, signaling
at least some circumnuclear star formation. Support for this hypothesis comes 
from the {\it L}-band spectroscopy of \citet{ima03}, where the 3.3~$\mu$m PAH emission
has been observed. Our NIR spectrum, is dominated by the emission lines of H\,{\sc i}, 
He\,{\sc i}, [S\,{\sc iii}], and [Fe\,{\sc ii}]. Pa$\alpha$ displays a conspicuous
broad component, not observed in any other permitted line. In {\it J}, a strong
broad absorption is seen to the right of Pa$\gamma$, which we associate with CN \citep{mar05}
while in {\it H}, several narrow absorptions lines of stellar origin are seen. Strong CO bandheads starting at
2.3$\mu$m are clearly present in {\it K}. No high-ionization lines were detected.

\item {\bf NGC\,34}. This is an infrared-luminous galaxy in an advanced
stage of a merger, with two nuclei separated by approximately 6~kpc \citep{aam00}.
It contains the most distant and one of the most luminous water vapor megamasers
so far observed in a Seyfert galaxy \citep{hen05}. The nature of 
its emission-line spectrum is highly controversial.
\citet{od83} claim that NGC~34 is an emission-line galaxy, but not a Seyfert one.
based on the optical emission-line ratios. However, Veron-Cetty and Veron
(1986) classify it as a Seyfert~2. \citet{aam00} argue that this galaxy is 
more properly classified as a starburst rather than a Seyfert.  
 To our knowledge, the only previous NIR spectroscopy reported 
in the literature was the {\it K}-band spectrum of \citep{ia04}.
It can be seen from our data (see Fig.~\ref{indiv1}), that
the continuum emission is dominated by absorption features of stellar origin, and it is indeed
very prominent in the {\it H}-band. NGC~34 is also one of the few sources where 
Pa$\beta$ and higher order Pashen lines appear in absorption. The {\it K}-band spectrum 
is dominated by the CO bandheads and H$_{2}$ emission lines. Overall, NGC~34 displays a 
poor emission-line spectrum, with weak [S\,{\sc iii}] emission, in contrast to
what is observed in the other Seyfert~2 spectra.  This suggests that NGC34 is not 
a genuine AGN or that it  has a buried nuclear activity at a level that is not observed at 
NIR wavelengths. Additional support for this conclusion comes from the lack of high-ionization lines in
its spectrum. Because of the above, we classified NGC 34 as a starburst
galaxy.

\item {\bf NGC\,262}. This CfA galaxy hosts a Sy~2 nucleus with strong emission lines. 
In polarized light, it presents a broad H$\alpha$ component (FWHM=8400~km~s$^{-1}$) and an underlying 
featureless continuum \citep{mg90}. 
A hard X-ray detection \citep{awa91} supports the idea that NGC~262
harbors an obscured Seyfert~1 nucleus. The {\it J-}band
spectroscopy of \citet{vgh97} reports the presence of a faint blue-wing
emission with FWHM$\approx$900-1600 \kms\ in Pa$\beta$ and He\,{\sc i}~10818~\AA,
which is not seen in the forbidden lines. 
The NIR spectrum of this source has been studied  
in separated works \citep{vgh97,sosa01,ia04}, but the one presented 
here is the first one covering the {\it JHK}-bands simultaneously. 
We find that the NIR continuum emission is essentially flat. A
rich emission line spectrum was detected with strong forbidden lines of 
[\ion{S}{iii}] and [\ion{Fe}{ii}], as well as [\ion{C}{i}]~9850\AA\ and 
[\ion{P}{ii}]\,11886\AA. High ionization lines such as 
[\ion{Si}{vi}]\,19630\AA, [\ion{S}{viii}]\,9912\AA\ and [\ion{Si}{x}]\,14300\AA,
are also present. No evidence of broad components or wings are found either in 
Br$\gamma$ (FWHM=520~\kms) or Pa$\beta$ (FHWM=400~\kms). We 
attribute the blue wing reported by \citet{vgh97} in Pa$\beta$
to [\ion{Fe}{ii}]~12788\AA, which contaminates the blue
profile of the former. For comparison, [\ion{S}{iii}] displays an FWHM of 560~\kms.
Faint absorption lines, mainly in the {\it H}-band, were detected.

\item {\bf Mrk\,993}. This CfA AGN \citep{hb92} has been classified as a 
Seyfert~1.5-2. Our spectrum  displays a conspicuous broad
component in \ion{He}{i}~10830\AA, with FHWM=4500~\kms, and in Pa$\beta$ (FHWM=3600~\kms). 
It means that in the NIR, this source can be considered a genuine Seyfert 1 object 
(see Fig.~\ref{indiv1}). The continuum emission decreases steeply towards longer
wavelengths. Absorption lines of stellar origin are seen across the spectrum, with a
prominent \ion{Ca}{ii} triplet to the blue edge and the 2300\AA\ CO bandheads in {\it K}.
It is worth mentioning that the narrow component of the \ion{H}{i} lines 
is seen mostly in absorption.
Its NIR spectrum is poor in emission lines. Besides \ion{He}{i} and
\ion{He}{i}, only [\ion{S}{iii}] and [\ion{Fe}{ii}] are detected (see Fig~\ref{indiv1}). 

\item {\bf NGC\,591}. This Seyfert 2 galaxy, also a radio source, displays 
water megamaser emission with high
velocity features that are approximately symmetrically spaced about the systemic velocity of
the galaxy, a possible signature of a nuclear disk \citep{bra04}. Optical spectroscopy
reported by \citet{dur94} reveals a moderately high excitation AGN, with an intrinsic 
E(B-V) of 0.6~mag. Previous published {\it K}-band spectroscopy for this source \citep{vgh97}, 
shows conspicuous Br$\gamma$, \ion{He}{i}, and H$_{2}$ lines. Here, we present the 
first simultaneous {\it JHK}-band spectroscopic observation of this galaxy.
The spectrum is rich in emission features, displaying bright lines of [\ion{S}{iii}],
\ion{He}{i}, \ion{H}{i}, and [\ion{Fe}{ii}]. We also detected [\ion{C}{i}]~9850\AA\ and 
[\ion{P}{ii}]\,11886\AA\,  as well as high ionization lines of
[\ion{Si}{vi}]\,19630\AA, [\ion{S}{viii}]\,9912\AA\, and [\ion{Si}{x}]\,14300\AA.
Strong molecular hydrogen lines are observed in {\it K}. 
No evidence of broad components or wings are observed in the
permitted lines, ruling out the hypothesis of a hidden BLR. In fact, both Bracket and 
Pashen lines are spectroscopically unresolved, while most forbidden lines have widths
varying between 500-600~\kms. Absorption lines are 
easily visible, mostly in the {\it H}- and {\it K}-bands, with the CO bandheads 
the most prominent ones.

\item {\bf Mrk\,573}. The CfA Seyfert~2 galaxy Mrk\,573 \citep{hb92} 
is a well-studied AGN with two 
ionization cones seen in [\ion{O}{iii}] maps \citep{pr95,fws98}. It is also 
known for the bright high-ionization emission lines displayed in its optical 
nuclear spectrum \citep{dur94}. Spectropolarimetric observations by \citet{nag04} 
show prominent scattered broad H$\alpha$ emission and various narrow forbidden
emission lines, the degree of polarization of the latter ones correlated with the 
ionization potential of the corresponding line. They interpret this correlation
in terms of obscuration of the stratified NLR by the optically and geometrically
thick dusty torus. The NIR spectrum of Mrk\,573 presented in Fig.~\ref{indiv1},  is very similar to 
that of NGC~591. The strong emission lines of [\ion{S}{iii}], \ion{He}{i}, and 
\ion{H}{i} dominates the {\it J}-band. Also, strong high-ionization lines 
were detected, including those of [\ion{S}{ix}]~12520\AA, [\ion{Si}{x}]~14300\AA,
and [\ion{Ca}{viii}]~23218\AA. The line profiles are narrow, with an FWHM typically
of 400~\kms. The most prominent stellar absorption features are the \ion{Ca}{ii} 
triplet at the blue end and the 2.3$\mu$m CO bandhead in {\it K}.

\item {\bf NGC\,1097}.  Classified originally as LINER on the basis 
of its optical spectrum \citep{kee83}, NGC~1097 was later 
reclassified by \citet{sb97} as a Seyfert~1 galaxy after observing the appearance of
broad Balmer line emission and a featureless blue continuum. It has a bright
star-forming ring of diameter $\approx$20", with the nucleus 
contributing negligibly to the integrated H$\alpha$ and Br$\gamma$ emission, 
as well as to the total MIR emission \citep{kot00}. The NIR spectrum resembles
anything but a Seyfert~1 galaxy. From 0.8~$\mu$m to 1.7~$\mu$m, a poor emission line
spectrum is detected, with [\ion{S}{iii}]~9068, 9531\AA\ the most prominent ones,
although intrinsically weak. Absorption bands and lines dominate the NIR region,
confirming that the nucleus contributes little to the integrated emission
line spectrum. In the {\it K}-band, the only emission lines detected
are those of molecular hydrogen. The 2.3~$\mu$m  CO absorption
bandheads dominate the red edge of the spectrum. Similar results are
found from the 1.5--2.5~$\mu$m spectrum reported by \citet{rkp02}.

\item {\bf NGC\,1144}. A CfA Seyfert 2, with NGC~1143 forms an
interacting pair separated by 0.7''. Our NIR spectrum is very similar to that of 
NGC\,1097. It displays a steep blue continuum in the interval 
0.8~$\mu$m--2.4~$\mu$m, dominated by stellar absorption features. The emission line
spectrum is rather poor, with only [\ion{S}{iii}]\,$\lambda\lambda$\,9069, 9531$\AA$, 
\fe2\,1.257~$\mu$m and \h2.2033, 2.121\,$\mu$m being detected.

\item {\bf Mrk\,1066}. \citet{rm99} describe this Seyfert~2 galaxy as a dusty 
object with a single broad dust lane dominating its morphology. It is an FIR luminous
galaxy containing a double nucleus \citep{gim04}. Recently, water vapor maser
emission was detected \citep{hen05}, with two components bracketing the 
systemic velocity of its parent galaxy.  The non-simultaneous {\it J-} and {\it K}-band 
spectroscopy of \citet{vgh97} shows strong [\ion{Fe}{ii}] and Pa$\beta$, with weak
excess of emission seeing at the sides of both lines. They put stringent constraints on the
flux of broad Br$\gamma$ and Pa$\beta$. Our {\it JHK} spectrum display a flat 
continuum from 0.8~$\mu$m to 1.3~$\mu$m. Redwards, it steeply decreases with wavelength. 
The emission line spectrum is strong and bright. [\ion{S}{iii}], \ion{He}{i},
\ion{H}{i}, [\ion{Fe}{ii}] and H$_{2}$ are the most conspicuous emission features.
Weak high-ionization lines of [\ion{Si}{vi}]~19630\AA\ and [\ion{Ca}{viii}]~23218\AA\
were detected. The line profiles are narrow, with FWHM $\approx$400-500~\kms.
No evidence of broad components in the permitted lines was found. The most
prominent stellar absorption features are the \ion{Ca}{ii} triplet
in the blue end and the 2.3~$\mu$m CO bandhead in {\it K}.

\item {\bf NGC\,1275}. One of most widely studied objects of our sample,
NGC\,1275 is a giant elliptical galaxy at the core of the Perseus cluster, with an optically
luminous nucleus, currently classified as a Seyfert 1.5/LINER \citep{sosa01}. It is
also a strong radio-source in the center of a strong cooling flow and
two systems of low-ionization filaments, one of which is probably the
remnants of a recent merger \citep{zin00}. In the NIR, it was studied in the 
({\it HK}-bands) by \citet{kra00}, who found that its NIR
properties can be described best as a combination of dense molecular gas, ionized 
emission line gas, and hot dust emission concentrated on the nucleus. They also argue 
that there is no evidence of a nuclear stellar continuum and that at a 
distance of $\sim$ 1 Kpc from the nucleus the emission is totally dominated by an
old normal stellar population. Recent NIR integral-field spectroscopy by \citet{wej05} shows that the observed
H$_{2}$ is part of a clumpy disk rotating about the radio-jet axis. Our spectrum,
the first to simultaneously cover the 0.8~$\mu$m--2.4~$\mu$m interval,
shows an outstanding emission line spectrum with strong \ion{He}{i}, [\ion{S}{iii}],
[\ion{Fe}{ii}], and H$_{2}$ lines. \ion{He}{i}~10830 displays a conspicuous
broad component, with FHWM $\approx$4700~\kms, not reported before in the
literature. It displays the richest H$_{2}$ emission line spectrum of the sample,
with up to the S(7)1-0 line present in the {\it H}-band. 
Note that high-ionization lines are 
totally absent in the nuclear spectrum. The continuum emission is steep, and 
decreasing in flux with wavelength. Stellar absorption lines
are almost absent. Only the 2.3~$\mu$m CO band-heads in {\it K} are 
barely visible.

\item {\bf NGC\,1614}. This is a strongly interacting galaxy in a late stage of a
merging process with spectacular tidal features. It is one of the
four starburst galaxies of our sample, also cataloged as a
luminous infrared galaxy \citep[LIRG][]{alm02}. The {\it HST}/NIR 
camera and multiobject spectrometer (NICMOS) observations reported by \citet{alh01}
show deep CO stellar absorption, tracing a starburst nucleus about 45~pc in diameter 
surrounded by a $\sim$600~pc diameter ring of supergiant \ion{H}{ii} regions. 
The luminosities of these regions are extremely high, an order of magnitude 
brighter than 30 Doradus. The spectrum presented in 
Fig.~\ref{indiv2}  agrees with the starburst nature of this source.
Only narrow nebular emission lines are detected, all spectroscopically 
unresolved. The molecular H$_{2}$ spectrum is particularly weak. 
The continuum is dominated by stellar absorption features, with strong CO bandheads 
in {\it K} and numerous CO absorptions in {\it H}.  

\item {\bf MCG-5-13-17}. This Seyfert~1 is a strongly perturbed galaxy, with the 
strongest [\ion{O}{iii}] emission concentrated in the nucleus and showing an 
extension to the southeast, suggestive of a conical morphology. The optical
spectrum is dominated by broad permitted lines and narrow permitted and 
forbidden lines \citep{ara00}. Our NIR spectroscopy is the first one
carried out on this source. In {\it J}, broad \ion{H}{i} and \ion{He}{i} lines
are strong, with an FWHM of $\approx$4500~\kms\ and $\approx$5400~\kms, respectively.
Numerous forbidden lines, including high-ionization lines of[\ion{S}{ix}], 
[\ion{Si}{x}], and [\ion{Si}{vi}] are
present. The Br$\gamma$ and H$_{2}$ molecular emission lines are observed in
{\it K}, although they are intrinsically weak. The nature of the continuum 
emission is clearly composite, with stellar absorption features of \ion{Ca}{ii}, CO 
in {\it H}, and the 2.3~$\mu$m CO bandheads in {\it K} on top of a steep 
power-law like continuum. 

\item {\bf NGC\,2110}. It was initially classified by \citet{bra78} as a 
narrow-line X-ray galaxy with sufficient column of dust to the nucleus 
to obscure the broad-line region, thus leading to a Seyfert~2 classification 
of the optical spectrum, but with an insufficient gas column to attenuate the 
2-10~keV emission. In fact, its hard X-ray luminosity is comparable 
to those of Seyfert~1 galaxies \citep{wea95}. In the NIR, it has been the 
subject of numerous studies \citep{vgh97,sosa01,rkp03}. Our NIR spectrum reflects
the low-ionization nature of this source. The most striking characteristic is
the strength of the [\ion{Fe}{ii}] emission line spectrum with 
[\ion{Fe}{ii}]~12570\AA/Pa$\beta \approx$6, three times higher than the
typical values observed on Seyferts \citep[0.6-2][]{rrp05} and the detection
of intrinsically weak lines such as [\ion{Fe}{ii}]~12950\AA\ and 13212\AA. The
Bracket and Pashen \ion{H}{i} lines are weak. Strong molecular lines are also
seen in {\it K}. The continuum emission in the {\it J-} and {\it H}-bands is 
nearly flat and steep towards the red in {\it K}. The \ion{Ca}{ii} triplet in 
absorption, as well as the CO bandheads at 2.3~$\mu$m and the CO bands in 
{\it H}, are the most conspicuous stellar absorption features detected.
No coronal lines are found from our data.

\item  {\bf ESO\,428-G014}. As the host of a Seyfert~2 nucleus \citep{bjo86}, 
the NLR of this object has
been the subject of investigation because of the many individual, thin strands that 
are very closely related to the radio jet and that produce a highly complex, yet ordered, 
structure \citep{fal96}. It also displays a two-side jet with a double helix of 
emission-line gas. In the NIR, ESO\,428-G014  has been studied, among others, by 
\citet{vgh97} and \citet{rkp03}. The latter authors reported bright, extended 
(up to $\approx$ 320~pc) [\ion{Fe}{ii}], Br$\gamma$, and H$_{2}$ emission, parallel 
to the cone. Our NIR spectrum, of larger wavelength coverage, shows that
the strongest nuclear emission lines are those of [\ion{S}{iii}] and \ion{He}{i}.
We also report the first detection of [\ion{S}{viii}]~9912\AA\ and 
[\ion{Si}{x}]~14300\AA\, as well as lines of [\ion{P}{ii}], 
[\ion{S}{ii}] and [\ion{Ca}{i}]. The continuum emission smoothly decreases in
flux with wavelength. 
The \ion{Ca}{ii} triplet and numerous CO bands (in both {\it H} and {\it K}) are clearly 
detected. We also confirm the detection of strong [\ion{Si}{vi}]~19630\AA as
previously reported by \citet{rkp03}. A hint of [\ion{Ca}{viii}]~23218 emission was
seen but it is strongly affected by the CO bandheads at 2.3~$\mu$m.

\item  {\bf Mrk\,1210}. Optically classified as a Seyfert~2 galaxy, this object 
shows broad polarized lines in H$\beta$ and H$\alpha$ \citep{tran92}.  
The NIR nuclear spectrum, studied in detailed by \citet{mza05}, 
is dominated by \ion{H}{i} and \ion{He}{i} 
recombination lines as well as [\ion{S}{ii}], [\ion{S}{iii}] and 
\fe2\ forbidden lines. Coronal lines of [\ion{S}{viii}], 
[\ion{S}{ix}], [\ion{Si}{vi}], [\ion{Si}{x}], and [\ion{Ca}{viii}]
in addition to molecular H$_{2}$ lines are also detected.
The analysis of the emission line profiles, both allowed and forbidden, 
shows a narrow (${\rm FWHM} \sim 500$~\kms) line on top of a broad 
(${\rm FWHM} > 1000$~\kms) component, ruling out the presence of a hidden 
BLR claimed to be present in earlier NIR observations \citep{vgh97}
and confirming the results of \citet{lut02}. 
\citet{mza05} reports extended emission of [\ion{S}{iii}] and
\ion{He}{i}, up to a distance of 500~pc from the center. The
continuum is steep, decreasing in flux with wavelength.
Absorption lines of CO, both in {\it H} and {\it K}, are observed,
indicating the presence of starlight contribution to the nuclear
integrated spectrum.

\item  {\bf Mrk\,124}. Optically classified as a NLS1 galaxy by \citet{deg92},
the spectrum shown in Fig.~\ref{indiv3} is the first one published on
this source in the NIR region. It is dominated by bright permitted Pa$\alpha$,
\ion{He}{i} and \ion{O}{i} and forbidden [\ion{S}{iii}] lines. 
Forbidden high-ionization lines of [\ion{S}{viii}] and [\ion{Si}{vi}] 
are also detected. Pa$\alpha$ shows a broad component of FWHM$\approx$2150~\kms,
while \citet{ver01} report a broad H$\beta$ component of FWHM in the range
1050-1400~\kms. It may indicate that dust obscuration may hide a large
fraction of the BLR contribution. Pa$\beta$ is severely affected by the 
atmospheric cutoff at the red edge of the {\it J}-band, so the presence
of such a broad permitted component cannot be fully confirmed. Permitted
\ion{Fe}{ii} lines in the region around 1$\mu$m were detected. The continuum
emission is featureless and flat, with a small excess of emission in the 
{\it H} and {\it K} bands. No evidence of stellar population is observed.

\item  {\bf Mrk\,1239}. Classified as a NLS1 galaxy by \citet{oter85}, this 
object displays a highly polarized optical spectrum \citep{goo89} and
one of the steepest X-ray spectra found in AGNs, with $\alpha_x$=+3.0
based on ROSAT PSPC data \citep{gmk04}. \citet{sm04}
modeled the polarization nature of this object and find that 
it is one of the rare cases of Seyfert~1 galaxies that appear
to be dominated by scattering in an extended region along the 
poles of the torus. The continuum emission of this source is the
most outstanding of all the objects in our sample. The detailed study by 
\citet{rom06} shows that the NIR is dominated by a strong bump of emission 
peaking at 2.2~$\mu$m, with a strength not reported before in an AGN. The bump follows a 
simple blackbody curve at T$\sim$1200~K. It suggests that we may be 
observing direct evidence of dust heated near to the sublimation 
temperature, probably produced by the putative torus of the unification 
model. The emission line spectrum shows numerous permitted and forbidden 
lines, with \ion{He}{i}~1.083$\mu$m the strongest. 
Permitted \ion{Fe}{ii} transitions, some of them in the 9200~\AA\ 
region and attributed to Ly$\alpha$ fluorescence, are clearly identified. 
A conspicuous NLR spectrum is detected, with strong [\ion{S}{iii}], as 
well as high-ionization lines of [\ion{Si}{vi}], [\ion{Si}{x}], [\ion{S}{viii}],
and [\ion{Ca}{viii}]. The last lines display a blue asymmetric 
profile with their peak centroid blueshifted relative to the systemic
velocity of the galaxy. Because of its extreme properties, the line
spectrum of Mrk~1239 is discussed in a separate paper (in preparation).

\item  {\bf NGC\,3227}. Because of its proximity (15.6~Mpc), NGC~3227 is a 
well-known and studied Seyfert 1/1.5 galaxy in virtually all wavelengths 
intervals. It displays all the possible ingredients found in an AGN:
variability in its nucleus \citep[both in the line and continuum][]{win95}, 
a radio jet \citep{kuk95}, an ionization cone in [\ion{O}{iii}] \citep{mun95}, 
a circumnuclear starburst \citep{gdp97}, strong X-ray emission \citep{rei85},
an inner warp molecular disk \citep{qui99}.  The NIR  images of this galaxy 
reveal an unresolved nuclear source in the K band and  a nuclear stellar 
cluster that is slightly resolved in the J and H bands, this  cluster contributes to about 40-65\% 
of the total emission continuum \citep{schinnerer01}. The NIR properties of this
object have been studied, among others, by \citet{rkp03}, \citet{schinnerer01}   and \citet{qui99}. 
Our spectrum shows a rich emission line spectrum with strong 
broad permitted lines in \ion{He}{i} and Pa$\beta$. [\ion{S}{iii}] is the 
brightest forbidden narrow line. [\ion{Fe}{iii}]~1.257, 1.644$\mu$m are also 
strong in the spectrum. High-ionization lines of [\ion{S}{viii}]~9912\AA\ 
and [\ion{Si}{vi}]~1.963$\mu$m, although weak, were detected. The continuum 
is steep, decreasing towards longer wavelengths. It displays CO absorption bands 
in {\it H} and {\it K}, as well as the \ion{Ca}{ii} triplet at the blue 
edge of the spectrum.  

\item  {\bf H\,1143-182}. This classical Seyfert~1 galaxy has been  studied mostly
in the X-rays and UV region. The NIR  spectrum is dominated by broad permitted lines
of \ion{H}{i}, \ion{He}{i}, \ion{O}{i}, and \ion{Fe}{ii}, with intrinsic
FWHM of $\approx$3800~\kms. The NLR spectrum is
rather weak with only [\ion{S}{iii}], [\ion{S}{ix}], [\ion{Si}{x}], and 
[\ion{Si}{vi}] detected. No narrow components of the permitted lines were
identified. The continuum emission is featureless of a power-law
form. From 10000~\AA\ bluewards, a small excess of emission over the
underlying power-law is observed. No evidence of absorption stellar features
were found. 
   
\item  {\bf NGC\,3310}. One of the four starburst galaxies of our sample, NGC~3310, is
thought to have merged with a companion galaxy \citep{wg05}. It has been 
extensively studied in the UV and optical regions because of its peculiar
properties. It is one of the bluest spiral galaxies in the \citet{vau91}
catalog and its far-infrared luminosity (L$_{\rm ir}$ = 1.1$\times$
10$^{10}$~L$\odot$) indicates that the starburst in this galaxy is
comparable to that of the ``prototypical'' starburst galaxy M82 \citep{smi96}.
Our  {\it JHK} spectrum of this object
(see Fig.~\ref{indiv3}) displays a continuum that decreases in flux with wavelength.
The emission line spectrum shows lines of only a few species,
\ion{H}{i}, \ion{He}{i}, \ion{He}{ii}, [\ion{S}{ii}] [\ion{S}{iii}] and [\ion{Fe}{ii}],
all spectroscopically unresolved. Molecular H$_{2}$
lines at {\it K} are barely detectable. The most prominent absorption features
are the \ion{Ca}{ii} triplet and the CO bandheads in {\it H} and {\it K}. 

\item  {\bf PG~1126-041}. The quasar PG~1126-041 is a strong X-ray source that shows
clear signs of warm absorption \citep{wbb96}. Its optical spectrum \citep{rb84}
is characterized by conspicuous \ion{Fe}{ii}
emission line complexes and broad Balmer lines. NIR information on this object
is scarce, but recently \citet{cre04} reported adaptive optics assisted 
{\it K}-band spectroscopy at a spatial resolution of $\sim$0.08'', allowing them
to spatially resolve the Pa$\alpha$ emission within the nuclear 100~pc. The 
comparison with higher excitation lines suggests that the narrow Pa$\alpha$ 
emission is due to nuclear star formation. Our NIR spectrum is dominated by the
classical broad permitted lines of \ion{H}{i}, \ion{He}{i}, \ion{O}{i}, and 
\ion{Fe}{ii}, emitted by the BLR. We clearly detect the forbidden lines of [\ion{Fe}{ii}],
[\ion{S}{iii}], [\ion{Si}{x}], and [\ion{Si}{vi}], the last line
also detected by \citet{cre04}. We also found evidence of the presence of H$_{2}$ 1.957~$\mu$m, 
but it is strongly blended with Br$\delta$ to the blue and [\ion{Si}{vi}] to the red. The latter set 
of lines (forbidden ones and molecular) allow us to favor the existence of a
classical NLR. We propose that part of the narrow flux found by \citet{cre04} 
in Pa$\alpha$ should come from the NLR emission.
No absorption lines were detected in our spectrum. The continuum is featureless, and has
power-law form with a similar excess of emission blueward of 10000~\AA\
described in H\,1143-182 (see above).

\item  {\bf NGC\,4051}. 
One of the most well-studied AGN of our sample, NGC\,4051 is classified as
a NLS1 galaxy. Our NIR spectrum reveals a large variety of spectroscopic
features, from low ionization forbidden lines such as, [\ion{C}{i}] and
[\ion{N}{i}], to high-ionization lines of [\ion{S}{ix}] and [\ion{Si}{x}]. 
The last two are particularly strong, compared to the other emission 
lines observed. The NLS1 nature of NGC\,4051 is revealed well by the
width of permitted lines. The FWHM of \ion{O}{i}~11287~\AA, an exclusive 
BLR feature free of contamination from the NLR, is only 940~\kms. 
Moreover, the broad component of \ion{H}{i} and \ion{He}{i} is 1200~\kms.
We also report the detection of several permitted \ion{Fe}{ii} lines,
including the ones near 9200~\AA, attributed to Ly$\alpha$ fluorescence
processes \citep{sp98,sp03}. The continuum emission is clearly composite and steep, 
decreasing in flux towards longer wavelengths. Underlying stellar population
is observed, as can be seen from absorption lines of CO, including the bandheads at 
2.3~$\mu$m. Towards the blue edge of the spectrum, a small excess of
emission is seen.

\item  {\bf NGC\,4151}. 
NGC\,4151 is probably the best-studied Seyfert~1 galaxy in the literature,
to which we owe much of our understanding about the AGN phenomenon.
Its nuclear continuum and BLR emission are highly variable 
\citep[see, e.g.,][]{mao91,kas96}, during its low-luminosity state may display 
characteristics of a Seyfert~2 nucleus \citep{pp84}.
Observations of this source span the full electromagnetic spectrum, including
the NIR. Simultaneous {\it JHK} spectroscopy was previously 
reported by \citet{th095}. Thomson's resolution was high enough to 
identify the prominent [\ion{Fe}{ii}] spectrum displayed by this
object as well as numerous permitted lines of \ion{H}{i}, \ion{He}{i},
\ion{He}{ii}, and \ion{O}{i}. A comparison of our NIR spectrum with 
Thompson's allowed us to conclude that his was
observed during a lower-luminosity state. The higher S/N of our data
allowed us to detect broad components in the lines of Pa$\delta$, 
\ion{He}{ii}, and Br$\gamma$. We also report the detection of
\ion{O}{i}~11287~\AA, not observed in the Thompson's spectrum. 
Moreover, our spectrum includes some small spectral regions not 
covered in the Thompson's data, leading to the first detection of
[\ion{Si}{x}]~14300~\AA, for instance. We also added new NLR features
such as [\ion{C}{i}]~9850\AA, [\ion{N}{i}]~1.04~$\mu$m, and 
[\ion{P}{ii}]~1.146~$\mu$m, 1.188~$\mu$m, not observed before. 
The continuum emission is featureless
and very steep in the region 0.8~$\mu$m--1.2~$\mu$m. Redwards, it decreases smoothly
in flux with wavelength. The CO stellar absorption features were 
detected in {\it H} but the bandheads at 2.3~$\mu$m are completely 
absent. 
 
\item  {\bf Mrk\,766}. Classified by \citet{oter85} as NLS1, this barred 
SBa galaxy displays a number of interesting features. The
HST images of this object show filaments, wisps and irregular dust lanes
around an unresolved nucleus \citep{mvt98}.
Radio observations at 3.6\,cm, 6\,cm and 20\,cm
\citep{uag95,nag99} show that the
radio source appears to be extended in both P.A. $\sim$27$\degr$
(on a scale of 0$\arcsec$.25) and P.A. 160$\degr$ (on
a scale of 0$\arcsec$3, \citealt{nag99}). In the optical, the
emission is extended \citep{gp96,mwt96} through a region of a total size greater
than that of the radio source. The NIR spectrum, described well
by \citet{rcv05}, is characterized by numerous
permitted lines of H\,{\sc i}, He\,{\sc i}, He\,{\sc ii}, and
Fe\,{\sc ii}, and by forbidden lines of [S\,{\sc ii}], [S\,{\sc iii}]
and [Fe\,{\sc ii}] among others. High
ionized species such as [Si\,{\sc ix}], [Si\,{\sc x}],
[S\,{\sc ix}] and [Mg\,{\sc vii}] were also observed. The continuum emission
has a complex shape, with contributions of the central engine,
circumnuclear stellar populations and dust. This last component is
shown by the presence of an excess of emission, similar in form
to what is reported above for Mrk\,1239 but with a much lower strength, 
peaking at 2.25$\mu$m, well-fitted by a blackbody function with 
$T_{\rm bb}$=1200~K \citep{rcv05}.

\item  {\bf NGC\,4748}. Classified as NLS1 by \citet{goo89}, this object had been 
previously cataloged as a Seyfert 1/1.5 \citep[see for example,][]{odr85}.
It resides in an interacting pair, in contact, as evidenced by the 
H$\alpha$+[\ion{N}{ii}] image of \citet{mwt96}. It displays considerable
soft X-ray emission \citep{rus96}. The {\it L}-band spectroscopy of \citet{iw04}
shows 3.3~$\mu$m PAH emission in this object, confirming the presence of
circumnuclear star formation. Our NIR spectrum reveals conspicuous emission
lines, where [\ion{S}{iii}]~9531~\AA\ and \ion{He}{i}~10830~\AA\ are
the strongest ones observed. From the width of the permitted lines, we
confirm the NLS1 classification of this object: \ion{O}{i}~11287~\AA\ and 
Pa$\beta$ show FWHM of $\approx$ 1740~\kms\ and 1950~\kms, respectively. 
The spectrum is dominated by BLR features, including fluorescent lines of 
\ion{Fe}{ii}. The continuum emission is composite by a power-law-like form and 
absorption lines of CO both in {\it H} and {\it K}. Molecular emission
lines of H$_{2}$ were also detected. 

\item  {\bf Ton\,0156}. This radio-quiet quasar is the most distant object of
the sample ($z$=0.549).
Because of its redshift, our NIR spectrum includes both Balmer and
Pashen \ion{H}{i} lines. As such, it is dominated by H$\alpha$, with an
FWHM of 6260~\kms. On the other hand, P$\beta$ displays a broad component
of FWHM $\sim$ 3200, while \ion{O}{i} and \ion{Fe}{ii} show an FWHM of
2000~\kms. No evidence of forbidden emission lines was found. The
continuum is featureless and rather blue shortward of 1.2$\mu$m. Redward, 
it becomes rather flat with wavelength increasing.  
 
\item  {\bf Mrk\,279} 
Our NIR spectrum  is dominated by broad 
permitted features of \ion{H}{i} and
\ion{He}{i}, with FWHM $\approx$ 4200~\kms. In contrast, \ion{O}{i}~11287\AA,
also emitted by the BLR, reaches only 2700~\kms. Forbidden lines of 
[\ion{S}{iii}], [\ion{Fe}{ii}], [\ion{C}{i}], and [\ion{S}{ii}] were
detected. We also found strong molecular lines of H$_2$ in 
the {\it K}-band. The continuum emission is rather blue shortward of
1.2~$\mu$m. At this position, there is a subtle change in the 
inclination, becoming less steep, as is typical of Type~1 objects. 
CO absorption bands, as well as the \ion{Ca}{ii} triplet in absorption
are also observed.

\item  {\bf NGC\,5548}.
The NIR spectrum of this well-studied and known Seyfert~1 is dominated by 
the \ion{He}{i}~1.083~$\mu$m
and [\ion{S}{iii}]~9531~\AA\ lines. It also shows very broad components in the permitted
lines (\ion{H}{i} and \ion{He}{i}), of complex structure, similar to that 
observed in double-peaked profiles. It is worth to mention that
similar reports of double-peak lines exist in the optical region on this object. 
In fact, due to the high
variability shown by NGC~5548, the double-peak in the optical lines is
not always detected. The coronal emission line spectrum is remarkably strong, 
with [\ion{S}{viii}]~9912~\AA, [\ion{S}{ix}]~12520~\AA, 
[\ion{Si}{x}]~14300~\AA, [\ion{Si}{vi}]~19630~\AA\, and 
[\ion{Ca}{viii}]~23210 clearly detected. The continuum emission displays
a clear turnover at 13000~\AA, which is rather steep to the blue. Redwards of 
the turnover, the continuum emission rises steeply to the red, very
likely due to dust emission, similar to the one reported in Mrk~1239.
The stellar contribution to the observed spectrum seems rather low, with
only a few weak CO bands in {\it H} detected

\item  {\bf PG1415+451}. One of the quasars of the sample, PG1415+451 has the 
appearance of a naked Type~1 object, as no evidence at all of forbidden 
emission lines was found on its spectrum. This agrees with the optical/NIR
imaging work of \citet{sur01}, who reports that no distinguishable 
features from the host galaxy are seen in the images of this source.
In contrast, \citet{eva01} report the detection of an edge-on CO molecular disk.
UV/Optical spectroscopy on this source presented
by \citet{cor96} shows that \ion{Fe}{ii} and \ion{Mg}{ii}
are the two most conspicuous features in its spectrum. Our NIR 
spectrum is totally dominated by emission from the BLR, 
with broad permitted lines of \ion{H}{i}, \ion{He}{i}, \ion{O}{i}, and 
\ion{Fe}{ii} detected. Spikes at the expected position of 
H$_{2}$~1.957~$\mu$m and H$_{2}$~2.121~$\mu$m are visible in the {\it K}-band 
but data with better S/N is required to confirm this detection. The 
observed continuum is very steep, of power-law form. 
  
\item  {\bf Mrk\,684}. \citet{oter85} classified this object as a NLS1. Indeed,
optical spectroscopy reveals prominent \ion{Fe}{ii} emission 
and broad Balmer lines with FHWM of only 1300~\kms\ \citep{per88}. The NIR spectrum
is dominated by permitted lines of \ion{H}{i} and \ion{He}{i} with 
FWHM $\approx$ 1150~\kms. Permitted lines of \ion{Fe}{ii}, \ion{O}{i} and \ion{Ca}{ii}
(seen in emission) are also detected but with slightly narrower FWHM (900~\kms),
suggesting that they are formed in the outer portions of the BLR. The
NLR emission is almost absent, while [\ion{S}{iii}]~9531~\AA\ is likely to be present
but is strongly blended with Pa~8 to make a secure detection. The continuum 
emission is rather steep and featureless. No evidence of stellar absorption features
was found. Like  PG1415+451, it has the appearance of a naked Type~I source.

\item  {\bf Mrk\,478}. Mrk~478 is an NLS1 with a steep soft X-ray spectrum \citep{gon94}.
Imaging work of \citet{sur01} detected evidence of faint shells or arms attributed
to its host galaxy. \citet{per88} reports \ion{Ca}{ii} lines with FWHM of 2500~\kms\ 
and H$\beta$ with an FWHM of 1250~\kms. Later, \citep{gru04} measured an FWHM of 1630~\kms\
on this same line. It is also a strong \ion{Fe}{ii} emitter with 
optical \ion{Fe}{ii}/H$\beta \approx$0.97 \citep{gru04}.  \citet{rudy01} 
report the 1~$\mu$m \ion{Fe}{ii} lines as well as emission from \ion{H}{i},
\ion{He}{i}, \ion{Ca}{ii} and \ion{O}{i}. We confirm the presence of all
emission line features shown in the Rudy et al. spectrum. Moreover, 
the FWHM of the \ion{Ca}{ii} lines are of 1350~\kms, nearly half the value found by
\citet{per88}, while Pa$\alpha$ displays a broad component of FWHM $\approx$
1950~\kms, more in accord with \citet{gru04} results. Forbidden emission from [\ion{S}{iii}] and
[\ion{N}{i}] as well as molecular hydrogen lines in {\it K} were clearly detected, 
evidencing the presence of a NLR/host galaxy.  The continuum emission flux from 1~$\mu$m
redwards increases slightly with wavelength up to 1.8~$\mu$m, from which it 
becomes steep. From 1~$\mu$m bluewards, it is rather steep, displaying a
large NIR excess. 

\item  {\bf NGC\,5728}. This is one of the Seyfert~2 galaxies that displays prominent 
co-linear biconical emission line cones, separated by a dark band \citep{wil93}. 
NGC\,5728 is also known for having two nested bars \citep{woz95}. In the NIR, the
{\it K}-band 
spectroscopy has been published by \citet{vgh97} and \citet{sosa01}. They both show 
a spectrum dominated by molecular H$_{2}$  and weak Br$\gamma$ emission.
Our composite {\it JHK} spectroscopy reveals a poor emission line spectrum, 
dominated by unresolved [\ion{S}{iii}]~9531~\AA\ and \ion{He}{i}~1.083~$\mu$m.
The \ion{H}{i} spectrum is very weak, showing strong reddening in the
direction of the NLR gas. We report the first detection
of [\ion{Si}{vi}]~19630~\AA, in the {\it K}-band, implying the existence of high-ionization 
gas. Moreover, [\ion{Si}{vi}] is the only line
spectroscopically resolved, with an FWHM of $\approx$ 430~\kms.
The continuum emission is steep and almost featureless. The only absorption
lines detected are the CO bandheads at 2.3~$\mu$m and Na~2.207~$\mu$m in {\it K}.

\item  {\bf PG\,1448+273}. This Palomar-Green quasar is also an NLS1 galaxy and a strong
\ion{Fe}{ii} emitter. \citet{gru04} reported an FWHM of H$\beta$ of 1330~\kms\ and
an \ion{Fe}{ii}/H$\beta$ ratio of 0.94. No previous spectroscopic observations
of this source were found in the literature. Our NIR spectroscopy shows a
conspicuous emission line spectrum, with broad permitted lines displaying
FWHM values in the range 800~\kms-2300~\kms. The lower values correspond to those
measured in the \ion{Fe}{ii}, \ion{Ca}{ii}, and \ion{O}{i} lines, while 
the large ones were found in Pa$\alpha$. High ionization lines of [\ion{S}{viii}] 
and [\ion{S}{iii}], as well as [\ion{S}{iii}] emission were detected. In the 
{\it K}-band, molecular H$_{2}$ is clearly present. The
secure detection of these molecular lines supports 
the idea for a starburst component on this source because of the
possible Wolf-Rayet features found in the optical region by \citet{lip03}.
The continuum emission is steep and featureless of a power-law type. 

\item  {\bf Mrk\,291}. Was classified as an NLS1 galaxy in \citet{goo89}. So far 
there has been no report of NIR spectroscopy on this object. At
first glance, the NIR spectrum only shows very narrow permitted lines, similar to 
those observed in Seyfert~2 galaxies. Weak broad components are seen in 
Br$\gamma$ (FWHM$\sim$1600~\kms),
Pa$\alpha$ (FWHM$\sim$2200~\kms), and Pa$\beta$ (FWHM$\sim$1800~\kms).
\ion{O}{i}, \ion{Ca}{ii}, and \ion{Fe}{ii} lines, all emitted only by
the BLR, are seen in the spectrum. The FWHM of these lines, however,
reaches only $\sim$1000~\kms. The forbidden line spectrum is rather poor, with
only [\ion{S}{iii}] and [\ion{Fe}{ii}] detected. Molecular lines of H$_{2}$
are seen in the {\it K}-band. The continuum emission is steep and featureless
of a power-law type.

\item  {\bf Mrk\,493}. Classified by \citet{oter85} as NLS1, Mrk\,493 displays
very narrow optical broad permitted lines. \citet{per88}, for instance, reports
an FWHM$\approx$650~\kms\ for \ion{Ca}{ii} and 450~\kms\ for both \ion{O}{i} and 
\ion{H}{i}. \citet{wbb96} note that the ROSAT PSPC spectrum is very 
steep and cannot be fitted by a power law with cold absorption. A power 
law with either a warm absorption or a soft black-body component fits
the data equally well. The only NIR spectrum reported to date for this source
is the {\it K}-band spectroscopy of \citet{gt02}. It shows 
Br$\gamma$ in emission, but this appears rather weak.
Our {\it JHK} spectrum displays numerous emission lines, mostly
permitted ones. Pa$\alpha$, \ion{He}{i}~1.083~$\mu$m, and 
\ion{O}{i}1.1287~$\mu$m were the brightest emission lines.
There is also a conspicuous \ion{Fe}{ii} emission line spectrum,
with primary fluorescent lines at 9200~\AA. The BLR 
profiles display a large interval in FWHM. Exclusive BRL emission 
features like \ion{O}{i} and \ion{Fe}{ii} show FWHM of $\approx$700~\kms\,
while the broad component of Pa$\beta$ has an FWHM of $\sim$1900~\kms. 
Forbidden emission lines of [\ion{S}{iii}] and [\ion{Fe}{ii}]
are detected but are weak. In the {\it K}-band, H$_{2}$ 21210~\AA\  
is relatively strong, indicating that hot molecular gas exists
in the circumnuclear region of this source.

\item  {\bf PG\,1519+226}. The NIR spectrum of this source, the first one reported in
the literature, is dominated by BLR line emission features. 
Pa$\alpha$, \ion{He}{i}~1.083~$\mu$m and \ion{O}{i}~8446~\AA\ are the brightest lines with
FWHM of $\approx$3500~\kms\ for the former two and $\approx$1700~\kms\ for the last.
Evidence of [\ion{S}{iii}]~9500~\AA\ was found. No other forbidden NLR features
were detected. The continuum emission displays the typical turnover at 1.2~$\mu$m. Bluewards,
it is very steep. Redwards, it becomes flat. No signs of stellar absorption
features were detected.

\item  {\bf NGC\,5929}.  This CfA Seyfert~2 galaxy \citep{hb92} resides in a strong interacting 
system and seems to share a common outer envelope with its neighbor, NGC 5930 \citep{nw99}.
The {\it K}-band spectroscopy for this object, published by \citet{ia04} 
and \citet{iva00}, shows a continuum dominated by CO absorption features and points out
towards a moderate circumnuclear starburst component. Our NIR
spectrum displays strong [\ion{S}{iii}]~9531~\AA, \ion{He}{i}, and [\ion{Fe}{ii}] 
emission and weak \ion{H}{i}. Conspicuous molecular hydrogen lines were detected in
the {\it K}-band. No high ionization lines were observed. The continuum is steep with
strong stellar CO absorption lines in {\it H} and {\it K}. 

\item  {\bf NGC\,5953}. This Arp-Madore galaxy is interacting with NGC~5954 
\citep{arp66}. It has a Seyfert~2 nucleus \citep{raf90}, 
surrounded by a ring of star formation with a radius of $\sim$4". Long-slit
spectroscopy shows that in the circumnuclear region a starburst coexists
with moderate-excitation gas ionized by the active nucleus \citep{gp96b}. The
{\it J}-band spectroscopy of \citet{alo00} shows a spectrum dominated by 
[\ion{Fe}{ii}]~1.257~$\mu$m and weak Pa$\beta$ emission. Our spectrum 
displays low- and moderate-ionization emission lines, 
on top of a steep continuum with numerous stellar absorption features. 
The \ion{Ca}{ii} triplet in absorption dominates the blue end of the spectrum,
while strong 2.3~$\mu$m CO bandheads are observed at the red edge. The CO absorption bands
are also strong in {\it H}. The highest (and most intense) forbidden 
line detected is [\ion{S}{iii}]. In addition, [\ion{Fe}{ii}] emission is 
prominent in the {\it J} and {\it H}. In {\it K}, the most conspicuous
emission features are from H$_2$. Br$\gamma$ is rather weak, and Pa$\alpha$
is severely affected by telluric absorption. A comparison of the NGC~5953
spectrum with the ones discussed in this section allows us to conclude that 
NGC~5953 is the edge of a Seyfert~2/LINER classification.

\item  {\bf PG\,1612+261}. Is a radio-quiet quasar that displays a very
interesting radio structure very likely associated to a one-sided jet \citep{kuk98}.
No previous NIR spectroscopy has been published on this object. Our spectrum
displays one of the flattest continuum of all Type~1 sources of the sample 
redwards of 1~$\mu$m. A small NIR blue excess was detected.  The contribution
of the NLR to the integrated spectrum is moderate. Strong
[\ion{S}{iii}] as well as high-ionization lines of [\ion{S}{viii}],
[\ion{S}{vi}], and [\ion{Si}{vi}] were detected. In addition, molecular
lines of H$_{2}$ are conspicuous in the {\it K}-band. The broad components
of the permitted lines display a large interval in FWHM. 
Pa$\alpha$, for instance, has an FWHM of $\sim$ 3800~\kms\, while
that of \ion{O}{i} is 1930~\kms. The \ion{Fe}{ii} emission is rather
weak.

\item  {\bf Mrk\,504}. This Palomar-Green galaxy was classified as a NLS1 by \citet{op87}.
It displays a moderate emission line spectrum, dominated by \ion{He}{i}~1.083~$\mu$m
\ion{H}{i} and [\ion{S}{iii}]. The broad components of the permitted lines
have FWHM consistent with its classification as an NLS1. Br$\gamma$, Pa$\alpha$ and
\ion{O}{i} display FWHM of 1860~\kms, 2560~\kms\ and 1330~\kms, respectively.
Overall, the NLR spectrum is rather weak. The continuum is very steep, with a
power-law form. The CO absorption bands in {\it H} and {\it K} are detected,
evidencing the presence of circumnuclear stellar population. 
   
\item  {\bf 3C\,351}. A very steep radio source and the second most distant object
of the sample, 3C~351 is a lobe-dominated, radio-loud
QSO (log$R$ = 2.81) with moderate-strength X-ray and ultraviolet absorption by
ionized gas \citep{bra00}. UV/optical spectroscopy published by \citet{cor96} shows a 
flat optical spectrum and a red UV continuum. In the optical, unusually strong narrow 
[\ion{O}{iii}] lines are observed on top of a very broad H$\beta$ (FHWM$\sim$6560~\kms) 
line. No \ion{Fe}{ii} emission was detected. Due to the redshift of this object,
our NIR spectrum covers from H$\alpha$ to the {\it H}-band, allowing the
simultaneous detection of Balmer and Pashen lines. In addition to the
extremely broad features in the permitted lines, which reach $\approx$12100~\kms\
of FWHM and 25000~\kms\ at FWZI in H$\alpha$, we found that the peak of the
broad component in Pa$\beta$ has a high blueshift, of 2340~\kms, relative
to the centroid position of the narrow component. This extremely high
blueshift is not detected in H$\alpha$. In this line, a shift of
only 365~\kms\ was measured between the peak position of the narrow and broad
components. Note that Pa$\beta$ is isolated from other
permitted lines, ruling out blending effects with nearby broad features to explain 
such a high blueshift. It cannot be due to errors in the wavelength calibration,
as the narrow component of Pa$\beta$ is located at the expected position. Narrow 
forbidden emission features of [\ion{N}{ii}] and [\ion{S}{iii}] were also 
detected, meaning that 3C\,351 has a noticeable NLR. The continuum emission is
featureless and very steep to the blue. Redwards of 1.2~$\mu$m, it
becomes rather flat. No evidence of absorption lines was found.

\item  {\bf Arp\,102B}. Arp 102B is the archetype double-peaked broad-line radio 
galaxy \citep{chf89}. \citet{corb98} note that the  H$\alpha$
profile is extremely broad and has two prominent peaks displaced
by 4970$\pm$150 and 7200$\pm$300 \kms\ to the blue and red,
respectively, of the narrow component of H$\alpha$. Though less prominent, a
third, central broad component can clearly be seen underlying the narrow
H$\alpha$ and [\ion{N}{ii}]~6548,~6583~\AA\ lines. Arp 102B is also
a low-luminosity radio galaxy  which at
6~cm exhibits a bright unresolved core with a relatively faint extended tail
\citep{corb98}. In our NIR spectrum no trace of double-peaked \ion{H}{i} lines
were detected. As judged by its overall appearance, and if we compare it
to the other spectra discussed in this section, in the NIR Arp\,102B 
is more similar to a LINER-like 
object than to a Seyfert~1. Moreover, the \ion{H}{i} lines, both 
the Bracket and Pashen series, are almost absent. Pa$\alpha$ and Pa$\beta$
are the only ones that are detected but the former is affected by telluric
absorption. In contrast, strong \ion{He}{i}, [\ion{S}{iii}], and [\ion{Fe}{ii}]
lines were observed. In addition, [\ion{C}{i}], [\ion{S}{ii}], and
molecular H$_{2}$ transitions are also found in the data. The most remarkable 
feature in the spectrum is a
large blue wing seeing in  \ion{He}{i}~1.083~$\mu$m that resembles a broad
component to this line. Moreover, the peak of the narrow component is
displaced to the red by 415~\kms. This displacement cannot be attributed to a bad
wavelength calibration of the spectrum, as all other emission features
appear at the expected position. The continuum emission is blue, decreasing
in flux towards longer wavelengths, with deep stellar absorption bands of CO in
{\it H} and {\it K}. 

\item  {\bf 1H\,1934-063}. The X-ray NLS1 galaxy 1H~1934-063 have been studied in detail in
the optical and NIR regions by \citet{ara00} and \citet{ara02}. It displays a very rich spectrum
with strong permitted lines of \ion{H}{i}, \ion{He}{i}, and \ion{O}{i} and strong
forbidden emission of [\ion{S}{iii}]. Forbidden high-ionization lines are also
conspicuous across the NIR. Emission lines of \ion{Fe}{ii}, probably produced by
Ly$\alpha$ fluorescence, are also conspicuous in the {\it J}-band. The continuum
emission is featureless of a power-law type. To the blue, a small excess of emission
over the power-law continuum is seen. No evidence of circumnuclear stellar
population was found in this object.  

\item  {\bf Mrk\,509}. The Seyfert~1 galaxy Mrk\,509 is known because of its high
X-ray variability as its continuum emission flux in the interval 2-10 keV 
changes by as much as a factor of 2,  and the iron line is detected in only 
five of 11 observations, for instance \citep{wea01}. It is a well-studyied
source, from radio to high-energies. The NIR spectrum is dominated
by emission from the BLR. The lines are broad, with FWHM varying from 
2800~\kms\ for \ion{O}{i} and \ion{Fe}{ii} up to $\sim$6300~\kms\ for the
\ion{H}{i} lines. The NLR emission from lines of [\ion{S}{ii}], [\ion{S}{iii}],
and [\ion{Si}{vi}] are clearly present, as well as emission from
the H$_{2}$ 21210~\AA\ line. The continuum emission is rather blue and
featureless, with a strong NIR excess of emission shortward of 10000~\AA\

\item  {\bf Mrk\,896}. Classified as an NLS1 galaxy by \citep{ver01}.
Our NIR spectrum is dominated by
permitted lines of \ion{He}{i}~1.083~$\mu$m, \ion{H}{i}, and \ion{O}{i}. Weak permitted
\ion{Fe}{ii} is also detected (e.g. $\lambda$10500$\AA$).
The FWHM of the broad components of these lines ranges from 1100~\kms\
to 1500~\kms, confirming its classification as an NLS1. Only medium to
high-ionization forbidden NLR features were detected ([\ion{S}{iii}], 
[\ion{Si}{vi}], and [\ion{Si}{x}]). Some molecular lines of H$_2$ are 
visible in the {\it K}-band. The continuum emission is steep, decreasing 
in flux with wavelength. The CO absorption lines, including the bandheads at
2.3~$\mu$m are distinguished in the spectrum, implying a young stellar component
in the circumunclear region. 
    
\item  {\bf 1H~2107-097}. This Seyfert~1 galaxy was the subject of a multiwavelength
study, from X-rays to radio, by \citet{gro96}. Among their most important
findings are the V band variability, which was observed 
to change by a factor of 1.8 in 6 weeks, and the strong coronal line
emission in the optical region. \citet{gro96} also called the attention to
the intrinsically weak blue bump even after correcting by
reddening. They concluded that weak blue bumps are, therefore, not always 
an artifact caused by extinction. Our NIR spectrum,  is
dominated by BLR features with strong \ion{He}{i} and \ion{H}{i} lines.
We also detected permitted transitions of \ion{Fe}{ii} and \ion{O}{i}. As
is the rule for Type~1 galaxies, the last two lines have smaller widths
(FWHM$\approx$1800~\kms) than the broad components of \ion{H}{i}
(FWHM$\approx$3600~\kms). The NLR
spectrum is dominated by [\ion{S}{iii}] and high-ionization lines of 
[\ion{Si}{vi}], [\ion{Si}{x}] and [\ion{Ca}{viii}]. The continuum 
emission is featureless with a power-law form, and a strong NIR bump in
the blue region. No stellar absorption features were
detected.

\item  {\bf Ark\,564}. Classified as an NLS1 galaxy by \citet{goo89}, Ark\,564 
is one of the best-studied objects in the X-ray because of its 
brightness in the 2-10 keV band \citep{coll01}. The results of an intensive 
variability campaign on several wavelength bands show 
that the optical continuum is not significantly 
correlated with  the X-ray \citep{shem01}.
The ionization state of the gas, as described by
\citet{cren02}, is relatively high.
They found that at least 85\% ~of the 
narrow emission-line flux comes from a region 
$\leq$ 95\,pc from  the nucleus and surrounded by a dust 
screen associated to a ``lukewarm'' absorber. Ark~564 is also known for the narrowness of its
permitted lines, with FWHM in the range 600\kms - 1030\kms. 
The NIR spectroscopic properties were analyzed in detailed by 
\citet{crv03}. The spectrum shown here was taken three years later at a better
S/N but essentially all spectroscopic features detected here have already been
described by \citet{ara02} and \citet{ara02c}. It displays a very rich
emission line spectrum, with bright high-ionization lines. The continuum
emission is featureless and well-described by a broken power-law.

\item  {\bf NGC\,7469}. This well-studied Seyfert~1 galaxy, is widely known because
the active nucleus is surrounded by a more or less complete ring of powerful
starburst activity \citep{mau94,mhh94,gen95}. 
The circumnuclear ring has a luminosity equivalent to two-thirds of 
the bolometric luminosity of the entire galaxy. It contains a number 
of supergiant star formation regions with a few 10$^4$ OB stars each.
It has been studied in the NIR by several authors, among them,
\citet{gen95}, \citet{tho96}, and \citet{sosa01}.
The NIR spectrum of \citet{gen95} display bright lines of Br$\gamma$, 
[\ion{Fe}{ii}], [\ion{Si}{vi}], H$_2$, \ion{He}{i}, and CO on a scale 
of less than a few hundred parsecs. The NIR spectrum shown 
in Fig.~\ref{indiv7}, to our knowledge the first one that 
simultaneously covers the interval 0.8~$\mu$m--2.4~$\mu$m, displays a wealth of emission
lines, with \ion{He}{i}~1.083~$\mu$m and [\ion{S}{iii}]~9531~\AA\ 
as the strongest ones. High-ionization lines of [\ion{S}{viii}],
[\ion{S}{ix}], [\ion{Si}{vi}], and [\ion{Si}{x}] were identified. The
continuum emission is steep, of a broken power-law type.
Stellar absorption features were detected mostly in the {\it H}- and
{\it K}-bands. In the latter, the 2.3~$\mu$m CO bandheads are prominent for a
Type~1 object.

\item  {\bf NGC\,7674}. Classified as a Seyfert~2 \citep{od83}, NGC\,7674 is a
CfA galaxy \citep{hb92} in interaction with UGC~12608 \citep{lwh01}. NGC~7674 
displays broad emission in polarized lines, which were first detected in this
object by \citet{mg90}. The extremely bright nuclear point
source compared to the other Seyfert~2 galaxies (with the exception of
NGC~1068) reinforces its interpretation as an obscured Seyfert~1.
Our NIR spectrum is ambiguous regarding the classification of this source.
It is a Seyfert~2 because of the absence of \ion{O}{i} and \ion{Fe}{ii}.
However, most \ion{H}{i} lines display a conspicuous
broad component, particularly strong in Pa$\alpha$ and Br$\gamma$, 
where it reaches an FWHM of $\sim$3000~\kms. This width is similar to
the one detected in polarized light for H$\beta$ \citep{tran95}. Note that the peak of
the broad component is blueshifted relative to the systemic velocity
of the galaxy  by $\sim$480~\kms\ in Pa$\alpha$ and $\sim$300~\kms\
in Br$\gamma$. Such a shift, of lower value ($\sim$100~\kms), is also
found in the polarized broad component reported by \citet{tran95}.
\citet{vgh97} suggests that the broad
component is not associated to the classical BLR because it is
similar in width to the one found in [\ion{O}{iii}]~5007~\AA. 
We checked the [\ion{S}{iii}]~9531 line, and indeed it has a
blue wing, well-fitted by a broad component of $\sim$1550~\kms,
half the value found in Pa$\beta$. Other lines, such as 
[\ion{Fe}{ii}]~1.257~$\mu$m, display a similar component to that of
sulfur. We therefore conclude that we are looking at 
a genuine BLR feature in NGC~7674. The continuum emission of this object is 
peculiar. It is nearly flat from the blue end up to $\sim$1.2~$\mu$m. Redwards, it 
increases with wavelength, with a shape very similar to that found in
Mrk~1239 and Mrk~766. In addition, NGC~7674 shows large
polarization, supporting the hypothesis that the excess of 
NIR continuum emission in {\it H} and {\it K} are very likely due 
to hot dust.

\item  {\bf NGC\,7682}. Is a CfA Seyfert~2 \citep{hb92} in interaction 
with NGC~7683 \citep{arp66}. Ionized gas in H$\alpha$+[\ion{N}{ii}] 
and [\ion{O}{iii}]~5007~\AA\ is detected
on scales of kiloparsecs on this object \citep{bro87,dur94}.
In the NIR, only {\it K}-band spectroscopy was previously reported
by \citet{ia04}. Our spectrum displays conspicuous emission lines
with bright [\ion{S}{iii}], \ion{He}{i}, and \ion{H}{i}.
Low-ionization lines such as [\ion{C}{i}], [\ion{S}{ii}], and
[\ion{Fe}{ii}] as well as high-ionization lines of [\ion{S}{viii}] 
and [\ion{Si}{vi}] were clearly detected. In the {\it K}-band,
molecular H$_{2}$ and Pa$\alpha$ were the brightest emission
features. All lines were spectroscopically unresolved or barely
resolved. The continuum emission is dominated by absorption 
features, including the \ion{Ca}{ii} triplet in the blue and 
numerous CO bands in {\it H} and {\it K}. 
  
\item {\bf NGC\,7714}. Is described by \citet{wee81} as an archetype of the 
starburst nucleus galaxies. According to \citet{kbc93}, the burst of star formation 
is thought to be caused by interaction with the companion NGC~7715. 
The central region of about 330~pc has been the site of active star
formation at a rate of about 1~M$\odot$ yr$^{-1}$ for some 10$^{8}$ years
\citep{bran04}. NGC~7714 is also classified as a Wolf-Rayet galaxy
because of its strong \ion{He}{ii}~4686~\AA\ line \citep{gon95}. 
Recent Spitzer observations of this object by \citet{bran04} show that it has an 
\ion{H}{ii} region-like spectrum with strong polycyclic aromatic 
hydrocarbon emission features. No evidence of an obscured AGN was found.  
With very little silicate absorption and a temperature 
of the hottest dust component of 340~K, NGC~7714 is defined by \citet{bran04}
as the perfect template for a young, unobscured starburst. However, we measured 
an E(B-V)=0.47 from our spectrum, based on three different indicators, implying 
the existence of dust along the line of sight to this source.
Although it has been extensively studied in the mid-infrared, 
\citet[see, e.g.,][]{dud99,oha00,bran04}, our SpeX spectrum is the first one
published with simultaneous {\it JHK} spectroscopy. Our data shows
unresolved emission lines of [\ion{S}{iii}], \ion{He}{i}, \ion{H}{i},
[\ion{C}{i}], [\ion{Fe}{ii}], and H$_{2}$. The continuum emission is steep, 
decreasing in flux towards longer wavelengths and dominated by absorption 
lines and bands across the whole NIR region, the strongest ones
being those of \ion{Ca}{ii} and  CO, including the bandheads at 23000~\AA.
\end{itemize}

\section{Final remarks}\label{s4}

We have presented the most extensive NIR spectral atlas of AGN to date. This
atlas offers a suitable database for studying the continuum and line emission 
properties of these objects in a region full of interesting features.
Ionization codes and models built to study the physical properties of AGNs need to 
include the constraints provided here
in order to fully describe the state of the emitting gas. 

The continuum and line emission properties of each subtype 
of active nucleus are described. In addition, we provide flux measurements 
of the lines detected in each of the 51 sources, distributed as follows: 
12 Seyfert~1, 13 narrow-line Seyfert 1, 7 quasars, 15 Seyfert~2 and 4 starburst. 

We found that the continuum of quasars, Seyfert~1s, and NLS1s are rather
similar and well-described by a broken power-law. At 1.2~$\mu$m, most objects
display a clear turnover in the continuum, changing from a steep blue continuum
shortwards of the breaking point to one being essentially flat or nearly flat
redwards. The steepness of the continuum bluewards of 1.2~$\mu$m changes from source
to source and we associate it to the extrapolation of the power-law that 
characterizes the UV/optical continuum of Type~1 sources. The exception to this
trend is Mrk~1239, which displays a remarkable bump of emission over the underlying
power-law, peaking at 2.2$\mu$m. This bump is accounted for by emission
from hot dust at T$\sim$1200~K. The Mrk~766 and NGC~7674 (which optically is
classified as Seyfert~2) show evidence of a much weaker but similar bump.
The continuum of all quasars is featureless. Some Seyfert~1s and NLS1s
show evidence of underlying stellar population
as told from the absorption features, mostly of CO, present in the {\it H} and 
{\it K} bands. 

In contrast to Type~1 objects, the continuum of the Seyfert~2s displays 
a strong young stellar component. In most objects, it decreases steeply in flux
with wavelength across the NIR, similar in shape to the continuum observed
in the starburst galaxies NGC~3310 and NGC~7714. Mrk~1066 and  NGC\,2110 are
somewhat peculiar as the continuum in the $z$+$J$ first increases in flux,
then becomes flat for a small wavelength interval and decreases in flux from 1.4~$\mu$m 
redwards, resembling the continuum seen in the
starburst galaxies NGC~1614 and NGC~34 . In NGC\,1275, NGC\,262, NGC\,7674, and Mrk\,1210, 
the continuum in the {\it K}-band rises with wavelength, suggesting the
presence of hot dust.  Strong absorption 
bands of CO were found in the {\it H} and {\it K} bands, except in the last four
Seyfert~2s. The \ion{Ca}{ii} triplet in absorption, as well as the CN band at
1.1~$\mu$m  are also seen in the vast majority of objects.  An atlas of 
absorption lines and a study of the stellar populations of these galaxies 
will be carried out in a separate publication (Riffel et al. 2006, in 
preparation). 

The NIR emission line spectrum varies from source to source and according 
to the type of activity. We found that [\ion{S}{iii}]~9531~$\AA$ and 
\ion{He}{i}~1.083~$\mu$m are, 
by far, the strongest lines in the NIR. They were detected in all 51 objects 
of the sample. Neutral oxygen, permitted \ion{Fe}{ii} transitions and the \ion{Ca}{ii} triplet in
emission are features seen only in Type~1 sources (Seyfert~1s, NLS1s and quasars).
These lines are absent in the spectra of the Sy~2, even in those objects that display
in the NIR genuine broad line components in the \ion{He}{i} lines. It
confirms previous suggestions that they are exclusive BLR features \citet[and 
references therein]{ara02b}. Therefore, they are useful indicators of the Seyfert 
type. Note, however, that the \ion{Fe}{ii} seems to be absent or rather weak
in radio-loud sources. 

Molecular H$_{2}$, as well as \fe2\ lines is present in almost all targets,
including quasars and Seyfert 1 galaxies. Moreover,
\h2 2.121\,$\mu$m is more intense relative
to Br$\gamma$ in Sy~2s than in starburst galaxies 
(see Figs.~\ref{plsy2} and \ref{plsb}),
suggesting that the AGN may play an important role in the excitation of the molecular 
gas in AGNs. Other NLR features worth mentioning are the forbidden high 
ionization lines, which were detected in the spectra of both Type~1 and Type~2
objects but are completely absent in the starburst galaxies. Therefore, their
detection is a clear signature of AGN activity.
The commonest coronal lines are [\ion{Si}{x}]\,1.43~$\mu$m and 
[\ion{Si}{vi}]\,1.963~$\mu$m. The presence of the 
coronal lines in the spectra of Sy~2 galaxies (i.e. Mrk\,573, NGC\,591, NGC\,1275 and NGC\,7674) 
suggests that the coronal line region is located very likely in the inner portion of
the NLR.

We found that the ratio Pa$\beta$/Br$\gamma$ rules out Case B recombination
values in some Type~1 sources and  it is very close to its
intrinsic value in a large fraction of these objects. This result shows 
that hydrogen recombination lines are not a suitable indicator of reddening 
for broad-line AGN. In contrast, the flux ratio between the
forbidden \fe2\ lines 1.257~$\mu$m/1.644~$\mu$m agrees, within errors, with 
the extinction measured by means of the Pa$\beta$/Br$\gamma$ in most Seyfert~2s,
allowing us to conclude that the former can also be applied, with confidence,
in Type~1 objects. We also found that the steepness of the continuum in 
Type~1 sources is not correlated with the extinction measured by means of the
\fe2\ lines. It suggest that the contribution of the BLR in the NIR
continuum is still larger than initially thought. In comparison, the form of 
the continuum in a fraction of Seyfert~2 galaxies appears to be related
to the amount of extinction measured for the NLR. 


\begin{acknowledgements}

This paper was partially supported by the Brazilian funding agency
CNPq(304077/77-1) to ARA. This research made use of the NASA/IPAC 
Extragalactic Database (NED), which is operated by the Jet Propulsion 
Laboratory, California Institute of Technology, under contract with 
the National Aeronautics and Space Administration. The
authors thank the anonymous referee for useful comments about
this manuscript.

\end{acknowledgements}


\setcounter{table}{0}
\begin{scriptsize}
\begin{longtable}{clcccccccc}
\caption{\label{obslogs} Observation log and basic galactic properties for the sample.} \\
\hline
\noalign{\smallskip}
&	    &	   &		 &  E(B-V)$_{\rm G}$ &  Date of     & Exposure  &	   &  PA     &     r	      \\
ID&Galaxy   & Type & $ z $	 &   (mag)	     &  observation & Time (s)  & Airmass  &  $\deg$ & (pc)$\rm ^a$  \\
 (1)& (2)   &  (3) & (4)	 &  (5) 	     &    (6)	    &	(7)	&   (8)    & (9)     &  (10)	      \\
\noalign{\smallskip}
\hline \hline
\noalign{\smallskip}
\endfirsthead
\caption{Continued.} \\
\hline
\noalign{\smallskip}
ID &Galaxy      & Type &  z         &  E(B-V)G     &   Date   & Exp. Time  & Airmass  &  PA  &  r  \\
\noalign{\smallskip}
\hline \hline
\noalign{\smallskip}
\endhead
\hline
\endfoot
1 &   Mrk\,334     & Sy1  &  0.021956   & 0.047	& 2003 Oct 23  & 1920	 & 1.00  & 303  & 340 \\ 
2 &   NGC\,34      & SB/Sy2&  0.019784   & 0.027	& 2003 Oct 24  & 1680	 & 1.19  & 0	& 230 \\
3 &   NGC\,262     & Sy2  &  0.015034   & 0.067	& 2003 Oct 24  & 1200	 & 1.03  & 160  & 175 \\
4 &   Mrk\,993     & Sy2  &  0.015537   & 0.060	& 2003 Oct 23  & 1920	 & 1.03  & 160  & 241 \\
5 &   NGC\,591     & Sy2  &  0.015167   & 0.046	& 2003 Oct 24  & 1800	 & 1.05  & 160  & 206 \\
6 &   Mrk\,573     & Sy2  &  0.017259   & 0.023	& 2003 Oct 24  & 1800	 & 1.10  & 40	& 267 \\
7 &   NGC\,1097    & Sy1  &  0.004253   & 0.027	& 2003 Oct 23  & 720	 &  1.6  &  0	&  58 \\
8 &   NGC\,1144    & Sy2  &  0.028847   & 0.072	& 2003 Oct 24  & 1500	 & 1.09  & 27	& 447 \\
9 &   Mrk\,1066    & Sy2  &  0.012025   & 0.132	& 2003 Oct 23  & 1920	 & 1.07  & 146  & 186 \\
10&   NGC\,1275    & Sy2  &  0.017559   & 0.163	& 2003 Oct 23  & 1440	 & 1.16  & 135  & 272 \\
11&   NGC\,1614    & SB   &  0.015938   & 0.154	& 2003 Oct 24  & 1800	 & 1.14  & 0	& 154 \\
12&   MCG-5-13-17  & Sy1  &  0.012642   & 0.017	& 2003 Oct 23  & 1440	 & 1.65  &  0	& 196 \\
  &   	           &      &  	        &	& 2003 Oct 24  & 1200	 & 1.69  &  0	& 196 \\
13&   NGC\,2110    & Sy2  &  0.007789   & 0.375	& 2003 Oct 23  & 1440	 & 1.14  &  20  & 121 \\
  &   	           &      &  	        &	& 2003 Oct 24  & 1200	 & 1.14  &  20  &  121 \\
14&  ESO\,428-G014 & Sy2  &  0.005664   & 0.197	& 2003 Oct 24  & 960	 & 1.56  & 345  &  88 \\
15&   Mrk\,1210    & Sy2  & 0.01406     & 0.030	& 2002 Apr 25  & 2700	 & 1.25  & 58	& 220 \\
16&   Mrk\,124     & NLS1 & 0.05710     & 0.015	& 2002 Apr 23  & 2640	 & 1.16  & 10	& 990 \\
17&   Mrk\,1239    & NLS1 & 0.01927     & 0.065	& 2002 Apr 21  & 1920	 & 1.08  & 0.0  & 335 \\
  &   	           &      &             &	& 2002 Apr 23  & 1920	 & 1.15  & 0.0  & 335 \\
18&   NGC\,3227    & Sy1  & 0.00386     & 0.023	& 2002 Apr 21  & 720	 & 1.00  & 158  & 67 \\
  &   	           &      &  	        &	& 2002 Apr 25  & 1080	 & 1.02  & 158  & 67 \\
19&   H1143-192    & Sy1  & 0.03330     & 0.039	& 2002 Apr 21  & 1920	 & 1.31  & 45	& 520 \\
20&   NGC\,3310    & SB   & 0.00357     & 0.022	& 2002 Apr 21  & 840	 & 1.21  & 158  & 56 \\
21&   PG1126-041   & QSO  &  0.060000   & 0.055 & 2002 Apr 24  & 2160	 & 1.12  & 10	& $>$1000 \\
  &   	           &      &  	        &	& 2002 Apr 23  & 1920	 & 1.14  & 21	& $>$1000 \\
22&   NGC\,4051    & NLS1 & 0.00234     & 0.013	& 2002 Apr 20  & 1560	 & 1.17  & 132  & 37 \\
23&   NGC\,4151    & Sy1  & 0.00345     & 0.028	& 2002 Apr 23  & 1800	 & 1.10  & 130  & 58 \\
24&   Mrk\,766     & NLS1 & 0.01330     & 0.020	& 2002 Apr 21  & 1680	 & 1.06  & 112  & 230 \\
  &   	           &      &  	        &	& 2002 Apr 25  & 1080	 & 1.02  & 112  & 230\\
25&   NGC\,4748    & NLS1 & 0.01417     & 0.052	& 2002 Apr 21  & 1680	 & 1.29  & 36	& 254 \\
  &                &      &  	        &	& 2002 Apr 25  & 1440	 & 1.21  & 36	& 254\\
26&   Ton\,0156    & QSO  &  0.549000   &  0.015& 2002 Apr 25  & 3600	 & 1.06  & 143  & $>$1000 \\
27&   Mrk\,279     & NLS1 & 0.03068     & 0.016	& 2002 Apr 24  & 3600	 & 1.54  & 0.0  & 480 \\
28&   NGC\,5548    & Sy1  & 0.01717     & 0.020	& 2002 Apr 23  & 1920	 & 1.05  & 112  & 298 \\
29&   PG1415+451   & QSO  &  0.114000   &  0.009&  2002 Apr 24 & 3960	 & 1.28  & 134  & $>$1000 \\ 
  &   	           &      &  	        &	& 2002 Apr 25  &  1440   & 1.14  & 143  & $>$1000 \\
30&   Mrk\,684    & Sy1  &  0.046079   &  0.021&  2002 Apr 21 & 1440	 & 1.02  & 172  & 980 \\  
31&   Mrk\,478     & NLS1 & 0.07760     & 0.014	& 2002 Apr 20  & 3240	 & 1.06  & 0.0  & 1200 \\
32&   NGC\,5728    & Sy2  & 0.01003     & 0.101	& 2002 Apr 21  & 960	 & 1.31  & 36	& 160 \\
33&   PG\,1448+273 & QSO  & 0.06522     & 0.029	& 2002 Apr 24  & 2160	 & 1.01  & 108  & 1020 \\
34&   Mrk\,291     & NLS1 & 0.03519     & 0.038	& 2002 Apr 21  & 2520	 & 1.04  & 84	& 550 \\
35&   Mrk\,493     & NLS1 & 0.03183     & 0.025	& 2002 Apr 20  & 1800	 & 1.07  & 0.0  & 500 \\
  &   	           &      &  	        &	& 2002 Apr 25  & 900	 & 1.04  & 0.0  & 500\\
36&   PG\,1519+226 & QSO  &  0.137000   & 0.043	& 2002 Apr 25  & 4000	 & 1.14  & 94	& $>$1000 \\
37&   NGC\,5929    & Sy2  &  0.008312   & 0.024	& 2004 Jun 01  & 1680	 & 1.38  & 116  &  193   \\
38&   NGC\,5953    & Sy2  &  0.006555   & 0.049	& 2004 Jun 02  & 2160	 & 1.18  & 0	&  165   \\ 
39&   PG\,1612+261 & QSO  & 0.13096     & 0.054	& 2002 Apr 2 3 & 2520	 & 1.10  & 107  & 2050 \\
40&   Mrk\,504     & NLS1 & 0.03629     & 0.050	& 2002 Apr 21  & 2100	 & 1.04  & 138  & 570 \\
41&   3C\,351      & BLRG &  0.371940   & 0.023 & 2002 Apr 24  & 3600	 & 1.36  & 170  & $>$1000\\  
42&   Arp\,102B    & Sy1  &  0.024167   & 0.024	& 2004 Jun 02  & 2880	 & 1.22  & 0	&  700 \\
43&   1H\,1934-063 & NLS1 & 0.01059     & 0.293	& 2004 Jun 02  &  2160   & 1.13  & 0    & 349 \\
44&   Mrk\,509     & Sy1  &  0.034397   &  0.057& 2003 Oct 23  & 1440	 & 1.16  &  0	& 730 \\
  &   	           &      &  	        &	& 2004 Jun 01  & 2160	 & 1.17  &  0	& 730 \\
45&    Mrk\,896    & NLS1 & 0.02678     & 0.045	& 2002 Apr 23  & 1440	 & 1.21  & 150  & 420 \\
  &   	           &      &             &	& 2002 Apr 24  & 1200	 & 1.18  & 150  & 420\\
  &   	           &      &  	        &	& 2002 Apr 25  & 1200	 & 1.17  & 150  & 420\\
46&    1H2107-097  & Sy1  & 0.026525    & 0.233	& 2003 Oct 24  & 1680	 & 1.15  &  338 & 565 \\
47&    Ark\,564    & NLS1 & 0.02468     & 0.060	& 2000 Oct 10  & 1500	 & 1.05  & 0.0  & 390 \\
  &   	           &      &  	        &	& 2003 Oct 24  & 2160	 & 1.05  & 250  & 383 \\
48&   NGC\,7469    & Sy1  &  0.016317   & 0.069	& 2003 Oct 23  & 1920	 & 1.03  & 303  & 253 \\
49&   NGC\,7674    & Sy2  &  0.028924   & 0.059	& 2003 Oct 23  & 2160	 & 1.02  & 303  & 448 \\
50&   NGC\,7682    & Sy2  &  0.017125   & 0.067	& 2003 Oct 24  & 2400	 & 1.06  & 314  & 179 \\
51&   NGC\,7714    &H\,II &  0.009333   & 0.052	& 2003 Oct 24  & 2400	 & 1.05  & 348  & 115 \\
\noalign{\smallskip}
\hline
\end{longtable}
$\rm ^a$ Radius of the integrated region
\end{scriptsize}

\begin{table*}

\begin{scriptsize}
\renewcommand{\tabcolsep}{0.70mm}
\caption{\label{flsy1_1} Observed fluxes, for Type~1 objects, in units of $\rm 10^{-15}\, erg \, cm^{-2} \, s^{-1}$.}    
\centering
\begin{tabular}{llccccccccccc}
\hline\hline
\noalign{\smallskip}
Ion  & $\rm \lambda_{lab}\,(\AA)$& Mrk\,334          &    NGC\,1097     & MCG\,-5-13-17    &    Mrk\,124         & Mrk\,1239          & NGC\,3227           &  H\,1143-182       & PG\,1126-041         & NGC\,4051           & NGC\,4151              \\
\noalign{\smallskip}
\hline \noalign{\smallskip}
\ion{O}{i}           & 8446     &	-	     &       -  	 &	 -	    &	13.59\pp1.35	 &  86.61\pp6.70       &       -	    &	    -		 &	     -  	& 79.93\pp4.61       & 130.94\pp15.86	      \\
\ion{Ca}{ii}	     & 8498	&	-	     &       -  	 &	-	    &	    -		 &	 -	       &       -	    &	    -		 &	     -  	&   -		     &      -		      \\
\lb\ion{S}{iii}\rb   & 9069	&   22.69\pp0.74     &  12.08\pp3.38	 & 18.33\pp0.46     &	4.37\pp0.30	 &  25.27\pp0.80       &  91.57\pp5.41      &  5.93\pp0.49	 &  11.15\pp1.00	& 26.92\pp2.38       & 266.76\pp6.10	      \\
\ion{Fe}{ii}         & 9127     &       -            &       -           &     -            &       -            &      -              &      -             &  1.52\pp0.38       &         -            &  6.70\pp1.67       &     -                  \\
\ion{Fe}{ii}         & 9177     &       -            &       -           &     -            &   2.46\pp0.61\dd   &      -              &      -             &  1.25\pp0.31       &         -            &  3.31\pp0.82       &  3.36\pp2.23\kk        \\
\ion{Fe}{ii}         & 9202     &       -            &       -           &     -            &   2.30\pp0.55      &      -              &      -             &  1.13\pp0.28       &         -            &  10.58\pp2.64      &     -                  \\
\ion{H}{i}	     & 9230	&	-	     &       -  	 &     -	    &	    -		 &	-	       &       -	    &     -     	 &      -       	&      -	     &  7.49\pp2.03	      \\
\ion{H}{i}	     & 9230(b)	&	-	     &       -  	 &     -	    &	    -		 &  66.21\pp2.83\ax\hh &       -	    &  54.60\pp8.61\ax	 &  65.96\pp4.16\ax\hh 	& 19.89\pp4.68\ax    & 39.71\pp7.40	      \\
\lb\ion{S}{iii}\rb   & 9531	&   56.91\pp0.52     &       -  	 & 38.71\pp0.79     &	17.55\pp0.24	 &  103.30\pp1.21      &  247.79\pp3.40     &  17.96\pp0.60	 &  46.28\pp3.71	& 75.91\pp1.71       & 706.19\pp7.23	      \\
\lb\ion{C}{i}\rb     & 9824	&  2.01\pp0.53       &       -  	 &     -	    &	    -		 &	 -	       &       -	    &	    -		 &	1.79\pp0.14  	&	  -	     &      -		      \\
\lb\ion{C}{i}\rb     & 9850	&   3.29\pp0.53      &       -  	 & 4.64\pp1.60      &	1.00\pp0.10	 &	 -	       &  13.30\pp1.74      &	    -		 &      1.15\pp0.11  	& 5.92\pp1.41	     & 54.04\pp5.63	      \\
\lb\ion{S}{viii}\rb  & 9910	&	-	     &       -  	 & 6.61\pp1.54      &	    -		 &  23.79\pp1.85       &  1.97\pp1.14	    &	    -		 &	2.58\pp0.21  	& 13.67\pp1.97       & 40.50\pp2.08	      \\ 
\ion{Fe}{ii}	     & 9999	&   2.04\pp0.51      &       -  	 &	-	    &   2.75\pp0.17 	 &	 -	       &       -	    &	    -		 &	     -  	& 11.21\pp1.63       &      -		      \\
\ion{H}{i}	     & 10049	&   2.06\pp0.22      &       -  	 &	-	    &	    -		 &   34.19\pp0.89      &     -  	    &  14.57\pp1.94	 &	  -		&  5.35\pp0.65       & 32.17\pp1.87	      \\
\ion{H}{i}	     & 10049(b) &   5.14\pp0.73      &       -  	 &	-	    &	    -		 &   46.26\pp1.64      &  43.84\pp7.83\ax   &  76.06\pp5.65	 &  104.83\pp3.76\ax	& 33.84\pp2.77       & 288.46\pp14.89$^i$	      \\
\ion{He}{ii}	     & 10122	&   1.69\pp0.36      &       -  	 &	-	    &	    -		 &   5.53\pp0.63       &	-	    &	    -		 &	     -  	&  3.11\pp0.63       & 28.55\pp1.51	      \\
\ion{He}{ii}	     & 10122(b) &   2.79\pp0.68      &       -  	 &	-	    &	    -		 &   59.28\pp2.85      &  12.68\pp2.22\ax   &  3.59\pp1.52\ax	 &	     -  	& 14.25\pp2.11       & 65.57\pp7.13$^j$	      \\
\ion{Fe}{ii}	     & 10171	&	-	     &       -  	 &	-	    &	    -		 &	 -	       &       -	    &	    -		 &	     -  	&    -  	     &      -		      \\
\lb\ion{S}{ii}\rb    & 10286	&    1.22\pp0.12     &       -  	 &	-	    &	    -		 &	 -	       &  13.28\pp2.68      &	1.14\pp0.31	 &	     -  	&  1.89\pp0.60       & 28.00\pp0.28	      \\
\lb\ion{S}{ii}\rb    & 10320	&    1.39\pp0.12     &       -  	 &	-	    &	    -		 &	 -	       &  13.73\pp2.68      &	1.27\pp0.31	 &	     -  	&  4.00\pp0.60       & 38.35\pp0.28	      \\
\lb\ion{S}{ii}\rb    & 10336	&    0.48\pp0.12     &       -  	 &	-	    &	    -		 &	 -	       &  13.40\pp2.68      &	0.82\pp0.31	 &	     -  	&  1.41\pp0.60       & 24.08\pp0.28	      \\
\lb\ion{S}{ii}\rb    & 10370	&    0.44\pp0.12     &       -  	 &	-	    &	    -		 &	 -	       &   2.78\pp1.43      &	  -		 &	     -  	&     - 	     & 10.40\pp0.28	      \\
\lb\ion{N}{i}\rb     & 10404	&    1.21\pp0.18     &       -  	 &	-	    &	    -		 &	 -	       &  13.44\pp4.13      &	  -		 &	     -  	& 2.71\pp0.59	     & 16.10\pp0.36	      \\
\ion{Fe}{ii}	     & 10500	&	-	     &       -  	 &	-	    &	    -		 &  18.66\pp0.91       &       -	    &	    -		 &    13.04\pp2.77	&   -		     &      -		      \\
\ion{He}{i}	     & 10830	&   30.95\pp0.70     &       -  	 &  57.80\pp1.88    &	 7.66\pp0.21	 &   98.38\pp1.71      &  163.98\pp1.95     &  73.86\pp1.66	 &	     -  	&  75.88\pp2.31      &  685.10\pp5.21	       \\
\ion{He}{i}	     & 10830(b) &   33.29\pp3.83     &       -  	 &  125.57\pp15.40  &	25.76\pp1.39	 &  173.81\pp4.28      &  298.21\pp7.37     &  255.46\pp5.22	 &  133.30\pp3.65\ax	& 115.37\pp7.55      & 1470.29\pp56.18         \\
\ion{H}{i}	     & 10938	&    7.91\pp0.58     &       -  	 &  2.01\pp0.06	    &	 1.13\pp0.17	 &   44.01\pp2.44      &   20.46\pp2.61     &  38.35\pp2.78	 &	   -		& 19.14\pp3.38       &  50.28\pp5.64	      \\
\ion{H}{i}	     & 10938(b) &    8.56\pp1.82     &       -  	 &  8.76\pp0.28	    &	 9.38\pp0.71	 &   87.04\pp7.32      &  122.01\pp7.43     &  98.31\pp6.45	 &  95.82\pp4.80\ax	& 41.47\pp7.66       & 295.02\pp50.94	      \\
\ion{Fe}{ii}	     & 11126	&	-	     &       -  	 &	-	    &	    -		 &	  -	       &       -	    &	    -		 &	     -  	&   -		     &      -		      \\
\ion{O}{i}	     & 11287	&   6.32\pp1.03      &       -  	 &	-	    &	10.93\pp0.79	 &  50.03\pp1.76       &       -	    &  27.93\pp0.89	 &   43.98\pp4.08	& 47.16\pp2.27       &      -		      \\
\lb\ion{P}{ii}\rb    & 11460	&	-	     &       -  	 &	-	    &	    -		 &	  -	       &       -	    &	    -		 &	     -  	&   -		     & 6.25\pp1.17	      \\
\lb\ion{P}{ii}\rb    & 11886	&   3.89\pp0.83      &       -  	 &	-	    &	    -		 &	  -	       &  14.76\pp3.23      &	    -		 &	     -  	&   -		     & 19.70\pp1.01	      \\
\lb\ion{S}{ix}\rb    & 12520	&	-	     &       -  	 &   1.97\pp0.17    &	    -		 &  15.51\pp2.84       &       -	    &	1.02\pp0.16	 &	     -  	& 11.13\pp0.77       & 39.71\pp0.91	      \\
\lb\ion{Fe}{ii}\rb   & 12570	&  8.45\pp0.21\ee    &       -  	 &   4.84\pp0.21\ee &	3.16\pp0.62\cc   & 7.90\pp0.55\cc      &  41.10\pp1.18\cc   &	2.54\pp0.45\cc   &	     -  	&  5.26\pp0.81\cc    & 58.80\pp2.94\cc	      \\
\ion{H}{i}	     & 12820	&  14.68\pp0.55\ee   &       -  	 &   2.14\pp0.65\ee &	     -  	 &  71.87\pp1.62       &   33.84\pp2.05     &	79.78\pp1.59	 &	     -  	& 20.71\pp0.55       &  108.26\pp0.99	      \\
\ion{H}{i}	     & 12820(b) &  18.37\pp2.05      &       -  	 &  41.69\pp6.11    &	14.6\dd\cc	 &  116.25\pp4.77      &  168.74\pp10.23    &	84.20\pp3.42	 & 101.83\pp3.70\ax\dd  & 66.61\pp1.67       &  712.48\pp8.77	      \\
\lb\ion{Fe}{ii}\rb   & 13201	&	-	     &       -  	 &	-	    &	     -  	 &	  -	       &  26.79\pp3.72      &	    -		 &	     -  	&	  -	     &  21.48\pp2.14	      \\
\lb\ion{Si}{x}\rb    & 14300	&	-	     &       -  	 &	-	    &	     -  	 &  26.15\pp1.85       &       -	    &  1.62\pp0.24	 &  5.12\pp1.90 	& 22.19\pp1.05       &  37.73\pp1.55	      \\
\lb\ion{Fe}{ii}\rb   & 16436	&   6.10\pp0.29\ee   &       -  	 &	-	    &  2.82\pp0.47\cc	 &  7.92\pp1.55\cc     &  41.0\pp3.60\cc    &  2.45\pp0.45\cc	 &	     -  	& 6.42\pp0.97\cc     & 56.2\pp1.96\cc	      \\
\ion{H}{i}	     & 18750	&   12.59\pp0.40     &       -  	 &	-	    &	15.20\pp0.53	 &   50.50\pp2.71      &	  -	    &	106.56\pp1.40	 &	    -		&	 -	     &       -  	      \\
\ion{H}{i}	     & 18750(b) &   40.24\pp1.02     &       -  	 &	-	    &  31.42\pp2.37	 &  204.60\pp8.74      & 212.11\pp4.38\ax\dd&	138.66\pp3.05	 &  139.53\pp1.84\ax	&111.49\pp4.27\ax\dd &581.50\pp12.04\ax\dd    \\
\ion{H}{i}	     & 19446	&       -            &       -  	 &	-	    &	     -  	 &	  -	       &	-	    &	1.44\pp0.52	 &	     -  	&	  -	     &   6.02\pp1.16??        \\
\ion{H}{i}	     & 19446(b) &  2.03\pp0.14\ax    &       -  	 &	-	    &	     -  	 &   31.14\pp4.07\ax   &       -	    &  21.38\pp1.08	 & 7.58\pp2.05\ax  	& 8.52\pp0.36\ax     &  17.92\pp4.32	      \\
$\rm H_2$  	     & 19570	&   3.80\pp0.23\ee   &  6.57\pp0.64\ee   & 2.88\pp0.36\ee   &  0.91\pp0.16\cc	 &	  -	       &  20.8\pp1.2\cc     &	     -  	 &   9.17\pp2.84 	& 7.51\pp0.62\cc     &   21.2\pp2.0\cc        \\
\lb\ion{Si}{vi}\rb   & 19641	&	-	     &       -  	 & 5.69\pp0.62      &	     -  	 &	  -	       &       -	    &  3.09\pp0.36	 &  1.23\pp0.37  	& 12.54\pp1.19       & 68.71\pp2.05	      \\
$\rm H_2$	     & 20332	&   1.35\pp0.08\ee   &  2.22\pp0.35\ee   & 0.59\pp0.12\ee   &  0.36\pp0.05\cc	 &	   -	       &  7.97\pp0.92\cc    &	      - 	 &	     -  	& 2.82\pp0.70\cc     &  6.84\pp0.42\cc        \\
\ion{He}{i}	     & 20580	&       -            &       -  	 &	-	    &	     -  	 &	  -	       &     -  	    &	     -  	 &	     -  	&   -		     &      -		      \\
\ion{He}{i}	     & 20580(b) &  2.09\pp0.02\ax    &       -  	 &	-	    &	     -  	 &	  -	       &  5.71\pp1.43\ax    &	     -  	 &	     -  	&   -		     &      -		      \\
$\rm H_2$	     & 21213	&   2.72\pp0.17\ee   &  3.89\pp1.16\ee   & 2.04\pp0.10\ee   &  0.90\pp0.12\cc	 &	   -	       &  17.7\pp1.00\cc    &	      - 	 &	     -  	& 5.81\pp0.37	     & 14.7\pp0.55\cc	      \\
\ion{H}{i}	     & 21654	&  4.69\pp0.15\ff\ee &       -  	 &	-	    &	       -	 &	 -	       &  6.51\pp1.34	    &	 2.49\pp0.37	 &   -  		&      -	     & 12.42\pp0.63	      \\
\ion{H}{i}	     & 21654(b) &      -             &       -  	 & 0.78\pp0.06\ax\ee&  2.49\pp0.55\cc\ax &  32.00\pp1.50\cc\ax &  21.54\pp6.31      &	22.46\pp1.02	 &   13.33\pp0.73\ax	&13.1\pp0.84\cc\ax   & 31.41\pp2.64	      \\
$\rm H_2$ 	     & 22230	&	-	     &        -  	 &	 -	    &	 0.28\pp0.08\cc  &   -  	       &  3.67\pp0.66\cc    &	      - 	 &	     -  	& 1.52\pp0.40\cc     & 5.04\pp0.62\cc	      \\
$\rm H_2$	     & 22470	&	-	     &        -  	 &	 -	    &	   -		 &   -  	       &  2.32\pp0.21\cc    &  0.24\pp0.08\cc	 &	     -  	&   -		     &      -		      \\
\lb\ion{Ca}{viii}\rb & 23218	&	-	     &       -  	 &	 -	    &	   -		 & 5.66\pp1.61	       &       -	    &	      - 	 &	     -  	& 2.36\pp0.44	     &   15.09\pp1.02       \\
\noalign{\smallskip}
\hline
\multicolumn{3}{l}{a Total Flux of the line.} \\
\multicolumn{3}{l}{b Broad Component of the line.} \\
\multicolumn{3}{l}{c \citet{ara04}} \\
\multicolumn{3}{l}{d Teluric absorption} \\
\multicolumn{3}{l}{e \citet{rrp05}} \\
\multicolumn{3}{l}{f Total flux. Line without broad component.} \\
\multicolumn{3}{l}{h Blend with \ion{Fe}{ii} $\lambda$9177$\AA$ and \ion{Fe}{ii} $\lambda$9202$\AA$.} \\
\multicolumn{3}{l}{i Blend with \ion{Fe}{ii}$\lambda$ 9999$\AA$.} \\
\multicolumn{3}{l}{j Blend with \ion{Fe}{ii}$\lambda$ 10171$\AA$.} \\
\multicolumn{3}{l}{k Blend with \ion{Fe}{ii}$\lambda$ 9202$\AA$.} \\
\end{tabular}
\end{scriptsize}
\end{table*}

\begin{table*}

\begin{scriptsize}
\renewcommand{\tabcolsep}{0.70mm}
\caption{\label{flsy1_2} Observed fluxes, for Type~1 objects, in units of $\rm 10^{-15}\, erg \, cm^{-2} \, s^{-1}$.}    
\centering
\begin{tabular}{llccccccccccc}
\hline\hline
\noalign{\smallskip}
Ion  & $\rm \lambda_{lab}\,(\AA)$& NGC\,4748     & Mrk\,279         & NGC\,5548        & PG\,1415+451        & Mrk\,684          &  Mrk\,478        &  PG\,1448+273      & Mrk\,291        & Mrk\,493         & PG\,1519+226 \\
\noalign{\smallskip}
\hline \noalign{\smallskip}
\ion{O}{i}           & 8446	&	  -	  & 32.16\pp7.67\bl &  12.52\pp1.46	&	    -	     &         -	  &	    -	     &  17.17\pp0.80	  &  1.74\pp0.26     & 20.69\pp1.18	& 14.58\pp1.58    \\
\ion{Ca}{ii}	     & 8498	&	  -	  &	   -	    &  11.45\pp1.62	&	    -	     &         -	  &	    -	     &   9.27\pp1.53	  &	    -	     & 12.51\pp1.64	& 9.14\pp2.10	  \\
\lb\ion{S}{iii}\rb   & 9069	&  30.67\pp1.50   & 7.24\pp0.85     &  12.60\pp0.51	&	    -	     &        - 	  &  5.22\pp0.37     &   4.15\pp0.40	  &  6.43\pp1.04     & 4.84\pp0.60	&     - 	  \\
\ion{Fe}{ii}         & 9127	&	 -	  &	   -	    &	  -		&	    -	     &         -	  &	   -	     &    -		  &  3.65\pp0.79     &  		&     - 	  \\   
\ion{Fe}{ii}         & 9177	&  2.57\pp0.61\kk &	   -	    &	   -		&	    -	     &         -	  &	   -	     &  1.95\pp0.23	  &  1.98\pp0.40     &  		&     - 	  \\
\ion{Fe}{ii}         & 9202	&	 -	  &	   -	    &	  -		&	    -	     &         -	  &	   -	     &  1.58\pp0.26	  &  1.57\pp0.28     &  		&     - 	  \\
\ion{H}{i}	     & 9230	&	 -	  &	   -	    &	    -		&	    -	     &        - 	  &	   -	     &   2.78\pp0.29	  &	    -	     &     -		&    -  	  \\
\ion{H}{i}	     & 9230(b)  & 38.41\pp8.92\ax &	   -	    &	    -		&	    -	     & 19.23\pp3.06\ax\hh &11.52\pp1.84\ax\hh&   9.77\pp0.88	  &	    -	     &21.84\pp2.45\ax\hh&11.00\pp1.85\ax\hh \\
\lb\ion{S}{iii}\rb   & 9531	&  90.23\pp0.67   & 39.98\pp0.55    &  33.69\pp0.81	&	    -	     &  10.58\pp2.14	  &  27.25\pp0.68    &  18.58\pp0.66	  &  6.83\pp0.40     & 16.70\pp1.10	& 8.84\pp1.73\dd  \\
\lb\ion{C}{i}\rb     & 9824	&	 -	  &	   -	    &	    -		&	    -	     &         -	  &	    -	     &  2.07\pp0.20       &	   -	     &     -		&     0.35\pp0.07\dd \\
\lb\ion{C}{i}\rb     & 9850	&	 -	  & 2.47\pp0.24     &	    -		&	    -	     &         -	  &	    -	     &  1.00\pp0.09       &	   -	     &     -		&    0.86\pp0.17    \\
\lb\ion{S}{viii}\rb  & 9910	&   8.23\pp2.52   &	 -	    &  5.99\pp0.44	&	    -	     &         -	  &	     -       &   0.66\pp0.06      &	    -	     &     -		&     0.67\pp0.12  \\ 
\ion{Fe}{ii}	     & 9999	&   9.15\pp1.45   & 5.16\pp0.29\dd  &	    -		&	    -	     &  8.34\pp1.02	  &   2.74\pp0.42    &  3.08\pp0.50	  &	   -	     & 4.33\pp0.74	&     - 	  \\
\ion{H}{i}	     & 10049	&	  -	  &	 -	    &	1.19\pp0.21	&	    -	     &       -  	  &   6.31\pp0.55    &   6.60\pp0.40	  &	   -	     & 6.86\pp0.69	&      -	  \\
\ion{H}{i}	     & 10049(b) & 45.61\pp2.69\ax &	 -	    &  35.30\pp3.13	&	    -	     &  12.08\pp1.31\ax   &  35.54\pp2.02    &  18.13\pp2.16	  &	    -	     & 5.47\pp2.25	& 5.81\pp0.85\ax  \\
\ion{He}{ii}	     & 10122	&	 -	  &	 -	    &	   -		&	    -	     &        - 	  &	     -       &   1.08\pp0.28	  &	    -	     &     -		&     - 	  \\
\ion{He}{ii}	     & 10122(b) & 13.98\pp1.47\ax &	 -	    &  4.00\pp0.37\ax	&	    -	     &        - 	  &	     -       &   0.57\pp0.20	  &	    -	     &     -		&     - 	  \\
\ion{Fe}{ii}	     & 10171	&    -  	  &	 -	    &	    -		&	    -	     &        - 	  &	     -       &   3.11\pp0.39	  &	    -	     & 1.68\pp0.08	&     - 	  \\
\lb\ion{S}{ii}\rb    & 10286	&    -  	  &	 -	    &	1.18\pp0.14	&	    -	     &        - 	  &	     -       &  	 -	  &	    -	     &     -		&     - 	  \\
\lb\ion{S}{ii}\rb    & 10320	&    -  	  &	 -	    &	1.72\pp0.14	&	    -	     &        - 	  &	     -       &  	 -	  &	    -	     &     -		&     - 	  \\
\lb\ion{S}{ii}\rb    & 10336	&    -  	  &	 -	    &	0.83\pp0.14	&	    -	     &        - 	  &	     -       &  	 -	  &	   -	     &     -		&     - 	  \\
\lb\ion{S}{ii}\rb    & 10370	&    -  	  &	 -	    &	0.23\pp0.14	&	    -	     &        - 	  &	     -       &  	 -	  &	   -	     &     -		&     - 	  \\
\lb\ion{N}{i}\rb     & 10404	&    -  	  &	 -	    &	1.04\pp0.21	&	    -	     &        - 	  &   2.26\pp0.17    &  	 -	  &	   -	     &     -		&     - 	  \\
\ion{Fe}{ii}	     & 10500	&    -  	  &	 -	    &	    -		&  1.36\pp0.21       &  7.18\pp1.77	  &  7.96\pp0.56     &  3.11\pp0.39	  &	   -	     & 2.78\pp0.27	&   2.90\pp0.87   \\
\ion{He}{i}	     & 10830	&   46.22\pp1.17  &	  -	    &	 69.88\pp1.56	&	 -	     &        - 	  &  44.38\pp1.71    &   16.87\pp0.19	  &  3.96\pp0.32     & 10.87\pp0.28	&   12.61\pp0.44  \\
\ion{He}{i}	     & 10830(b) &  101.76\pp4.81  & 134.83\pp5.24\ax&	131.51\pp11.66  &  31.77\pp1.60\ax   &  39.46\pp2.40\ax   &  60.57\pp5.11    &   26.74\pp0.55	  &  5.63\pp2.40     & 27.31\pp0.84	&   26.34\pp1.34  \\
\ion{H}{i}	     & 10938	&   20.90\pp2.22  &	   -	    &	   -     	&	  -	     &        - 	  &  18.67\pp1.57    &   12.01\pp0.22	  &	   -	     &  7.41\pp0.28	&   4.91\pp0.39   \\
\ion{H}{i}	     & 10938(b) &   19.89\pp2.52  & 27.05\pp3.98\ax & 5.57\pp1.76\ax	&   10.19\pp1.21\ax  &  24.96\pp2.36	  &  22.72\pp3.76    &   15.57\pp0.79	  &  2.34\pp0.38\ax  & 14.58\pp0.95	&   9.65\pp1.06   \\
\ion{Fe}{ii}	     & 11126	&    -  	  &	 -	    &	   1.84\pp0.42	&	    -	     &        - 	  &  3.52\pp0.46     &  1.64\pp0.17	  &  0.85\pp0.20     & 1.86\pp0.10	&     - 	  \\
\ion{O}{i}	     & 11287	&  23.17\pp2.13   &	 -	    &	    -		&   4.31\pp0.96      &  9.16\pp1.83	  & 23.03\pp0.36     &  11.33\pp0.53	  &	   -	     & 10.28\pp0.54	&  7.04\pp0.62\dd \\
\lb\ion{P}{ii}\rb    & 11460	&    -  	  &	 -	    &	    -		&	    -	     &        - 	  &	 -	     &  	 -	  &	   -	     &     -		&     - 	  \\
\lb\ion{P}{ii}\rb    & 11886	&    -  	  &	 -	    &	    -		&	    -	     &        - 	  &	 -	     &  	 -	  &	   -	     &     -		&     - 	  \\
\lb\ion{S}{ix}\rb    & 12520	&    -  	  & 1.84\pp0.48     &  2.79\pp0.37	&	    -	     &        - 	  &	 -	     &  	 -	  &	   -	     &     -		&     - 	  \\
\lb\ion{Fe}{ii}\rb   & 12570	&  7.39\pp0.37\cc & 7.22\pp0.46\cc  &  1.71\pp0.26\cc	&	    -	     &        - 	  &	  -	     &  	 -	  &  0.91\pp0.24\cc  & 1.96\pp0.34\cc	&     - 	  \\
\ion{H}{i}	     & 12820	&   7.73\pp0.62   &  1.43\pp0.26    &	5.34\pp0.30	&	    -	     &  14.61\pp0.49	  &	 -	     &  	 -	  &  2.56\pp0.22     & 12.89\pp0.52	&   10.16\pp0.50  \\
\ion{H}{i}	     & 12820(b) &   60.12\pp2.20  &  55.87\pp2.51   &  49.30\pp2.91	&   19.28\pp0.84\ax  &  13.48\pp1.17	  & 82.70\cc\dd      &  	 -	  &  2.76\pp0.85     & 16.10\pp1.82	&   13.06\pp1.21  \\
\lb\ion{Fe}{ii}\rb   & 13201	&  3.08\pp0.22    &	 -	    &	    -		&	    -	     &        - 	  &	 -	     &  	 -	  &	   -	     &     -		&     - 	  \\
\lb\ion{Si}{x}\rb    & 14300	&    -  	  &	 -	    &  5.57\pp0.38	&	    -	     &        - 	  &	 -	     &  	 -	  &	   -	     &     -		&     - 	  \\
\lb\ion{Fe}{ii}\rb   & 16436	&  7.85\pp0.61\cc & 5.36\pp0.61\cc  &  1.30\pp0.14\cc	&	    -	     &        - 	  &	 -	     &  	 -	  &   0.80\pp0.09\cc & 1.40\pp0.27\cc	&     - 	  \\
\ion{H}{i}	     & 18750	&  43.24\pp0.57   &  -  	    &	    -		&	    -	     &   24.76\pp0.46	  &  31.15\pp0.60    &  26.98\pp1.17	  &  9.44\pp0.20     & 23.83\pp0.51	&    9.80\pp0.39  \\
\ion{H}{i}	     & 18750(b) &  78.26\pp2.49   & 17.45\pp0.50\ax &	  78.87\ax\dd		& 21.35\pp0.73\ax    &   17.21\pp1.46	  &  55.63\pp1.70    &   2.40\pp0.10	  &  4.04\pp1.25     & 14.18\pp1.71	&    17.79\pp1.06 \\
\ion{H}{i}	     & 19446	&  2.24\pp0.30    &	 -	    &	    -		&	    -	     &        - 	  &	-	     &        - 	  &	   -	     &     -		&     - 	  \\
\ion{H}{i}	     & 19446(b) &  8.90\pp0.99    &	 -	    &	    -		&	    -	     &        - 	  & 7.63\pp0.63\ax   &  2.40\pp0.10\ax    &	   -	     &     -		&     - 	  \\
$\rm H_2$  	     & 19570	&  1.90\pp0.42\cc &  0.67\cc\dd     &	    -		&	    -	     &        - 	  &  3.73\pp0.40\cc  &  0.93\pp0.07\cc    &  0.30\pp0.12\cc  &     -		&     - 	  \\
\lb\ion{Si}{vi}\rb   & 19641	&  8.96\pp0.29    &	 -	    &  9.97\pp0.86	&	    -	     &        - 	  &	 -	     &  1.00\pp0.07	  &	   -	     &     -		&     - 	  \\
$\rm H_2$	     & 20332	&	-	  & 1.04\pp0.16\cc  &	     -  	&	    -	     &        - 	  &	 -	     &  0.37\pp0.08\cc    & 0.22\pp0.07\cc   & 0.32\pp0.07\cc	&     - 	  \\
\ion{He}{i}	     & 20580	&     - 	  &	 -	    &	    -		&	    -	     &        - 	  &	 -	     &      -		  &	    -	     &     -		&     - 	  \\
\ion{He}{i}	     & 20580(b) &  1.43\pp0.08\ax &	 -	    &	    -		&	    -	     &        - 	  &	 -	     &  0.42\pp0.10\ax    &  0.47\pp0.05\ax  &     -		&     - 	  \\
$\rm H_2$	     & 21213	&  1.59\pp0.20\cc & 1.89\pp0.23\cc  &  0.80\pp0.11\cc	&	    -	     &        - 	  &  1.26\pp0.15\cc  &  0.45\pp0.08\cc    & 0.47\pp0.10\cc   & 1.04\pp0.20\cc	&     - 	  \\
\ion{H}{i}	     & 21654	&   1.41\pp0.11   & 0.63\pp0.36     &	    -		&	    -	     &  1.68\pp0.36	  &  1.13\pp0.23     &  1.45\pp0.16	  & 0.36\pp0.04      & 0.85\pp0.12	&     - 	  \\
\ion{H}{i}	     & 21654(b) &   9.65\pp0.49   & 8.86\pp2.28     & 16.27\pp2.0\cc\ax &	    -	     &  3.66\pp1.01	  &  6.29\pp0.62     &  2.63\pp0.38	  & 0.91\pp0.13      & 1.77\pp0.25	&     - 	  \\
$\rm H_2$ 	     & 22230	&   -		  &  0.62\pp0.15\cc &	    -		&	    -	     &        - 	  &	 -	     &       -  	  &	   -	     &     -		&     - 	  \\
$\rm H_2$	     & 22470	&  0.40\pp0.13\cc &  0.56\pp0.15\cc &	    -		&	    -	     &        - 	  &	 -	     &       -  	  &	   -	     &     -		&     - 	  \\
\lb\ion{Ca}{viii}\rb & 23218	&     - 	  &  -  	    & 1.66\pp0.38	&	    -	     &        - 	  &	 -	     &  	 -	  &	   -	     &     -		&     - 	  \\
\noalign{\smallskip}
\hline
\multicolumn{3}{l}{a Total Flux of the line.} \\
\multicolumn{3}{l}{b Broad Component of the line.} \\
\multicolumn{3}{l}{c \citet{ara04}} \\
\multicolumn{3}{l}{d Teluric absorption} \\
\multicolumn{3}{l}{e \citet{rrp05}} \\
\multicolumn{3}{l}{f Total flux. Line without broad component.} \\
\multicolumn{3}{l}{$\xi$ Blend 8446+8498} \\
\multicolumn{3}{l}{h Blend with \ion{Fe}{ii} $\lambda$9177$\AA$ and \ion{Fe}{ii} $\lambda$9202$\AA$.} \\
\multicolumn{3}{l}{k Blend with \ion{Fe}{ii}$\lambda$ 9202$\AA$.} \\

\end{tabular}
\end{scriptsize}
\end{table*}


\begin{table*}

\begin{scriptsize}
\renewcommand{\tabcolsep}{0.70mm}
\caption{\label{flsy1_3} Observed fluxes, for Type~1 objects, in units of $\rm 10^{-15}\, erg \, cm^{-2} \, s^{-1}$.}    
\centering
\begin{tabular}{llccccccccccc}
\hline\hline
\noalign{\smallskip}
Ion  & $\rm \lambda_{lab}\,(\AA)$&PG\,1612+261       & Mrk\,504         &  Arp\,102B         &1\,H\,1934-063     & Mrk\,509               & Mrk\,896          &  1\,H\,2107-097  & Ark\,564        & NGC\,7469 \\
\noalign{\smallskip}
\hline \noalign{\smallskip}
\ion{O}{i}           & 8446	&    16.40\pp1.20    & 4.57\pp1.24	 &	 -	      &  40.57\pp1.49	  &   1946.39\pp92.08\bl  & 21.78\pp2.79      &  35.16\pp4.53	  &   47.35\pp1.30  & 67.80\pp3.36	     \\
\ion{Ca}{ii}	     & 8498	&	  -	     &       -  	 &	 -	      &  29.18\pp2.42	  &	    -		  &	  -	      &   6.29\pp3.67	  &   39.22\pp2.50  &	    -		     \\
\lb\ion{S}{iii}\rb   & 9069	&   7.86\pp0.21      & 2.97\pp0.51	 & 9.49\pp1.51        &  27.25\pp0.85	  &   276.46\pp12.18	  &	  -	      &        -	  &   18.97\pp0.52  &  43.49\pp0.24	     \\
\ion{Fe}{ii}         & 9127	&     - 	     &     -		 &	-	      &  4.07\pp0.69	  &	    -		  &	  -	      &  5.63\pp0.39	  &  4.23\pp0.38    & 3.43\pp0.21	     \\
\ion{Fe}{ii}         & 9177	&     - 	     &     -		 &	-	      &  8.77\pp0.96	  &	    -		  &	  -	      &  4.73\pp0.33	  &  5.18\pp0.47    & 3.33\pp0.18	     \\
\ion{Fe}{ii}         & 9202	&     - 	     &     -		 &	-	      &  3.12\pp0.34	  &	    -		  &	  -	      &  16.12\pp1.13	  &  5.30\pp0.48    & 3.38\pp0.19	     \\
\ion{H}{i}	     & 9230	&   -		     &       -  	 &	 -	      &      -  	  &	    -		  &	  -	      &        -	  &	 -	    &	 -		     \\
\ion{H}{i}	     & 9230(b)  &13.36\pp1.40\ax\hh  &       -  	 &	 -	      &  25.56\pp2.04\ax  & 320.59\pp12.18\ax\hh  &	  -	      &  21.68\pp1.51\ax  & 52.78\pp4.75\ax &  40.73\pp3.90	     \\
\lb\ion{S}{iii}\rb   & 9531	&   26.16\pp0.92     & 7.17\pp0.45	 &   16.07\pp0.68     &  82.95\pp1.18	  &   379.31\pp14.90	  & 7.85\pp0.68       &  16.84\pp1.06	  &   55.39\pp0.82  &  139.79\pp3.18	     \\
\lb\ion{C}{i}\rb     & 9824	&  0.39\pp0.02	     &       -  	 &   1.77\pp0.53      & 	-	  &	    -		  &	  -	      &        -	  &	  -	    &  3.34\pp0.52	     \\
\lb\ion{C}{i}\rb     & 9850	&  0.56\pp0.03\dd    & 0.33\pp0.05	 &   3.68\pp0.53      &  2.75\pp0.75	  &	    -		  &	  -	      &        -	  &   0.76\pp0.16   &  9.81\pp0.52	     \\
\lb\ion{S}{viii}\rb  & 9910	&  1.08\pp0.04       & 0.59\pp0.05	 &	 -	      &  9.76\pp1.10	  &	    -		  &	  -	      &        -	  &   5.94\pp0.35   &  5.62\pp0.83	     \\ 
\ion{Fe}{ii}	     & 9999	&    4.54\pp0.64     &       -  	 &	 -	      &  23.70\pp1.18	  &	    -		  & 4.14\pp0.75       &        -	  &  20.36\pp0.58   &	     -  	     \\
\ion{H}{i}	     & 10049	&	 -	     &  1.11\pp0.10	 &	 -	      &        -	  &	    -		  &	-	      &        -	  &   16.13\pp0.37  &	5.12\pp0.37	     \\
\ion{H}{i}	     & 10049(b) &   12.98\pp0.70\ax  & 12.14\pp0.61	 &	-	      & 28.41\pp0.79\ax   &	    -		  & 6.98\pp0.70\ax    &  104.42\pp6.07\ax &   27.74\pp1.12  &  96.93\pp2.30	     \\
\ion{He}{ii}	     & 10122	&	 -	     &      -		 &	 -	      &  18.67\pp1.24	  &	    -		  &	  -	      &        -	  &    6.94\pp0.43  &	8.72\pp0.52	     \\
\ion{He}{ii}	     & 10122(b) &  7.39\pp0.86\ax    & 0.98\pp0.13\ax	 & 0.68\pp0.38\ax     & 	-	  &	   -		  & 2.49\pp1.04\ax    &  2.45\pp1.03\ax   &   12.73\pp1.28  &  13.09\pp1.57	     \\
\ion{Fe}{ii}	     & 10171	&	   -	     &       -  	 &	  -	      &   3.94\pp0.77	  &	    -		  &	  -	      &        -	  &   3.50\pp0.58   &	     -  	     \\
\lb\ion{S}{ii}\rb    & 10286	&    0.94\pp0.06     &       -  	 &	-	      &   1.35\pp0.19	  &	    -		  &	  -	      &        -	  &    0.88\pp0.21  &  4.48\pp0.12	     \\
\lb\ion{S}{ii}\rb    & 10320	&    0.85\pp0.06     &       -  	 &	-	      &   1.61\pp0.19	  &	    -		  &	  -	      &        -	  &    1.81\pp0.21  &  4.51\pp0.12	     \\
\lb\ion{S}{ii}\rb    & 10336	&    0.24\pp0.06     &       -  	 &	-	      &   0.80\pp0.19	  &	   -		  &	  -	      &        -	  &    1.46\pp0.21  &  1.35\pp0.12	     \\
\lb\ion{S}{ii}\rb    & 10370	&    0.42\pp0.06     &       -  	 &	-	      &   0.32\pp0.19	  &	   -		  &	  -	      &        -	  &    0.92\pp0.21  &  1.24\pp0.12	     \\
\lb\ion{N}{i}\rb     & 10404	&	   -	     &       -  	 &	-	      & 	-	  &	    -		  &	  -	      &        -	  &    2.54\pp0.40  &  2.82\pp0.14	     \\
\ion{Fe}{ii}	     & 10500	&	   -	     &       -  	 &	  -	      & 	-	  &   59.66\pp2.98	  & 2.05\pp0.30       &  10.86\pp2.67	  &   13.05\pp0.58  &	9.88\pp3.00	     \\
\ion{He}{i}	     & 10830	&    22.59\pp0.63    & 12.22\pp0.39	 &    9.97\pp1.59     &  52.96\pp0.53	  &	   -		  & 5.47\pp0.37       &  55.50\pp1.06	  &   43.85\pp0.68  &  115.90\pp1.38	     \\
\ion{He}{i}	     & 10830(b) &    63.54\pp2.59    & 26.67\pp1.15	 &   33.50\pp7.88     &  70.28\pp1.58	  &   6355.76\pp69.69\ax  & 8.09\pp1.11       &  196.79\pp4.16    &  85.96\pp2.85   &  327.62\pp5.69	     \\
\ion{H}{i}	     & 10938	&    17.26\pp1.39    &  1.30\pp0.27	 &	  -	      &  25.32\pp0.63	  &	   -		  &	   -	      &        -	  &   23.16\pp0.69  &	30.83\pp2.23	     \\
\ion{H}{i}	     & 10938(b) &    23.41\pp4.09    & 16.10\pp1.15	 &	  -	      &  28.32\pp1.57	  &   2205.06\pp64.87\ax  & 3.70\pp1.49\ax    &  71.67\pp2.75\ax  &   39.27\pp2.04  &  160.91\pp6.88	     \\
\ion{Fe}{ii}	     & 11126	&	   -	     & 0.98\pp0.08	 &	  -	      &  3.82\pp0.76	  &	   -		  &	  -	      &        -	  &   5.57\pp0.40   &	     -  	     \\
\ion{O}{i}	     & 11287	&   7.08\pp0.51      & 3.39\pp0.15	 &	  -	      &  26.83\pp1.24	  &	    -		  & 7.49\pp0.41       &  17.32\pp1.61	  &   34.66\pp0.36  &  38.87\pp0.42	     \\
\lb\ion{P}{ii}\rb    & 11460	&	   -	     &       -  	 &	  -	      &  7.74\pp0.39	  &	    -		  &	  -	      &        -	  &	 -	    &  4.75\pp0.25	     \\
\lb\ion{P}{ii}\rb    & 11886	&	   -	     &       -  	 &	  -	      &        -	  &	    -		  &	  -	      &        -	  &   1.87\pp0.29   &  7.08\pp1.03	     \\
\lb\ion{S}{ix}\rb    & 12520	&  0.96\pp0.11       &       -  	 &	  -	      &  7.83\pp0.40	  &	    -		  &	  -	      &        -	  &   6.88\pp0.38   &	6.10\pp1.06	     \\
\lb\ion{Fe}{ii}\rb   & 12570	&  3.56\pp0.25\cc    &       -  	 &   9.66\pp0.81\ee   &  3.48\pp0.23	  &	   -		  &	  -	      &        -	  &   3.87\pp0.44\cc&  12.44\pp0.69\ee       \\
\ion{H}{i}	     & 12820	&    10.70\pp0.56    &       -  	 &	-	      & 43.62\pp0.61	  &   1125.26\pp30.16	  &  3.43\pp0.26      &   71.66\pp1.97    &   43.16\pp0.45  &	31.66\pp1.74\ee      \\
\ion{H}{i}	     & 12820(b) &    38.32\pp1.72    & 21.8\pp1.22\cc\ax &	-	      & 36.15\pp1.81	  &   1824.67\pp77.87	  & 17.95\pp0.80      &   67.50\pp5.69    &   59.03\pp1.54  &  153.08\pp7.27	     \\
\lb\ion{Fe}{ii}\rb   & 13201	&	   -	     &       -  	 &	-	      &  1.35\pp0.43	  &	   -		  &	  -	      &        -	  &	    -	    &	     -  	     \\
\lb\ion{Si}{x}\rb    & 14300	&     1.14\pp0.16    &       -  	 &	-	      &  8.57\pp1.10	  &	    -		  &	  -	      &   2.46\pp0.63	  &   17.77\pp0.54  &  11.39\pp1.30	     \\
\lb\ion{Fe}{ii}\rb   & 16436	&	   -	     &       -  	 & 7.25\pp0.41\ee     &  5.46\pp1.12	  &	    -		  &	  -	      &        -	  &   3.90\pp0.65\cc&  9.87\pp0.68\ee	     \\
\ion{H}{i}	     & 18750	&    13.23\pp0.23    &  9.57\pp0.37	 &	 -	      &  72.88\pp0.52	  &	   -		  &   2.71\pp0.26     &    90.90\pp1.82   &   55.03\pp0.86  &	25.84\pp1.55	     \\
\ion{H}{i}	     & 18750(b) &    63.88\pp1.05    & 16.37\pp1.00	 &  7.13\ax\dd        &  28.11\pp1.81	  & 3721.14\pp30.16\ax\dd &  31.38\pp1.21     &    78.76\pp4.63   &   50.74\pp2.82  &  206.48\pp11.56	     \\
\ion{H}{i}	     & 19446	&   4.10\pp0.45      &       -  	 &	 -	      &      -  	  &	   -		  &	  -	      &        -	  &   3.07\pp0.18   &	1.85\pp0.25	     \\
\ion{H}{i}	     & 19446(b) &	   -	     &       -  	 &	 -	      &    5.77\pp0.67\ax &	    -		  &	  -	      &        -	  &   5.90\pp0.59   &	2.46\pp0.44	     \\
$\rm H_2$  	     & 19570	&   -		     &       -  	 & 1.71\pp0.20\ee     & 3.67\pp0.78	  &	     -  	   & 0.41\cc\dd        &	-	   &   0.80\pp0.22\cc&  16.04\pp0.61\ee       \\
\lb\ion{Si}{vi}\rb   & 19641	&    3.16\pp0.20     &       -  	 &	     -        & 7.64\pp0.61	  &	     -  	   &	   -	       &     1.49\pp0.61   &   6.16\pp0.30   &  12.36\pp0.46	      \\
$\rm H_2$	     & 20332	&	  -	     & 0.15\pp0.04\cc	 &  0.66\pp0.11\ee    &        -	  &	   -		  & 0.42\pp0.13\cc    &        -	  &   0.53\pp0.18\cc&  3.28\pp0.23\ee	     \\
\ion{He}{i}	     & 20580	&	   -	     &       -  	 &	-	      &      -  	  &	   -		  &	  -	      &        -	  &   - 	    &  1.42\pp0.29	     \\
\ion{He}{i}	     & 20580(b) &	   -	     &       -  	 &	-	      &    2.45\pp0.33\ax &	    -		  &	  -	      &        -	  &   - 	    &  4.95\pp1.78	     \\
$\rm H_2$	     & 21213	&   0.76\pp0.17\cc   & 0.35\pp0.06\cc	 &  1.77\pp0.16\ee    &  0.59\pp0.25	  &	    -		  & 0.41\pp0.10\cc    &        -	  &   1.24\pp0.27   &  8.74\pp0.28\ee	     \\
\ion{H}{i}	     & 21654	&	   -	     &      -		 &	-	      &       - 	  &	   -		  & 0.37\pp0.09       &   14.57\pp1.16    &   4.76\pp0.16   &	4.52\pp0.28\ee       \\
\ion{H}{i}	     & 21654(b) &	   -	     &18.06\pp0.72\cc\ax &	-	      &   9.63\pp0.60\ax  &  381.96\pp33.25\ax    & 2.78\pp0.43       &    7.93\pp3.09    &   5.70\pp0.43   &  20.66\pp1.41	     \\
$\rm H_2$ 	     & 22230	&	   -	     & 0.20\pp0.06\cc	 &   0.31\pp0.09\ee   &        -	  &	   -		  &	  -	      &        -	  &	  -	    &	2.02\pp0.53\ee       \\
$\rm H_2$	     & 22470	&	   -	     &       -  	 &    - 	      &        -	  &	    -		  & 0.20\pp0.06\cc    &        -	  &	  -	    &	1.55\pp0.53\ee       \\
\lb\ion{Ca}{viii}\rb & 23218	&	   -	     &       -  	 &	-	      &  2.55\pp0.37	  &	   -		  &	  -	      &     1.46\pp0.30   &   3.09\pp0.54   & $>$ 1.90   	     \\
\noalign{\smallskip}
\hline
\multicolumn{3}{l}{a Total Flux of the line.} \\
\multicolumn{3}{l}{b Broad Component of the line.} \\
\multicolumn{3}{l}{c \citet{ara04}} \\
\multicolumn{3}{l}{d Teluric absorption} \\
\multicolumn{3}{l}{e \citet{rrp05}} \\
\multicolumn{3}{l}{f Total flux. Line without broad component.} \\
\multicolumn{3}{l}{h Blend with \ion{Fe}{ii} $\lambda$9177$\AA$ and \ion{Fe}{ii} $\lambda$9202$\AA$.} \\
\multicolumn{3}{l}{$\xi$ Blend 8446+8498} \\
\end{tabular}
\end{scriptsize}
\end{table*}


\begin{table*}

\begin{scriptsize}
\renewcommand{\tabcolsep}{0.70mm}
\caption{\label{flsy2} Observed flux for Type 2 and Starburst galaxies, in units of $\rm 10^{-15}\, erg \, cm^{-2} \, s^{-1}$.}    
\centering
\begin{tabular}{llcccccccccc}
\hline\hline
\noalign{\smallskip}
Ion & $\rm \lambda_{lab}\,(\AA)$& NGC\,34    &  NGC\,262      & Mrk\,993         &   NGC\,591       &  Mrk\,573      & NGC\,1144       & Mrk\,1066      & NGC\,1275      &  NGC\,1614        \\
\noalign{\smallskip}
\hline 
\noalign{\smallskip}
\lb\ion{S}{iii}]  &  9069    &16.57\pp2.90    & 47.76\pp0.68   &  7.11\pp0.98	 &  34.95\pp1.05   & 38.92\pp1.09    & 3.77\pp0.23    & 65.66\pp0.52    & 72.93\pp2.39    &  100.39\pp1.57    \\
\lb\ion{S}{iii}]  &  9531    &18.47\pp1.01    & 128.13\pp1.20  &  16.51\pp0.07	 &  92.63\pp0.95   & 108.65\pp0.69   & 7.11\pp0.16    & 178.79\pp1.02   & 188.73\pp1.55   & 278.66\pp1.22    \\ 
\lb\ion{C}{i}]    &  9824    &2.55\pp0.94     & 1.70\pp0.52    &  0.36\pp0.11	 &  3.38\pp0.70    & 0.59\pp0.05     &    -	      &  5.72\pp0.27	&  5.38\pp0.57    &	  -	     \\ 
\lb\ion{C}{i}]    &  9850    &7.40\pp0.94     & 4.13\pp0.52    &  2.10\pp0.20	 &  7.23\pp0.70    & 2.57\pp1.23     &    -	      & 12.54\pp0.27	& 17.26\pp0.59    &	  -	     \\ 
\lb\ion{S}{viii}] &  9910    & - 	      & 2.65\pp0.59    &  0.76\pp0.17    &  2.52\pp0.23	   & 6.10\pp0.10     &	 -	      &        -        &       -	  &	-	   \\ 
\ion{H}{i}        &  10049   &   - 	      & 2.26\pp0.52    &         -       &  2.54\pp0.58	   & 3.27\pp0.16     &	 -	      & 9.74\pp0.30     & 13.32\pp1.08	  & 17.28\pp0.67     \\ 
\ion{He}{ii}      &  10122   &   - 	      & 3.78\pp0.54    &	-        &  4.28\pp0.83	   &	 -	     &	 -	      & 5.36\pp0.55     &  4.00\pp0.78	  &	-	   \\ 
\lb\ion{S}{ii}]   &  10286   & - 	      & 5.85\pp0.55    &	-        &  1.98\pp0.57	   & 3.58\pp0.22     &	 -	      & 5.52\pp0.38     & 33.05\pp1.30	  &	-	   \\ 
\lb\ion{S}{ii}]   &  10320   & - 	      & 8.18\pp0.55    &	-        &  4.52\pp0.57	   & 2.78\pp0.22     &	 -	      & 7.88\pp0.38     & 42.16\pp1.30	  &	-	   \\ 
\lb\ion{S}{ii}]   &  10336   & - 	      & 4.02\pp0.55    &	 -       &  1.91\pp0.57	   & 1.67\pp0.22     &	 -	      & 3.76\pp0.38     & 34.65\pp1.30	  &	-	   \\ 
\lb\ion{S}{ii}]   &  10370   & - 	      & 2.06\pp0.55    &	 -       &  1.86\pp0.57	   & 1.22\pp0.22     &	 -	      & 2.16\pp0.38     & 17.64\pp1.30	  &	-	   \\ 
\lb\ion{N}{i}]    &  10404   & - 	      & 1.94\pp0.38    &	 -       &  1.99\pp0.73	   & 1.80\pp0.44     &	 -	      & 1.54\pp0.36     & 42.82\pp1.38	  &	-	   \\ 
\ion{He}{i}       &  10830   &20.71\pp1.30    & 66.98\pp0.61   &  32.41\pp1.89   &  53.20\pp0.75   & 50.95\pp0.40    &	 -	      & 102.69\pp1.39   & 588.47\pp5.14   & 238.51\pp4.60    \\ 
\ion{H}{i}        &  10938   &   - 	      &  5.82\pp0.62   &  15.39\pp2.46   &   5.70\pp0.56   &  6.11\pp0.41    &	 -	      &  25.53\pp1.20   &  83.53\pp4.25   & 53.67\pp4.79     \\ 
\lb\ion{P}{ii}]   &  11470   & - 	      & 1.71\pp0.36    &	 -       &     -  	   & 2.17\pp0.19     &	 -	      & 4.51\pp0.20     & 5.21\pp0.41	  &	-	   \\ 
\ion{He}{ii}      &  11626   &   - 	      & 0.76\pp0.19    &	  -      &     -  	   & 1.77\pp0.12\dd  &	 -	      & 1.51\pp0.30     & 3.24\pp0.39	  &	-	   \\ 
\lb\ion{P}{ii}]   &  11886   &8.85\pp2.33     & 3.68\pp0.39    & 	  -      &  5.35\pp0.77	   & 2.18\pp0.69\dd   &	 -	      & 16.31\pp1.45    & 18.95\pp0.50	  &	-	   \\ 
\lb\ion{S}{ix}]   &  12520   & - 	      & 2.28\pp0.58    &	  -      &     -  	   & 4.85\pp0.13     &	 -	      &  4.81\pp0.32    &       -	  &	-	   \\ 
\lb\ion{Fe}{ii}]  &  12570   &13.76\pp1.24\ee & 11.59\pp0.52\ee& 5.05\pp0.30\ee  & 17.19\pp0.75\ee & 5.90\pp0.14\ee  & 0.91\pp0.23\ee & 38.53\pp0.25\ee & 63.49\pp1.51\ee &35.02\pp1.22\ee    \\ 
\ion{H}{i}        &  12820   &12.83\pp1.13\ee & 10.75\pp0.33\ee& 14.90\pp2.66\ee & 15.14\pp0.49\ee & 9.58\pp0.17\ee  &	 -	      & 54.07\pp0.24\ee & 60.66\pp3.15\ee &129.28\pp1.14\ee   \\ 
\lb\ion{Fe}{ii}]  &  12950   & - 	      & 1.61\pp0.33    &	  -      &  2.15\pp0.40	   &	  -	     &	 -	      & 5.59\pp0.71     & 12.13\pp2.62	  &	-	    \\ 
\lb\ion{Fe}{ii}]  &  13209   & - 	      & 3.34\pp0.81    &	  -      &  6.41\pp1.09	   &	  -	     &	 -	      & 13.70\pp0.54    & 18.48\pp1.64	  &	-	    \\ 
\lb\ion{Si}{x}]   &  14300   & - 	      & 2.03\pp0.50    &	  -      &  3.16\pp0.42	   & 7.54\pp0.25     &	 -	      & 4.25\pp1.07     &       -	  &	-	   \\ 
\lb\ion{Fe}{ii}]  &  15342   & - 	      & 2.32\pp0.22    &	  -      &     -  	   &	   -	     &	 -	      & 6.61\pp0.64     & 8.82\pp0.94	  &	-	   \\ 
\lb\ion{Fe}{ii}]  &  16436   &11.00\pp2.87\ee & 9.09\pp0.31\ee & 	  -      & 14.19\pp0.73\ee & 4.11\pp0.09\ee  &	 -	      & 37.07\pp0.34\ee & 61.49\pp0.65\ee & 20.84\pp2.91\ee     \\ 
\lb\ion{Fe}{ii}]  &  16773   & - 	      &   -	       &	  -      &     -  	   &	  -	     &	 -	      & 3.97\pp0.50     & 8.97\pp0.53	  &	-	     \\ 
\ion{H}{i}        &  16806   &   - 	      &   -	       &	  -      &     -  	   &	  -	     &	 -	      & 3.98\pp0.55     &       -	  &	-	   \\ 
\ion{H}{i}        &  18750   &62.75\pp0.43    & 35.26\pp1.05   &	  -      &  60.10\pp0.77   & 45.57\pp0.28    &	 -	      & 145.74\pp1.60   &  145.14\pp2.52  &  295.45\pp6.86    \\ 
\ion{H}{i}        &  19446   &   - 	      &   -	       &	  -      &  0.93\pp0.08	   & 1.37\pp0.06     &	 -	      &  8.67\pp0.04    & 13.98\pp2.06	  & 20.50\pp3.48     \\ 
$\rm H_2$         &  19570   &12.43\pp1.02\ee & 2.09\pp0.18\ee &  1.63\pp0.10\ee &  6.82\pp0.63\ee & 1.44\pp0.17\ee  &	 -	      & 15.17\pp0.38\ee & 43.15\pp0.67\ee & 11.18\pp3.97\ee     \\ 
\lb\ion{Si}{vi}]  &  19641   & - 	      & 4.43\pp0.19    &	 -       &  8.07\pp0.76	   & 10.17\pp0.18    &	 -	      &  3.53\pp0.34    &       -	  &	-	   \\ 
$\rm H_2$         &  20332   &4.66\pp0.23\ee  & 0.60\pp0.09\ee & 0.92\pp0.12\ee  &  2.66\pp0.22\ee &  0.75\pp0.24\ee & 0.47\pp0.06\ee &  4.83\pp0.10\ee & 15.26\pp0.33\ee &	-	    \\ 
\ion{He}{i}       &  20580   &   - 	      &   -	       &	  -      &  1.42\pp0.19	   & 0.53\pp0.06     &	 -	      & 6.83\pp0.19     & 9.62\pp1.14	  & 25.99\pp0.56      \\ 
$\rm H_2$         &  21213   &12.15\pp1.27\ee & 1.61\pp0.05\ee &  1.01\pp0.17\ee &  6.52\pp0.14\ee & 1.82\pp0.19\ee  & 0.94\pp0.04\ee & 13.41\pp0.16\ee & 42.95\pp0.37\ee & 5.70\pp0.77\ee     \\ 
\ion{H}{i}        &  21654   &7.12\pp0.13\ee  & 1.09\pp0.11\ee &   -	         &  3.55\pp0.22\ee & 2.77\pp0.09\ee  &	 -	      & 14.16\pp0.22\ee & 9.77\pp0.41\ee  &42.45\pp0.74\ee     \\ 
$\rm H_2$         &  22230   &   - 	      &   -	       &   -	         &  1.77\pp0.16\ee &	   -	     &	 -	      &  2.66\pp0.06\ee & 11.81\pp0.30\ee &	-	    \\ 
$\rm H_2$         &  22470   &   - 	      &   -	       &   -	         &  0.70\pp0.16\ee &	   -	     &	 -	      &  1.77\pp0.03\ee & 4.51\pp0.33\ee  &	-	   \\ 
\lb\ion{Ca}{viii}]& 23218    &	-	      &   -	       &    -	         &     -	   & 3.18\pp0.29     &    -	      &	-	        &	 -	  &	  -	     \\ 
\hline
&&&&&&& & & & &  \\
\multicolumn{3}{l}{{\bf Table~\ref{flsy2}.} continued.}&&&& & & & &  \\
&&&&&&& & & & &  \\
\hline\hline
\noalign{\smallskip}
Ion & $\rm \lambda_{lab}\,(\AA)$& NGC\,2110        &  ESO\,428-G014  & NGC\,3310        & NGC\,5728       & NGC\,5929       & NGC\,5953        & NGC\,7674	 & NGC\,7682	  & NGC\,7714 \\
\noalign{\smallskip}
\hline
\noalign{\smallskip}
\lb\ion{S}{iii}]  &  9069    &   50.08\pp1.63	   &  143.16\pp0.89  & 35.21\pp1.82	&	 -	   &  14.70\pp1.22   &18.21\pp2.73     &  56.85\pp1.68   & 32.27\pp1.01   &  82.46\pp2.57  \\
\lb\ion{S}{iii}]  &  9531    & 125.39\pp0.88	   &  371.27\pp2.39  & 92.45\pp1.33	& 62.15\pp0.70     &  34.32\pp0.95   &34.85\pp2.32     &  143.23\pp5.73  & 85.25\pp0.35   &  202.45\pp1.54  \\ 
\lb\ion{C}{i}]    &  9824    &  8.38\pp0.64	   &  7.50\pp0.32    &        - 	&	 -	   &	     -       &       -         &     -  	 & -		  &  2.72\pp0.54    \\ 
\lb\ion{C}{i}]    &  9850    & 20.67\pp0.64	   &  12.76\pp0.32   &        - 	&	 -	   &	     -       &       -         &  6.84\pp0.64	 & -		  &  3.18\pp0.35     \\ 
\lb\ion{S}{viii}] &  9910    &   -		   &  6.03\pp0.31    &        - 	&	 -	   &	     -       &       -         &  4.03\pp0.98    & 1.69\pp0.71    &	      -      \\ 
\ion{H}{i}        &  10049   & 3.00\pp0.27	   &   9.19\pp0.63   &   3.44\pp0.51	& 5.45\pp1.42	   &	     -       &       -         &  8.38\pp0.67	 & 1.18\pp0.18    &    11.47\pp0.63  \\ 
\ion{He}{ii}      &  10122   &   -		   &  20.18\pp0.80   &  12.59\pp1.34	& 7.03\pp1.58	   &	     -       &4.9\pp1.05       &  10.21\pp0.67   & 3.62\pp0.31    &	      -      \\ 
\lb\ion{S}{ii}]   &  10286   & 16.89\pp1.16	   &   9.53\pp0.20   &        - 	&	 -	   &	     -       &       -         &  4.58\pp0.60	 & 3.91\pp0.77    &	      -      \\ 
\lb\ion{S}{ii}]   &  10320   & 19.75\pp1.16	   &  13.45\pp0.20   &        - 	&	 -	   &	     -       &       -         &  4.73\pp0.60	 & 5.39\pp0.77    &	      -      \\ 
\lb\ion{S}{ii}]   &  10336   & 13.50\pp1.16	   &   7.55\pp0.20   &        - 	&	 -	   &	     -       &       -         &  3.09\pp0.60	 & 2.31\pp0.77    &	      -       \\ 
\lb\ion{S}{ii}]   &  10370   &  8.40\pp1.16	   &   2.53\pp0.20   &        - 	&	 -	   &	     -       &       -         &  0.49\pp0.60	 & 1.96\pp0.77    &	      -       \\ 
\lb\ion{N}{i}]    &  10404   & 25.94\pp1.65	   &   0.91\pp0.12   &        - 	&	 -	   &	     -       &       -         &  6.26\pp1.45	 & 2.09\pp1.00    &	      -       \\ 
\ion{He}{i}       &  10830   & 102.30\pp1.00	   &  178.01\pp0.90  &        - 	&  32.69\pp1.38    &  20.70\pp0.69   & 18.73\pp1.63    &  56.48\pp0.57   & 49.09\pp1.50   &   128.83\pp2.39   \\ 
\ion{H}{i}        &  10938   &  12.66\pp1.03	   &   31.04\pp1.15  &        - 	&  3.80\pp1.05     &  3.79\pp0.27\dd &        -        &  13.87\pp1.31   &  4.98\pp1.12   &   26.70\pp2.12    \\ 
\lb\ion{P}{ii}]   &  11470   &   -		   &  4.55\pp0.71    &        - 	&	 -	   &	     -       &       -         &  1.52\pp0.13	 &    - 	  &	      -       \\ 
\ion{He}{ii}      &  11626   &   -		   &   2.92\pp0.51   &        - 	&	 -	   &	     -       &       -         &  0.65\pp0.08	 &    - 	  &	      -       \\ 
\lb\ion{P}{ii}]   &  11886   &  14.49\pp2.75	   &  15.18\pp0.74   &        - 	&	 -	   &	     -       &       -         &  5.69\pp0.30	 &    - 	  &	      -        \\ 
\lb\ion{S}{ix}]   &  12520   &   -		   &   5.15\pp0.80   &        - 	&  1.19\pp0.59     &	     -       &       -         &  4.11\pp1.48	 &    - 	  &	      -        \\ 
\lb\ion{Fe}{ii}]  &  12570   & 81.24\pp1.02\ee     & 30.24\pp0.62\ee & 10.70\pp0.87\cc  &  5.08\pp0.50\cc  &  8.44\pp0.36\ee &19.06\pp0.66\ee  & 10.81\pp0.92\ee & 4.58\pp0.74\ee &  17.35\pp0.53\ee	\\ 
\ion{H}{i}        &  12820   & 14.91\pp0.86\ee     & 45.26\pp0.57\ee &31.74\pp1.03\cc	& 7.37\pp1.16\cc   &  7.68\pp0.20\ee &       -         & 10.36\pp0.58\ee & 9.92\pp0.65\ee &   59.73\pp1.38     \\ 
\lb\ion{Fe}{ii}]  &  12950   & 13.03\pp1.18	   &  5.94\pp0.89    &        - 	&	 -	   &	     -       &       -         &      - 	 &    - 	  &	      - 	\\ 
\lb\ion{Fe}{ii}]  &  13209   & 30.76\pp0.63	   &  22.26\pp2.06   &        - 	&	 -	   & 3.50\pp0.76     &13.89\pp2.11     &     -  	 &    - 	  &	      - 	\\ 
\lb\ion{Si}{x}]   &  14300   &   -		   &	 -	     &        - 	&	 -	   &	     -       &       -         &  3.53\pp0.25	 &    - 	  &	      - 	\\ 
\lb\ion{Fe}{ii}]  &  15342   & 14.62\pp1.33	   &  7.82\pp0.82    &        - 	&	 -	   &	     -       &       -         &     -  	 &    - 	  &	      - 	\\ 
\lb\ion{Fe}{ii}]  &  16436   & 70.17\pp0.55\ee     & 28.47\pp1.05\ee &  11.12\pp1.57\cc & 4.96\pp0.61\cc   & 7.83\pp0.37\ee  & 15.21\pp1.89\ee &  9.74\pp0.47\ee &    - 	  &  14.84\pp2.08\ee	\\ 
\lb\ion{Fe}{ii}]  &  16773   & 8.80\pp1.18	   &   3.83\pp0.20   &        - 	&	 -	   &	     -       &       -         &  24.07\pp5.31   &    - 	  &	      - 	\\ 
\ion{H}{i}        &  16806   &   -		   &  8.22\pp0.29    &        - 	&	 -	   &	     -       &       -         &  25.66\pp4.52   &    - 	  &	      - 	\\ 
\ion{H}{i}        &  18750   & 32.95\pp4.50	   &  102.05\pp0.78  & 112.98\pp1.57	& 20.63\pp1.92     &	     -       &19.82\pp0.83     &  32.06\pp1.09   & 31.82\pp0.81   &  117.94\pp2.35	\\ 
\ion{H}{i}        &  19446   &   -		   &  3.00\pp0.49    &  4.04\pp0.83	&	 -	   &	     -       &       -         &  3.38\pp0.51	 &     -	  &	      - 	\\ 
$\rm H_2$         &  19570   & 11.35\pp0.28\ee     &  8.29\pp0.55\ee &        - 	& 6.04\pp0.56\cc   &  3.07\pp0.60\ee &6.67\pp1.85\ee\dd&  4.05\pp0.56\ee & 9.40\pp0.51\ee &	      - 	\\ 
\lb\ion{Si}{vi}]  &  19641   &   -		   &  21.09\pp0.66   &        - 	&  7.08\pp0.88     &	     -       &       -         &  10.71\pp1.11   & 2.96\pp0.63    &	      - 	\\ 
$\rm H_2$         &  20332   &  4.22\pp0.21\ee     &  3.06\pp0.27\ee &  1.24\pp0.25\cc  &  2.40\pp0.20\cc  & 1.33\pp0.23\ee  &2.28\pp0.72\ee   &  1.20\pp0.23\ee & 3.50\pp0.22\ee &	      - 	\\ 
\ion{He}{i}       &  20580   & 1.67\pp0.22	   &  2.14\pp0.08    &  6.69\pp0.52	&	 -	   &	     -       &       -         &  1.27\pp0.27	 &    - 	  &   8.83\pp0.31      \\ 
$\rm H_2$         &  21213   & 8.39\pp0.19\ee	   &  8.27\pp0.12\ee &  1.54\pp0.32\cc  & 6.28\pp0.12\cc   &  2.88\pp0.22\ee & 3.54\pp0.51\ee  &  2.12\pp0.12\ee & 9.78\pp0.14\ee &  3.31\pp0.43\ee    \\ 
\ion{H}{i}        &  21654   & 2.50\pp0.22\ee	   &  8.98\pp0.14\ee &  10.79\pp0.33\cc &  2.11\pp0.19\cc  &  1.35\pp0.25\ee &       -         &  3.13\pp0.25\ee & 1.94\pp0.11\ee &   15.47\pp0.28     \\ 
$\rm H_2$         &  22230   & 2.70\pp0.36	   &	 -	     &        - 	&  1.68\pp0.20\cc  & 0.80\pp0.26\ee  &       -         &      - 	 & 2.51\pp0.30\ee &	      -        \\ 
$\rm H_2$         &  22470   &   -		   &	 -	     &        - 	&  0.89\pp0.20\cc  &	     -       &       -         &  0.67\pp0.11\ee & -		  &	      -        \\ 
\lb\ion{Ca}{viii}]& 23218    &   -		  &	 -	     &         -	&	  -	   &	     -       &       -         & 1.45\pp0.14	 & -		  &	      -        \\ 
\hline
\multicolumn{3}{l}{a Total Flux of the line.} \\
\multicolumn{3}{l}{c \citet{ara04}} \\ 
\multicolumn{3}{l}{d Teluric absorption} \\
\multicolumn{3}{l}{e \citet{rrp05}} \\
\end{tabular}
\end{scriptsize}
\end{table*}

\setcounter{figure}{0}
\begin{figure}
\centering
\includegraphics[scale=0.8]{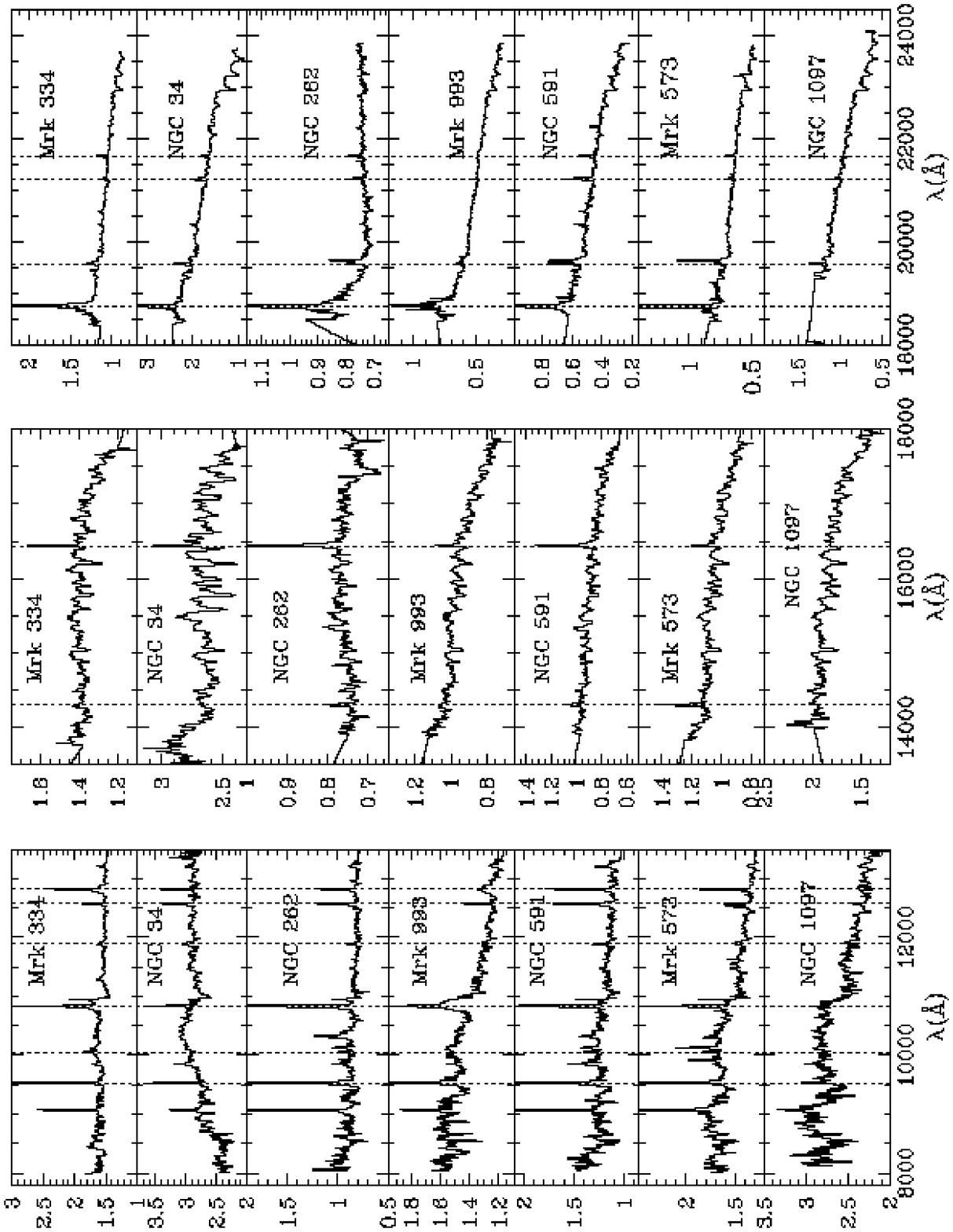}
\vspace{-0.5cm} 
\caption{Final reduced spectra in the Earth's frame. In the left panel 
we present the $z$+$J$ band, in the middle panel the $H$ band, and in 
the right panel the $K$ band. The abscissa is the flux in units of 
$\rm 10^{-15}\, erg \, cm^{-2} \, s^{-1}$. 
The dotted lines are: \lb\ion{S}{iii}] 0.9531\,$\mu$m, Pa$\delta$, 
\ion{He}{i} 1.0830\,$\mu$m, [\ion{P}{ii}] 1.1886\,$\mu$m, 
\fe2\ 1.2570\,$\mu$m, Pa$\beta$ (left panel), [\ion{Si}{x} 1.4300\,$\mu$m, 
\fe2\ 1.6436\,$\mu$m (middle panel), Pa$\alpha$ 
\h2\ 1.9570\,$\mu$m, \h2\ 2.1213\,$\mu$m, and Br$\gamma$ (right panel).}
\label{indiv1}
\end{figure}

\begin{figure}
\centering
\includegraphics[scale=0.8]{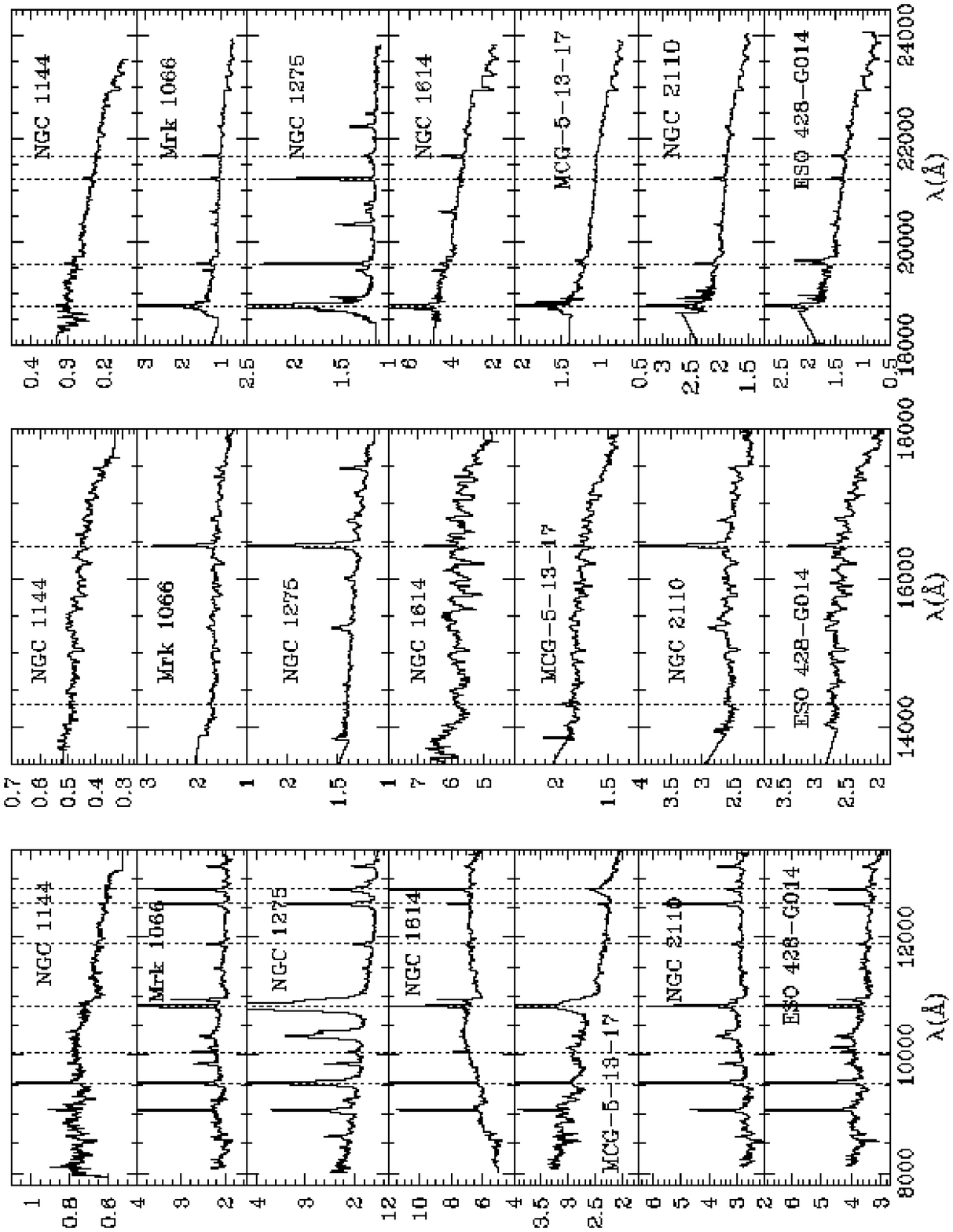}
\vspace{-0.5cm} 
\caption{Same as Fig.~\ref{indiv1}.}
\label{indiv2}
\end{figure}

\begin{figure}
\centering
\includegraphics[scale=0.8]{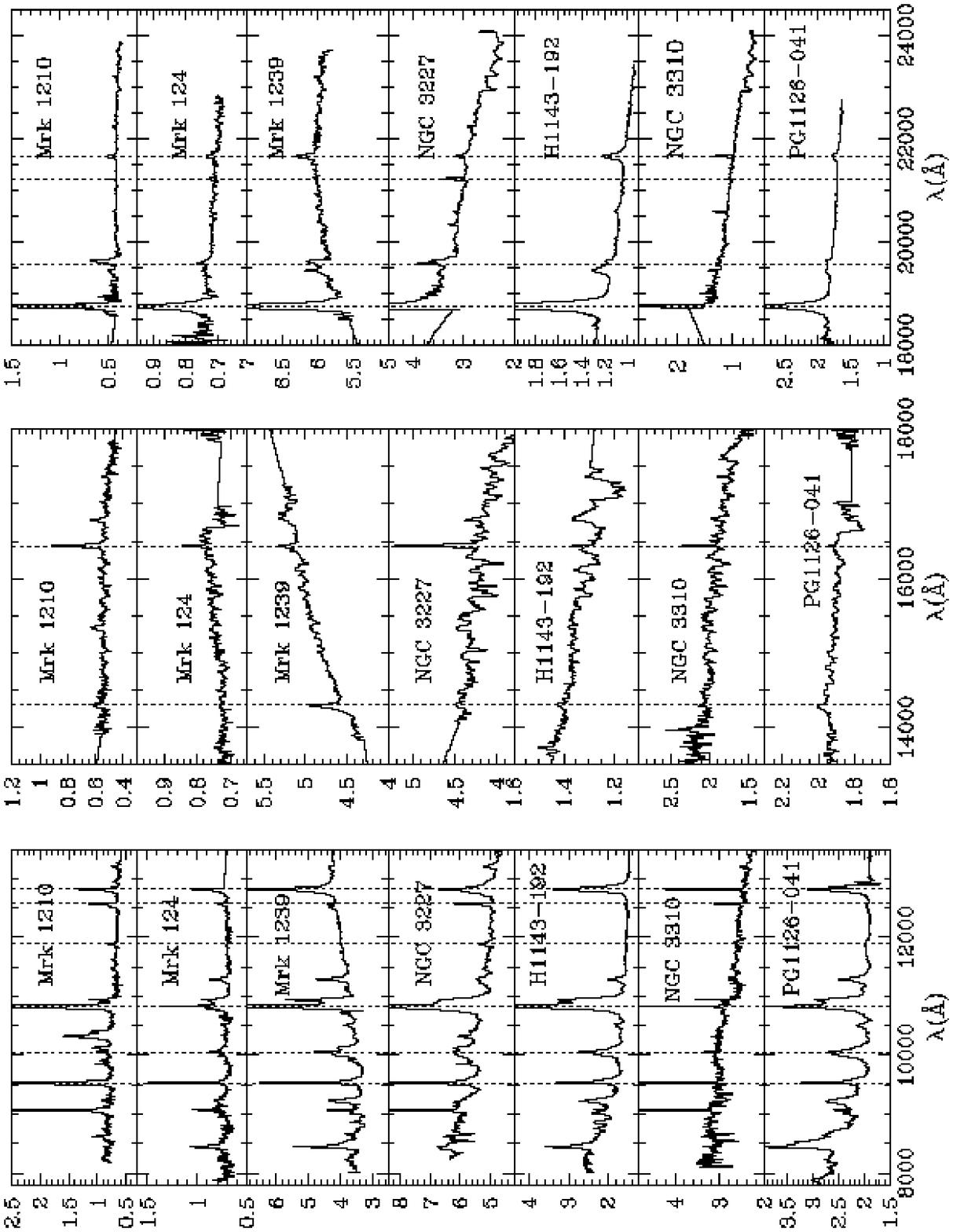}
\vspace{-0.5cm} 
\caption{Same as Fig.~\ref{indiv1}.}
\label{indiv3}
\end{figure}

\begin{figure}
\centering
\includegraphics[scale=0.8]{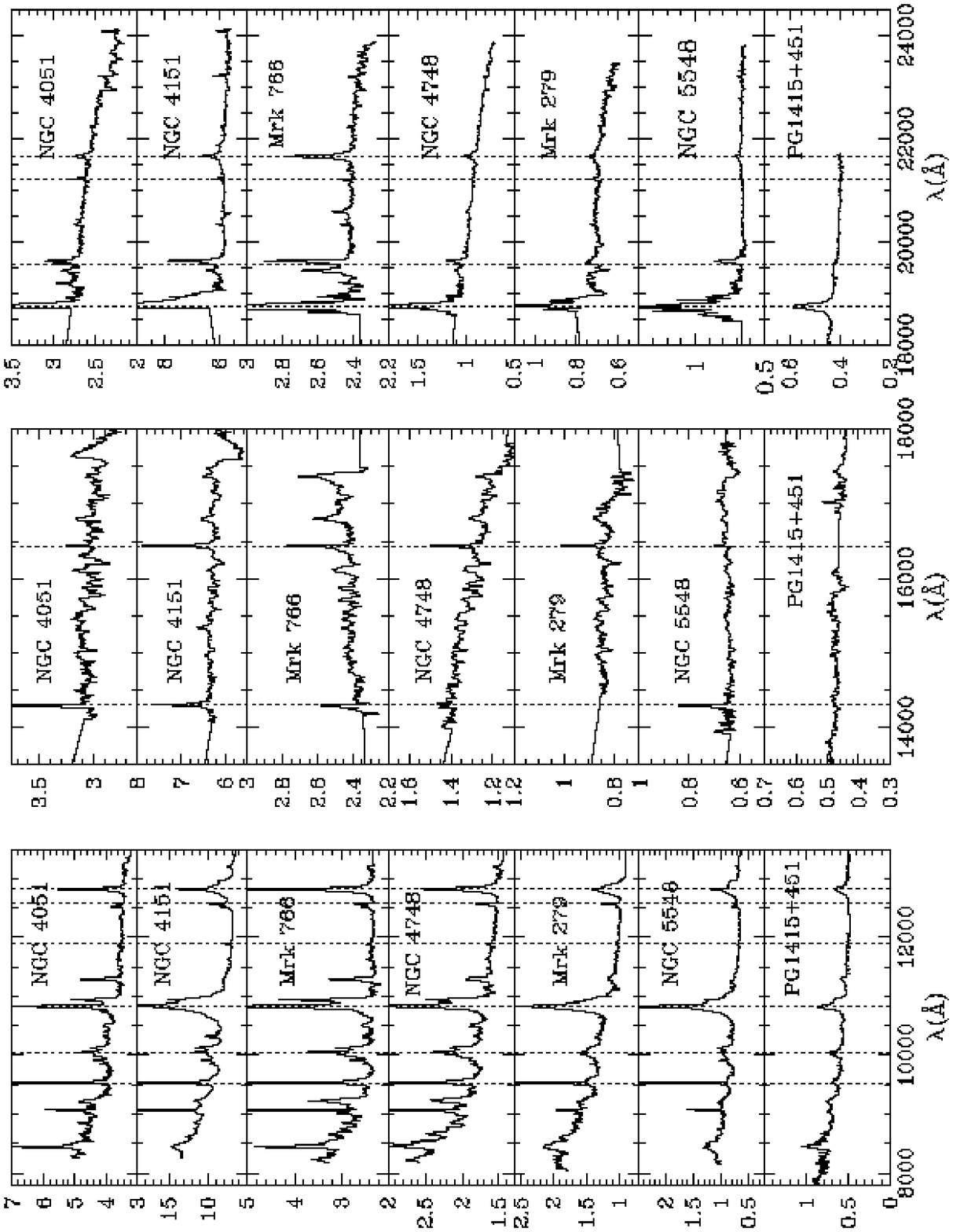}
\vspace{-0.5cm} 
\caption{Same as Fig.~\ref{indiv1}.}
\label{indiv4}
\end{figure}

\begin{figure}
\centering
\includegraphics[scale=0.8]{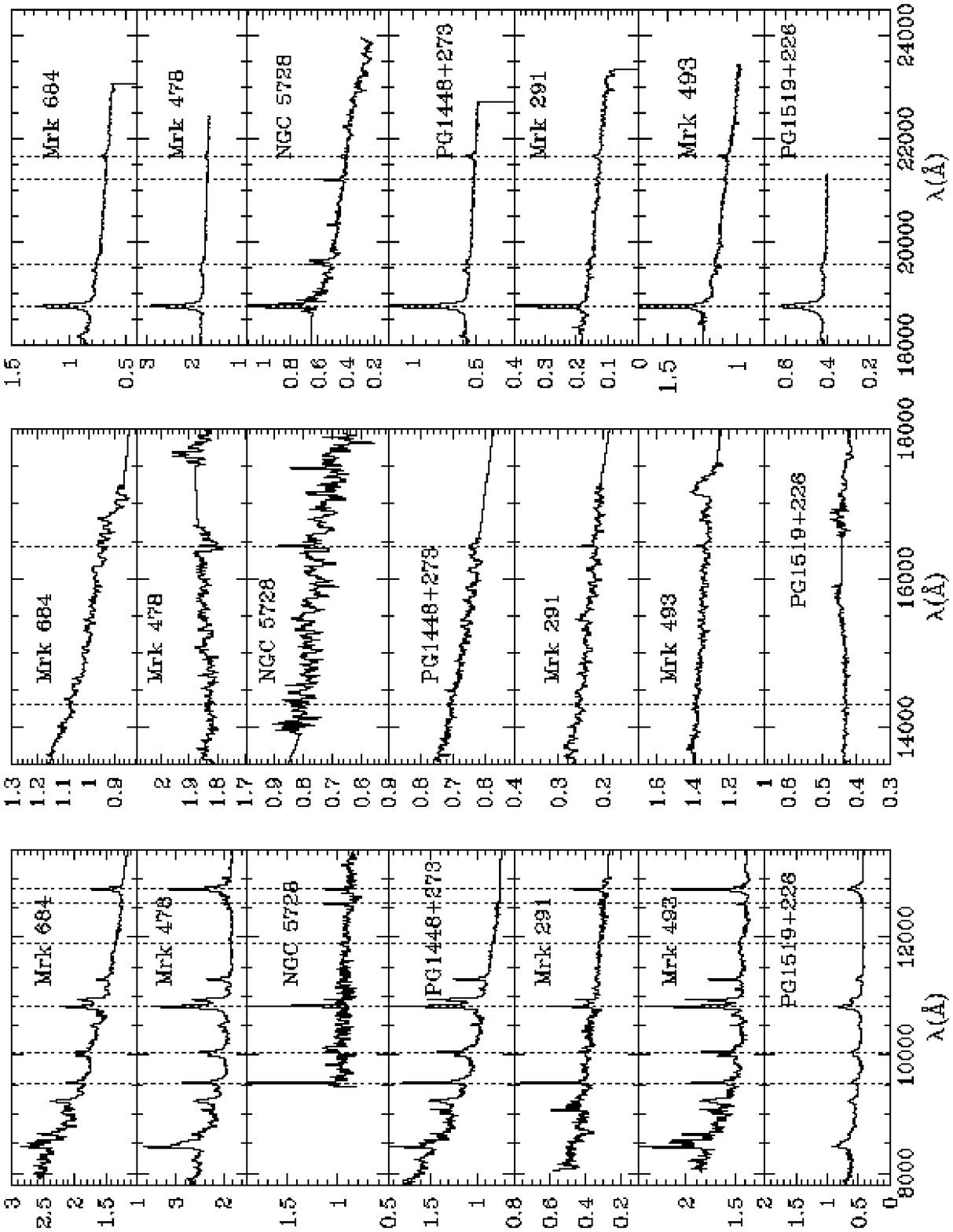}
\vspace{-0.5cm} 
\caption{Same as Fig.~\ref{indiv1}.}
\label{indiv5}
\end{figure}

\begin{figure}
\centering
\includegraphics[scale=0.8]{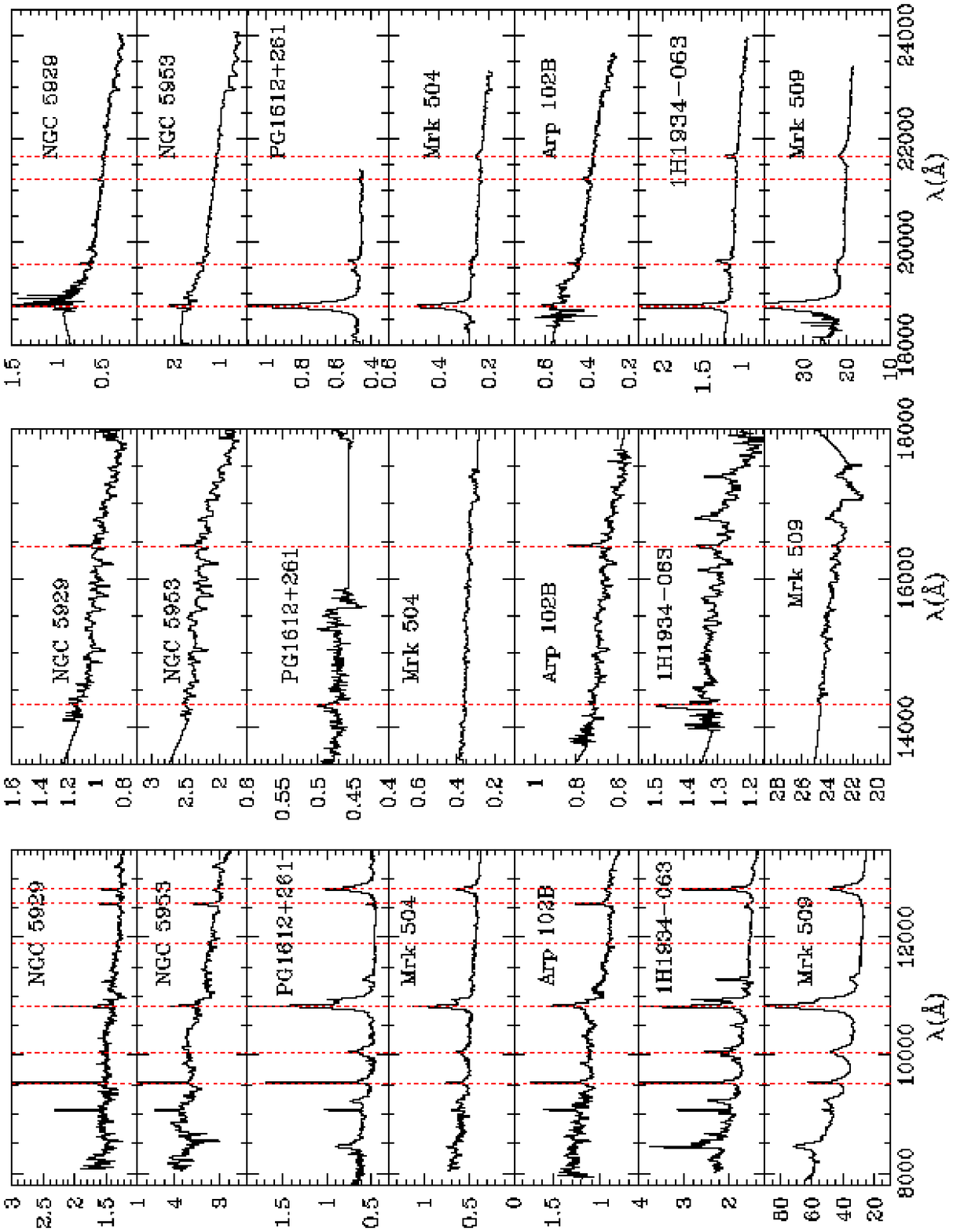}
\vspace{-0.5cm} 
\caption{Same as Fig.~\ref{indiv1}.}
\label{indiv6}
\end{figure}

\begin{figure}
\centering
\includegraphics[scale=0.8]{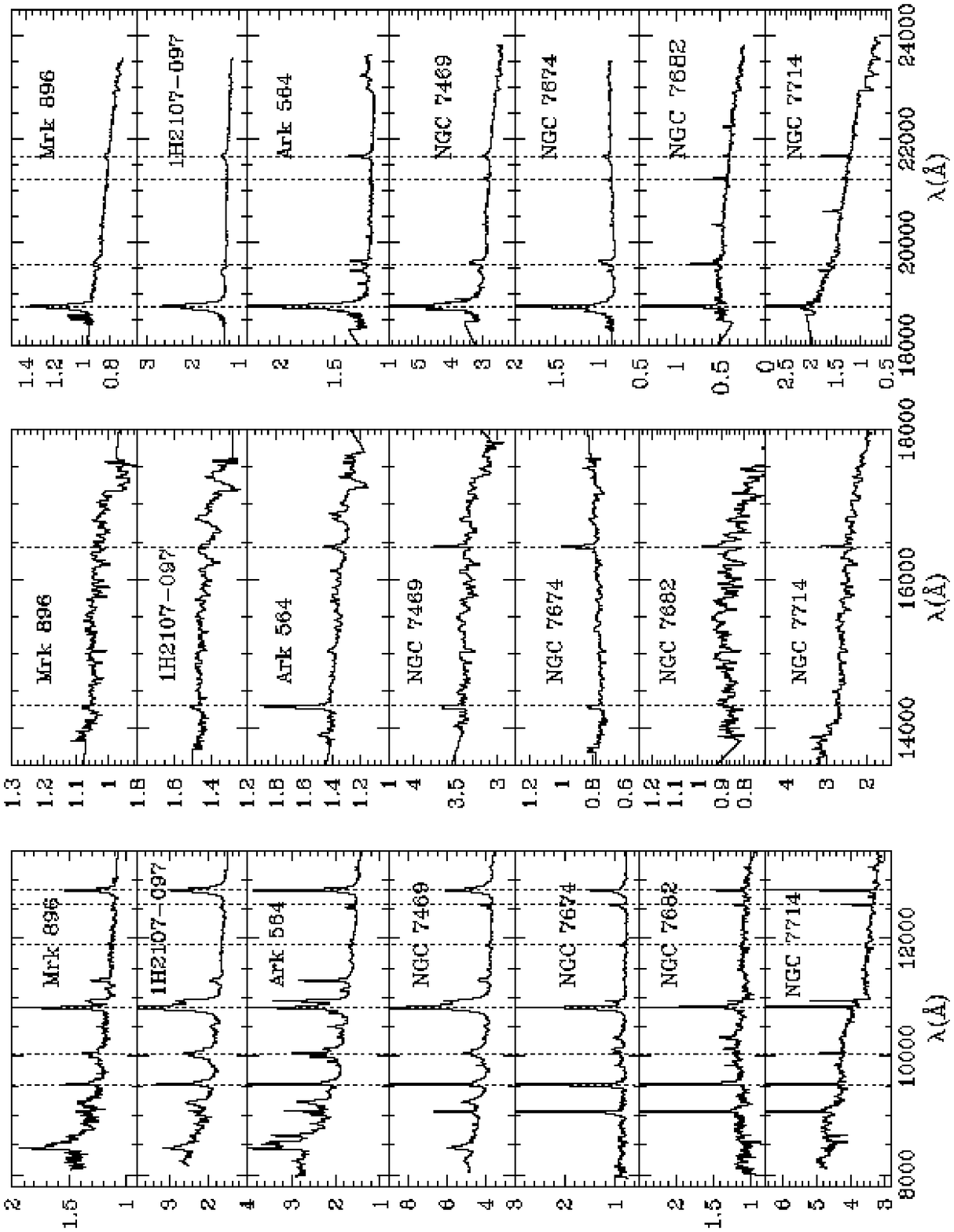}
\vspace{-0.5cm} 
\caption{Same as Fig.~\ref{indiv1}.}
\label{indiv7}
\end{figure}

\begin{figure}
\centering
\includegraphics[scale=0.8]{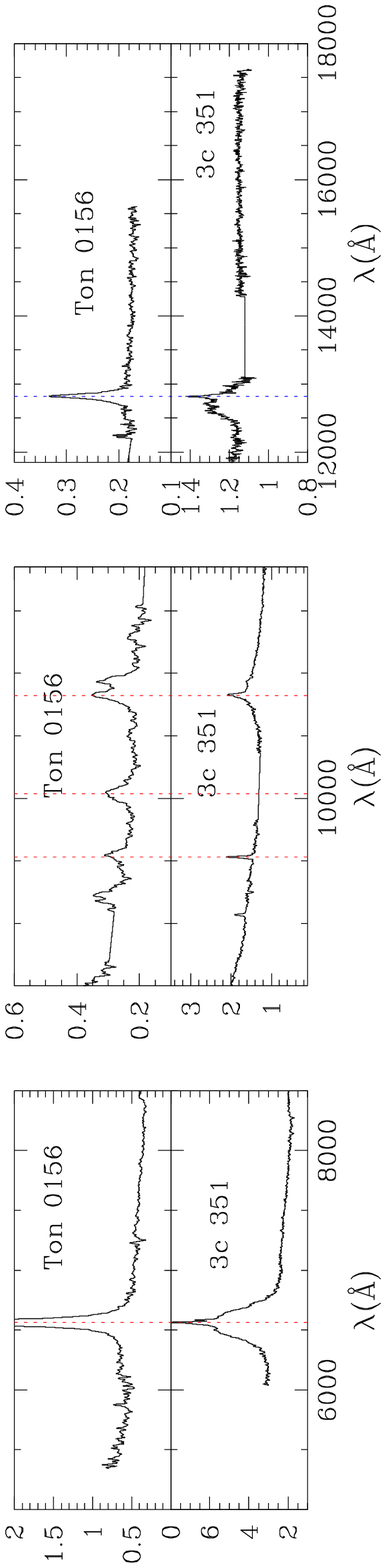}
\vspace{-0.5cm} 
\caption{Final reduced spectra for the two high-redshift galaxies in the 
Earth's frame. In the left panel we present the observed $z$+$J$ band, 
in the middle panel the observed $H$ band, and in the right panel the 
observed $K$ band. The abscissa is the flux in units of $\rm 10^{-15}\, 
erg \, cm^{-2} \, s^{-1}$. The dotted lines are: H$\alpha$ (left panel), 
\lb\ion{S}{iii}] 0.9531\,$\mu$m, Pa$\delta$, \ion{He}{i} 1.0830\,$\mu$m
(middle panel), and Pa$\beta$ (right panel).}
\label{indiv8}
\end{figure}

\begin{figure}
\centering
\includegraphics[scale=0.8]{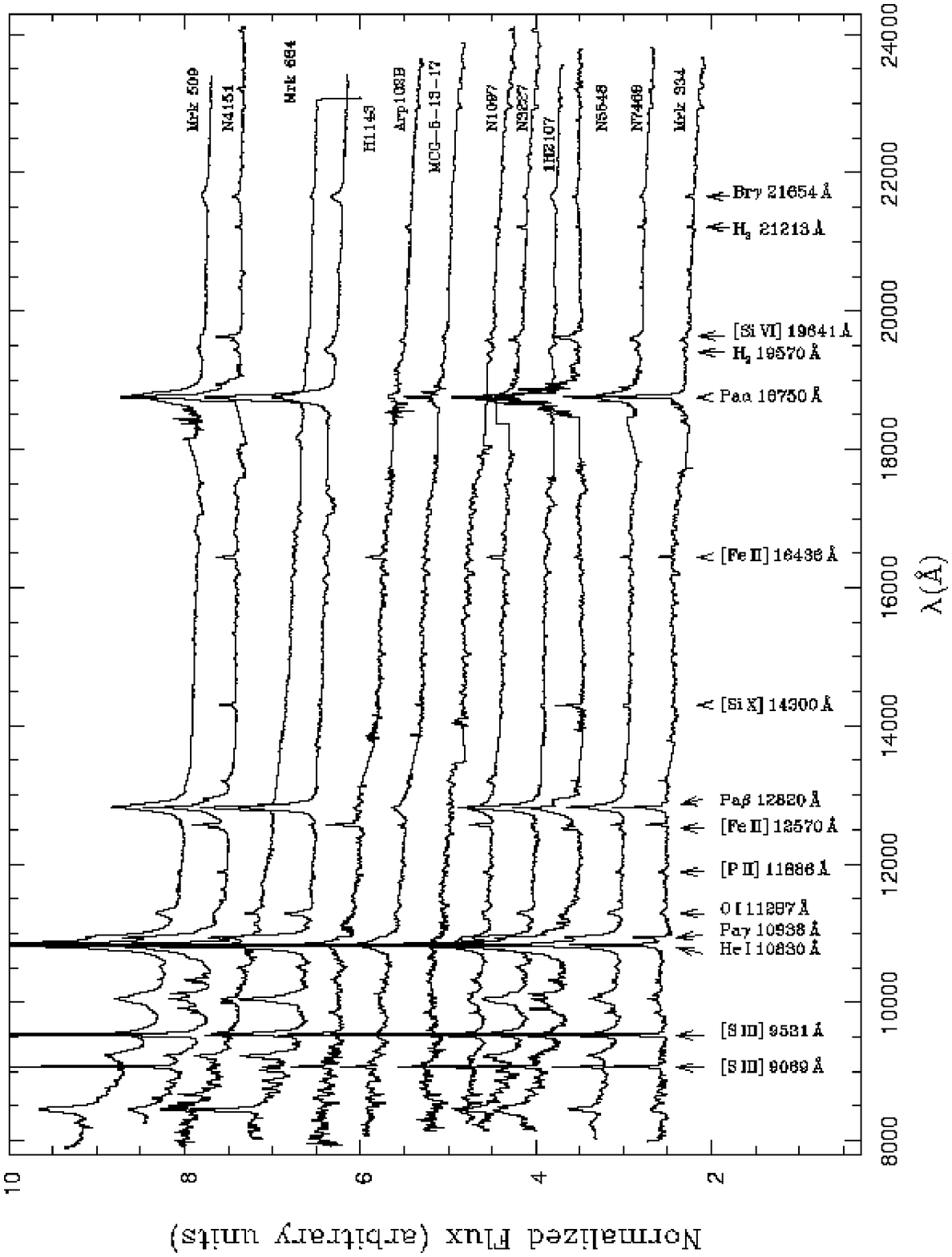}
\vspace{-0.5cm} 
\caption{Plot of normalized Sy~1's galaxies spectra ordered according to 
their shapes from a stepeer spectrum (top) to a flatter one  (bottom). 
Some emission lines are also identified. For more details see text.}
\label{plsy1}
\end{figure}

\begin{figure}
\centering
\includegraphics[scale=0.8]{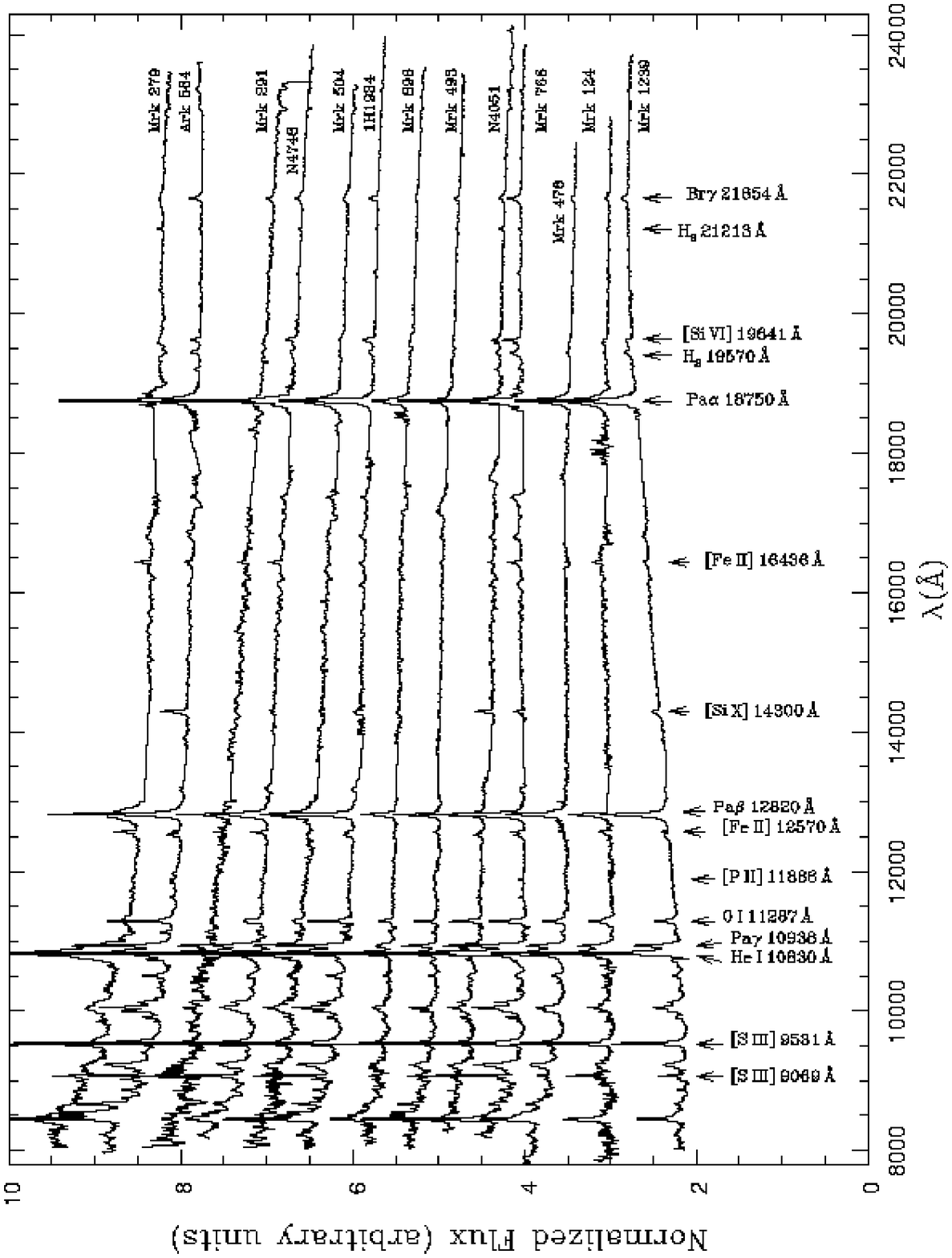}
\vspace{-0.5cm} 
\caption{Plot of normalized NLS1's galaxies spectra ordered according to their shapes 
from  a stepeer spectrum (top) to a flatter one  (bottom). 
Some emission lines are also identified. For more details see text.}
\label{plnls}
\end{figure}

\begin{figure}
\centering
\includegraphics[scale=0.8]{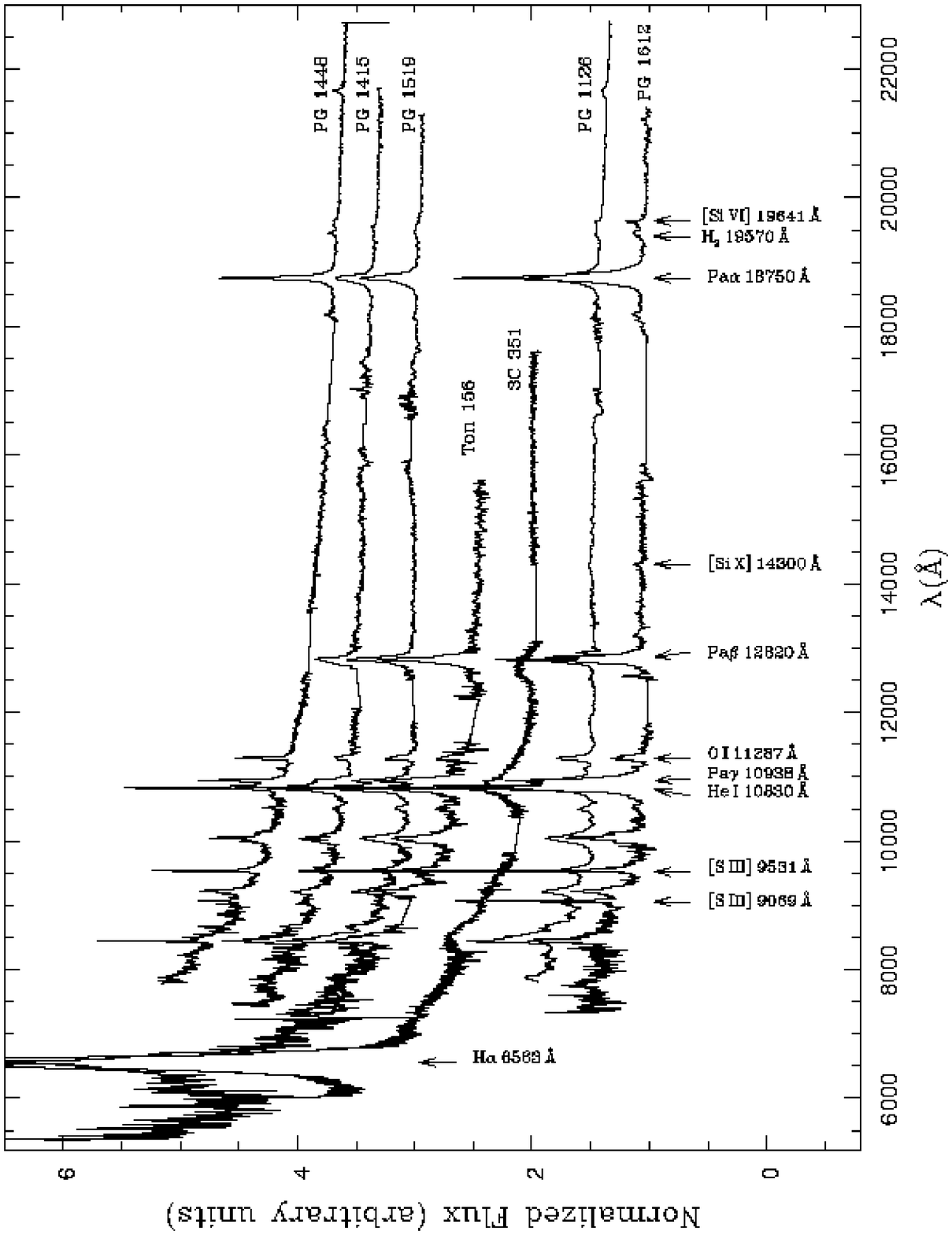}
\vspace{-0.5cm} 
\caption{Plot of normalized QSO's galaxies spectra ordered according to their shapes 
from a stepeer spectrum (top) to a flatter one  (bottom).
Some emission lines are also identified. For more details see text.}
\label{plqso} 
\end{figure}

\begin{figure}
\centering
\includegraphics[scale=0.8]{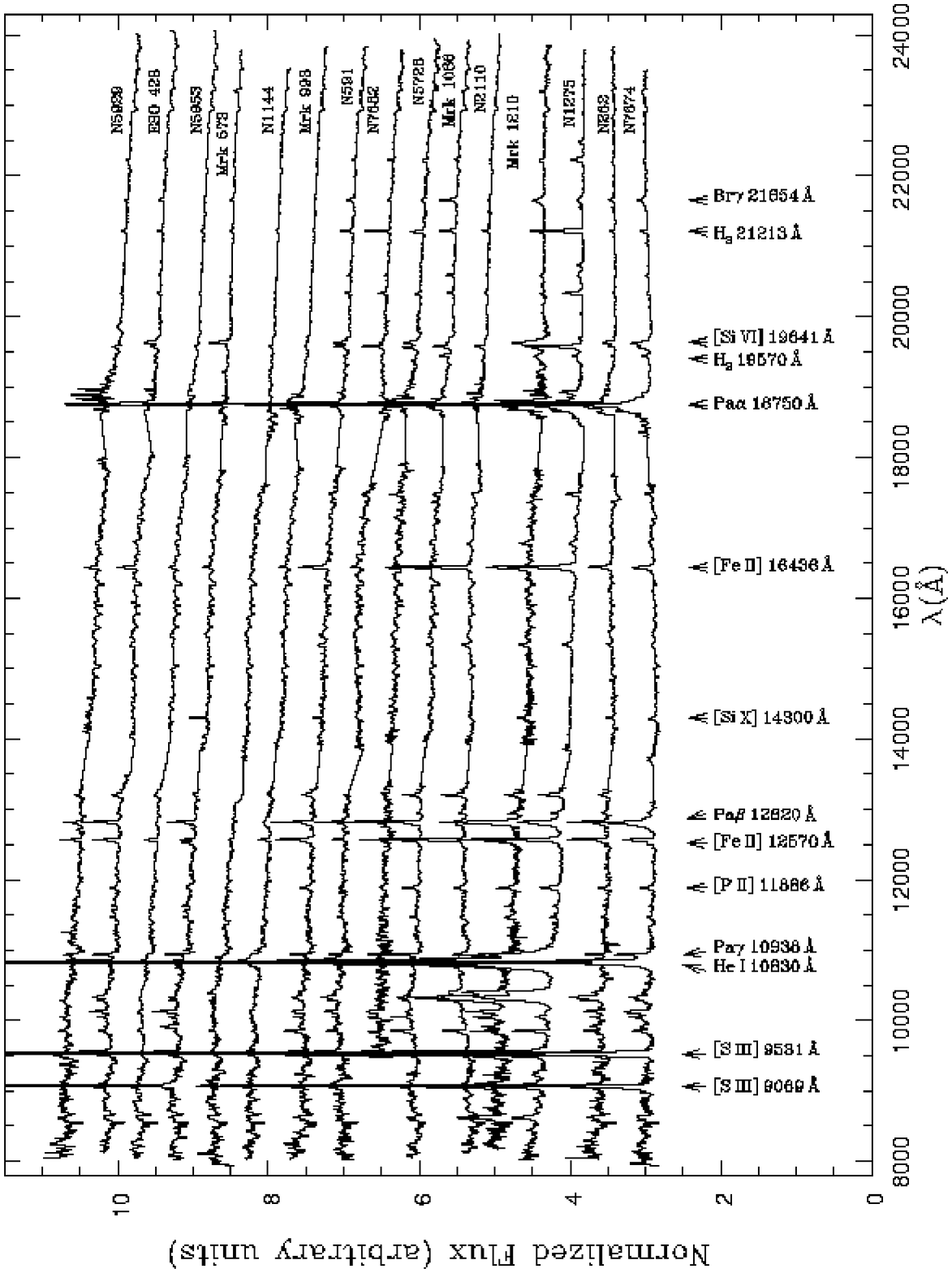}
\vspace{-0.5cm} 
\caption{Plot of normalized Sy~2's galaxies spectra ordered according to their shapes 
from a stepeer spectrum (top) to a flatter one  (bottom). 
Some emission lines are also identified. For more details see text.}
\label{plsy2}
\end{figure}

\begin{figure}
\centering
\includegraphics[scale=0.8]{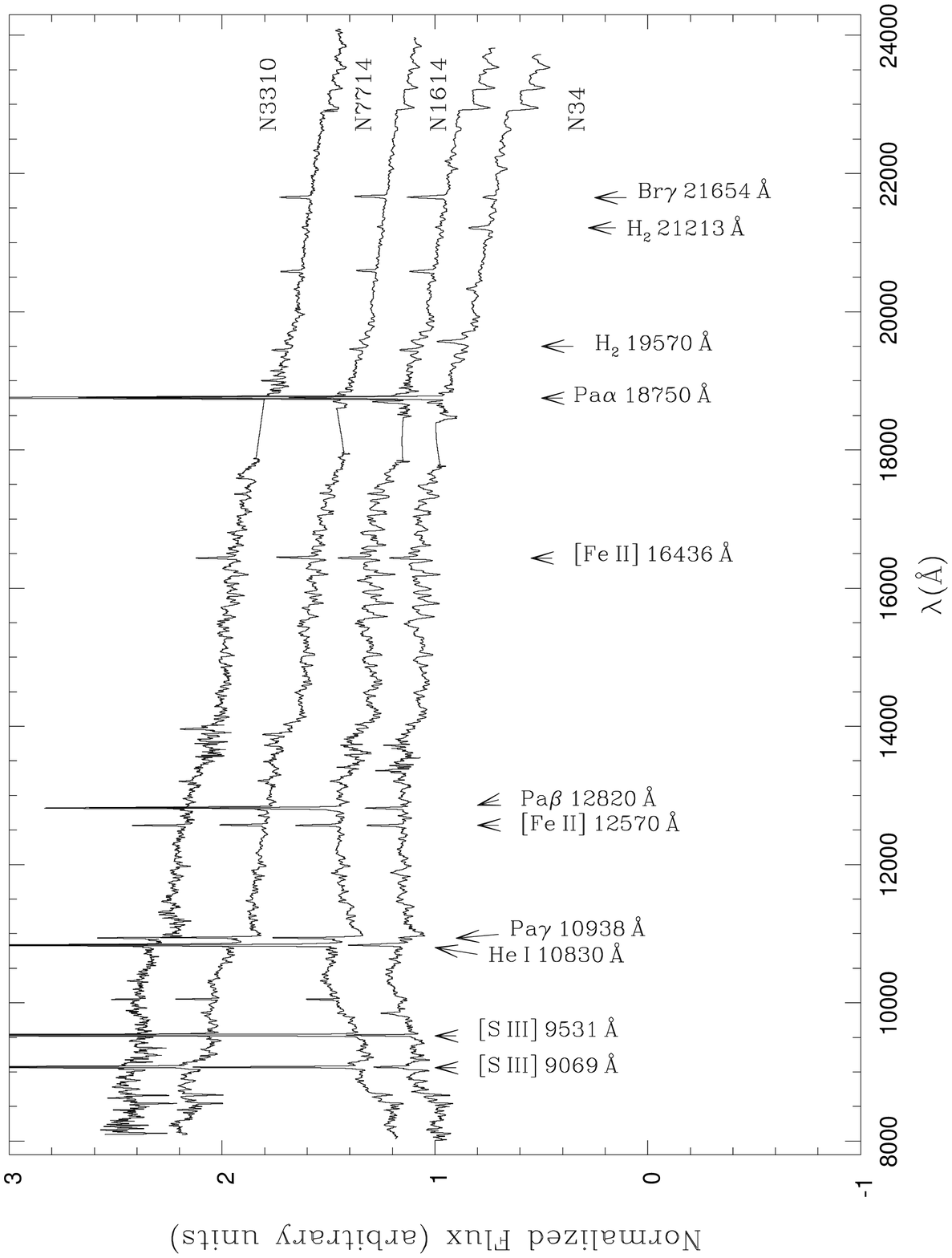}
\vspace{-0.5cm} 
\caption{Plot of normalized Starburst galaxies spectra ordered acording to their shapes 
from a stepeer spectrum (top) to a flatter one  (bottom). 
Some emission lines are also identified. For more details see text.}
\label{plsb}
\end{figure}

\end{document}